\documentclass[preprint,aps,floatfix,nofootinbib,a4paper]{revtex4-1}

\usepackage{amsmath, amssymb, amsfonts, amsthm, latexsym, epsfig, mathrsfs, xcolor, bbm, slashed}

\usepackage[inline]{enumitem}

\usepackage{setspace}
\usepackage[marginal, multiple]{footmisc}

\usepackage[colorlinks, allcolors=blue!70!black, linktocpage]{hyperref}

\usepackage[utf8]{inputenc}

\numberwithin{equation}{section}

\usepackage{cleveref}

\let\OLDthebibliography\thebibliography
\renewcommand\thebibliography[1]{%
	\setstretch{1.079} 
	\OLDthebibliography{#1}%
	\small %
	\setlength{\itemsep}{0.2\baselineskip} 
}

\let\OLDfootnote\footnote
\renewcommand\footnote[1]{%
	\setlength{\footnotesep}{0.75\baselineskip}%
	{\footnotesize \OLDfootnote{#1}}%
}

\setlist[enumerate]{noitemsep, label=(\arabic*), ref=(\arabic*)}

\renewcommand\thesection{\arabic{section}}
\renewcommand\thesubsection{\arabic{subsection}}

\makeatletter
\def\p@subsection{\thesection.}
\def\p@subsubsection{\thesection.\thesubsection.}
\makeatother 

\newlist{asslist}{enumerate}{1}
\setlist[asslist,1]{itemsep=0.1\baselineskip, label=(\arabic*), ref=(\arabic*)}
\crefname{asslisti}{Assump.}{Assumps.}

\newlist{propertylist}{enumerate}{1}
\setlist[propertylist,1]{noitemsep, label=(\arabic*), ref=(\arabic*)}
\crefname{propertylisti}{Property}{Properties}

\theoremstyle{plain}
\newtheorem{thm}{Theorem}
\newtheorem{lemma}{Lemma}[section]
\newtheorem{prop}{Proposition}[section]
\newtheorem{corollary}{Corollary}[section]

\theoremstyle{definition}
\newtheorem{definition}{Definition}[section]
\newtheorem{ass}{Assumptions}[section]
\newtheorem{property}{Properties}[section]

\theoremstyle{remark}
\newtheorem{remark}{Remark}[section]

\crefname{equation}{Eq.}{Eqs.}
\creflabelformat{equation}{#2#1#3}

\crefname{section}{\S}{\S}
\crefname{appendix}{\S}{\S}
\crefname{figure}{Fig.}{Figs.}

\crefname{definition}{Def.}{Defs.}
\crefname{prop}{Prop.}{Props.}
\crefname{lemma}{Lemma}{Lemmas}
\crefname{corollary}{Cor.}{Cors.}
\crefname{thm}{Theorem}{Theorems}
\crefname{remark}{Remark}{Remarks}

\crefname{ass}{Assumptions}{Assumptions}
\crefname{property}{Properties}{Properties}

\newcommand{\be}{\begin{equation}}
\newcommand{\ee}{\end{equation}}

\newcommand{\lb}{\left}
\newcommand{\rb}{\right}

\newcommand{\mc}{\mathcal}

\newcommand{\ms}{\mathscr}
\newcommand{\mf}{\mathfrak}
\newcommand{\bb}{\mathbb}

\newcommand{\eqsp}{\hspace{10pt};\hspace{10pt}} 

\newcommand{\hr}{\begin{center}* * *\end{center}}

\newcommand{\id}{\mathbbm 1} 

\newcommand{\union}{\cup} 
\newcommand{\inter}{\cap} 

\newcommand{\surgrav}{\kappa} 

\newcommand{\Lie}{\pounds} 
\newcommand{\defn}{\mathrel{\mathop:}=} 
\newcommand{\dfM}{\underline} 

\newcommand{\df}[1]{\boldsymbol{#1}}

\newcommand{\degf}[1]{{\rm deg}(#1)} 

\newcommand{\dirac}{\slashed D} 
\newcommand{\adj}{\overline} 

\begin{document}

\setstretch{1.2}


\title{The First Law of Black Hole Mechanics for Fields with Internal Gauge Freedom}

\author{Kartik Prabhu}
\email{kartikp@uchicago.edu}
\affiliation{Enrico Fermi Institute and Department of Physics,\\
The University of Chicago, Chicago, IL 60637, USA}

\begin{abstract}
	We derive the first law of black hole mechanics for physical theories based on a local, covariant and gauge-invariant Lagrangian where the dynamical fields transform non-trivially under the action of some internal gauge transformations. The theories of interest include General Relativity formulated in terms of tetrads, Einstein-Yang-Mills theory and Einstein-Dirac theory. Since the dynamical fields of these theories have some internal gauge freedom, we argue that there is no natural group action of diffeomorphisms of spacetime on such dynamical fields. In general, such fields cannot even be represented as smooth, globally well-defined tensor fields on spacetime. Consequently the derivation of the first law by Iyer and Wald cannot be used directly. Nevertheless, we show how such theories can be formulated on a principal bundle and that there is a natural action of automorphisms of the bundle on the fields. These bundle automorphisms encode both spacetime diffeomorphisms and internal gauge transformations. Using this reformulation we define the Noether charge associated to an infinitesimal automorphism and the corresponding notion of stationarity and axisymmetry of the dynamical fields. We first show that we can define certain potentials and charges at the horizon of a black hole so that the potentials are constant on the bifurcate Killing horizon, giving a generalised zeroth law for bifurcate Killing horizons. We further identify the gravitational potential and perturbed charge as the temperature and perturbed entropy of the black hole which gives an explicit formula for the perturbed entropy analogous to the Wald entropy formula. We then obtain a general first law of black hole mechanics for such theories. The first law relates the perturbed Hamiltonians at spatial infinity and the horizon, and the horizon contributions take the form of a ``potential times perturbed charge" term.  We also comment on the ambiguities in defining a prescription for the total entropy for black holes.
\end{abstract}

\maketitle
\tableofcontents

\section{Introduction}\label{sec:intro}
	In \cite{LW} Lee and Wald provided a construction of the phase space, Noether current and Noether charge for Lagrangian field theories where the dynamical fields can be viewed as tensor fields on spacetime (or more generally maps from spacetime into another finite-dimensional manifold). Using this construction, Iyer and Wald \cite{W-noether-entropy, IW-noether-entropy} have given a derivation of the first law of black hole mechanics for arbitrary (in particular non-stationary) perturbations off a stationary axisymmetric black hole in any diffeomorphism covariant theory of gravity where the gravitational dynamical field is a Lorentzian metric and the matter fields are smooth tensor fields on the spacetime manifold. They also identified the corresponding entropy as the integral over a horizon cross-section of the Noether charge associated with the action of diffeomorphisms generated by the horizon Killing field. This formulation of the Noether charges and first law has been useful in analysing solutions to Einstein-matter theories \cite{SW1, SW2, SchW, Sch, MC1, MC2}, and stability of black holes \cite{HW, PW} and perfect fluid stars \cite{GSW}.\\

	Even though the results of \cite{LW, W-noether-entropy, IW-noether-entropy} encompass a wide variety of theories, there are situations of physical interest where their analysis cannot be directly applied. In particular, we would like to derive a first law of black hole mechanics for gravity formulated in terms of orthonormal coframes (i.e. vielbeins), Einstein-Yang-Mills theory and Einstein-Dirac theory. In all these cases the dynamical fields of the theory have some internal gauge freedom under the action of a group. As we will discuss below this internal gauge freedom is the main reason we cannot directly use the results of \cite{LW, W-noether-entropy, IW-noether-entropy} to derive a first law. Since all the ``fundamental" fields in the Standard Model of particle physics have such gauge freedom, it is of interest to formulate the first law when such dynamical fields are present in the theory. The main obstructions to using the formalism of \cite{LW, W-noether-entropy, IW-noether-entropy} for charged fields are as follows.

	The first obstruction is that fields with internal gauge transformations cannot, in general, be represented as globally smooth tensor fields on spacetime. A typical example from Maxwell electrodynamics is when the source is a magnetic monopole. In the presence of a magnetic monopole the Maxwell gauge field (or vector potential) \(A_\mu\) cannot be chosen to be smooth everywhere (in any choice of gauge; see Problem 2. \S~Vbis \cite{CDD-book}). As is well-known, this ``singularity" in the Maxwell gauge field is an artefact of trying to make a global choice of gauge. Even in the absence of monopoles, the most ``convenient" choice of gauge might make the Maxwell vector potential singular; see \cite{Gao-YM} for a discussion of the Reissner-Nordstr\"om black hole where, in the traditional choice of gauge, the vector potential is singular on the bifurcation surface. Similarly, in Yang-Mills theories the dyanmical gauge fields \(A_\mu^I\) might not be representable as smooth tensor fields on spacetime. The analysis of \cite{LW} assumes from the outset that a global choice of gauge can be made to represent Yang-Mills gauge fields as tensor fields on spacetime. To derive the first law for Yang-Mills theory, Sudarsky and Wald \cite{SW1, SW2} also assume that a choice of gauge has been made so that the ``gauge-fixed" fields are smooth everywhere, and moreover, are stationary (i.e. \(\Lie_{t} A_\mu^I = 0\)) in that choice of gauge. Similarly, for a coframe formulation of gravity (and also to describe spinor fields), one introduces a Lorentz gauge field (or spin connection) \(\omega_\mu{}^a{}_b\) which might not be smooth everywhere in some chosen gauge (see \S~III. \cite{JM}). In fact, for non-parallelisable manifolds, there does not even exist a globally smooth choice of coframes \(e^a_\mu\). In all of these cases the obstruction is that a globally smooth choice of gauge can only be made under certain topological restrictions. Even when we can make a global gauge choice it is far from obvious that a gauge choice can be made such that the gauge-fixed fields are stationary in that gauge (see  \S~4 \cite{Gao-YM} for a related discussion).  Thus, it is of interest to have a formulation of the first law for theories like Yang-Mills and coframe gravity that does not require a choice of gauge a priori.

	The second obstruction arises in defining the action of diffeomorphisms of spacetime on dynamical fields with internal gauge transformations. To formulate a first law of black hole mechanics, we need a notion of a stationary solution to the equations of motion. When the dynamical fields are usual tensor fields on the spacetime \(M\), there is a natural action of the diffeomorphism group of the spacetime \(M\) on tensor fields which can be used to define stationarity by the action of the Lie derivative with respect to the corresponding vector fields. However, when the dynamical fields transform under some internal gauge transformation, we cannot distinguish the action of a diffeomorphism of the spacetime manifold from the action of a diffeomorphism along with an arbitrary (spacetime dependent) internal gauge transformation. That is, we only have a notion of ``diffeomorphisms up to a gauge transformations". Similarly in general, a stationary (axisymmetric) black hole will only be ``stationary (axisymmetric) up to an internal gauge transformation" i.e. \(\Lie_{t} \psi = gauge\) where \(\psi\) denotes the charged dynamical fields.\footnote{Such a notion of ``stationarity up to gauge" was already used in \cite{JM-YM, Keir, RS, HSSJ} where nevertheless, the dynamical fields are assumed to be tensor fields on spacetime.} Thus, the full group of transformations of the dynamical fields of the theory is not simply a product group of diffeomorphisms and internal gauge transformations. Without such a separation of diffeomorphisms and internal gauge transformations we have to consider the full group of transformations to define the appropriate notion of Noether charge to obtain a first law.

	There have been numerous attempts to sidestep this problem. One approach is to use the usual Lie derivative on spacetime, ignoring the internal gauge structure of the fields \cite{CWE, DKG, CRGV, CRGV2}. A straightforward computation shows that such a Lie derivative (acting on the Yang-Mills gauge fields \(A_\mu^I\) for instance) depends on the gauge choice made. Thus any definition of Noether charges, stationarity and first law using such Lie derivatives on charged fields will also depend on the choice of gauge used. In \cite{SW1,SW2} stationarity of the Yang-Mills gauge fields was defined by requiring a global choice of gauge so that the gauge-fixed fields \(A_\mu^I\) on spacetime are annihilated by the Lie derivative along the stationary Killing field. It is far from obvious that such a global choice of gauge exists, even when a global gauge choice can be made, and as we will show, this assumption is actually a restriction on the types of Yang-Mills fields considered in \cite{SW1,SW2}.

	One could alternatively attempt to define the infinitesimal action of a diffeomorphism of spacetime by a ``gauge covariant Lie derivative" and use the vanishing of this Lie derivative to define stationarity and axisymmetry. For example in Ch.10 \cite{YM50} the action of a diffeomorphism along a vector field \(\dfM X \equiv X^\mu \in TM\) is defined through a ``gauge covariant Lie derivative" of a Yang-Mills gauge field \(A_\mu^I\) (here the \(I\) index only refers to the Yang-Mills Lie algebra) as
	\be\label{eq:YM-gauge-Lie}
		\hat \Lie_{\dfM X} A^I_\mu \defn X^\nu F_{\nu\mu}^I
	\ee
	where \(F_{\mu\nu}^I\) is the curvature \(2\)-form for the gauge field \( A_\mu^I\). Similarly, to obtain a first law for a coframe formulation of gravity, Jacobson and Mohd \cite{JM} use a \emph{Lorentz-Lie derivative} for the coframes by \footnote{\cite{JM} define the Lorentz-Lie derivative to act on arbitrary tensors, including the Lorentz connection, carrying a representation of the Lorentz group, but we only present the coframes to be brief. They also use the symbol \(\mc K\) to denote the Lorentz-Lie derivative in honour of Kosmann.} (see also \cite{BG} and references in \cite{JM})
	\be\label{eq:Lor-Lie}\begin{split}
		\hat\Lie_{\dfM X} e^a_\mu & \defn \Lie_{\dfM X} e^a_\mu + \lambda^a{}_b e^b_\mu  \\
		\text{with}\quad \lambda^{ab} & \defn E^{\mu [a}\Lie_{\dfM X} e^{b]}_\mu = X^\mu \omega_\mu{}^{ab} + E^{\mu[a} e^{b]}_\nu \nabla_\mu X^\nu
	\end{split}\ee
	where the \(\Lie_{\dfM X}\) is computed ignoring the internal indices and \(\omega_\mu{}^{ab}\) is the Lorentz connection on spacetime. Likewise, there have been many attempts to define a Lie derivative for spinor fields (viewed as fields on spacetime). A definition of Lie derivative for spinors  with respect to a Killing field of the metric was put forth by Lichnerowicz \cite{Lich}, and then generalised for arbitrary vector fields by Kosmann \cite{Kos} by prescribing that one use the same formula as that given by Lichnerowicz but for any vector field (see also \cite{BG} and Supplement~2. \cite{CD-book}). The Kosmann prescription can be formalised on a principal bundle through the notion of a \emph{Kosmann lift} (see \cite{FF, GM} and the references therein).\footnote{Also note that the ``spinorial Lie derivative" prescription in \cite{LRW} annihilates the metric for any vector field --- making every vector field a ``Killing field" and all spacetimes stationary --- which is clearly not desirable.} The \emph{Lichnerowicz-Kosmann-Lie derivative} acts on Dirac spinor fields \(\Psi\) according to
	\be\label{eq:LK-Lie}
	\hat\Lie_{\dfM X} \Psi \defn X^\mu D_\mu \Psi - \tfrac{1}{8} \nabla_{[\mu}X_{\nu]}[\gamma^\mu, \gamma^\nu] \Psi
	\ee
	where \(D_\mu\) is the covariant spin derivative on Dirac spinor fields with respect to a torsionless spin connection. Even though these definitions of Lie derivative are gauge covariant, it is a straightforward computation to verify that
	\be\label{eq:Lie-comm-gauge}
		[\hat\Lie_{\dfM X}, \hat\Lie_{\dfM Y}] = \hat\Lie_{[{\dfM X},{\dfM Y}]} + gauge
	\ee
	and the \(gauge\)-term does not vanish except when
	\begin{enumerate*}
	\item \(\dfM{\df F}^I = 0\) for ``Lie derivative" in \cref{eq:YM-gauge-Lie}
	\item either \(X^\mu\) or \(Y^\mu\) is a conformal Killing field of the metric for the Lorentz-Lie derivative \cref{eq:Lor-Lie} and the Lichnerowicz-Kosmann-Lie derivative \cref{eq:LK-Lie} (see \cite{BG} for a proof).
	\end{enumerate*}
	Thus even though, the linear maps \(X^\mu \mapsto \hat{\Lie}_{\dfM X}\) in \cref{eq:YM-gauge-Lie}-\cref{eq:LK-Lie} are gauge-covariant, none of the above prescriptions for the Lie derivative implement the Lie algebra for the diffeomorphism group of spacetime and the Noether currents derived from these notions of a Lie derivative cannot be interpreted as Noether currents associated to diffeomorphisms.

	Further, if the dynamical fields \(\psi\) of the theory are chosen to be stationary with respect to these modified Lie derivatives, i.e. \(\hat{\Lie}_{\dfM X}\psi = 0 \) then they are obviously ``stationary up to internal gauge transformations". But it might not be possible to choose a globally smooth gauge representative of the dynamical fields so that they are stationary in this modified sense, particularly when we cannot make a global choice of gauge.\\

	The main aim of this work is to address these issues directly, by formulating physical theories with charged dynamical fields on a \emph{principal bundle} (see \cref{sec:bundles} for details). All the charged fields are legitimate (globally well-defined and smooth) tensor fields on the principal bundle. A smooth global choice of gauge exists only when the principal bundle is trivial, and only in that case can we write the charged fields as smooth tensor fields on spacetime. However, working directly on a principal bundle avoids the issue of making any choice of gauge, and thus, we can handle theories defined on non-trivial principal bundles where there is no way to represent the fields as smooth tensor fields on spacetime. We shall similarly formulate the Lagrangian, Noether currents and charges directly on the principal bundle without making any gauge choices. The principal bundle also provides the necessary structure to consider the full group of transformations of the dynamical fields as the group of automorphisms of the bundle manifold. The automorphisms of the bundle then encode both diffeomorphisms and internal gauge transformations, and we will not need to ``artificially" single out the action of just diffeomorphisms. Infinitesimal actions of these automorphisms are then generated by (standard) Lie derivatives with respect to vector fields on the bundle. Using this we define the appropriate notions of ``stationarity (and axisymmetry) up to gauge transformations" for charged dynamical fields as automorphisms of the bundle that project down to the stationary (axisymmetric) diffeomorphisms of spacetime (see \cref{def:stationary-axisymm}). We then generalise the constructions of \cite{LW} to define the symplectic form on arbitrary (non-stationary) perturbations of the dynamical fields, and the Noether current and Noether charge associated with any bundle automorphism. Next, we describe the main results of this paper, which include a derivation of a generalised zeroth law for bifurcate Killing horizons, and the first law of black hole mechanics for stationary and axisymmetric black holes which arise from solutions to the equations of motion, for theories with charged dynamical fields.\\

	We are interested in static or stationary-axisymmetric, asymptotically flat black hole spacetimes with bifurcate Killing horizons determined by dynamical fields which have non-trivial internal gauge transformations. We expect our results can be generalised to spacetimes with different asymptotics but we stick to the asymptotically flat case. We refer the reader to \cref{sec:zeroth-law} for a more detailed description of the spacetimes under consideration.

	As our first result we show in \cref{thm:zeroth-law} that on any bifurcate Killing horizon in spacetime (not necessarily a solution to any equations of motion) we can define certain \emph{potentials} \(\ms V^\Lambda\) at the horizon which are constant along the entire bifurcate Killing horizon. This can be viewed as a generalised zeroth law for bifurcate Killing horizons. These potentials are defined solely in terms of the dynamical gauge fields of the theory (for instance, a Yang-Mills gauge field or the Lorentz gauge field for gravity) and get no direct contributions from any other matter fields. In \cref{cor:potential-charge}, we show that the perturbed Hamiltonian \(\delta H_K\) associated to the horizon Killing field \(K^\mu\) at the bifurcation surface can be put into a ``potential times perturbed charge" form where the \emph{charges} are determined by the dependence of the Lagrangian on the curvatures of the dynamical gauge fields of the theory. Then, in \cref{thm:temp-entropy}, we provide a new perspective on the temperature \(T_{\ms H}\) and perturbed entropy \(\delta S\) of the black hole by identifying them with the potential and perturbed charge, respectively, corresponding to the Lorentz connection in a first-order formulation of gravity. Thus, the temperature and perturbed entropy can be viewed on the same footing as any other potentials and perturbed charges of any matter gauge fields (like Yang-Mills gauge fields) in the theory. This also gives us an explicit formula for the perturbed entropy in direct parallel with the \emph{Wald entropy} formula \cite{W-noether-entropy, IW-noether-entropy}.

	Our main result (\cref{thm:first-law-gen}) is a general formulation of the first law of black hole mechanics for theories with charged dynamical fields, where the dynamical fields solve the equations of motion obtained from a gauge-invariant Lagrangian. The first law is obtained as an equality between the perturbed boundary Hamiltonian \(\delta H_K\) associated to the horizon Killing field \(K^\mu\) evaluated at the bifurcation surface and at spatial infinity, and takes the form
	\be
	T_{\ms H}\delta S + \ms V'^\Lambda \delta\ms Q'_\Lambda = \delta E_{can} - \Omega_{\ms H}^{(i)}~ \delta J_{(i),can}
	\ee
	for any perturbation which solves the linearised equations of motion off a stationary, axisymmetric (up to internal gauge transformations) black hole background which solves the equations of motion. The left-hand-side terms are the potentials and charges of the black hole on the bifurcation surface defined in \cref{thm:zeroth-law} and \cref{cor:potential-charge}. The first term consists of the temperature and perturbed entropy of the black hole, identified with the gravitational potential and perturbed charge (\cref{thm:temp-entropy}), while the second term is the contribution of the non-gravitational gauge fields (such as Yang-Mills fields). The quantities on the right-hand-side are the perturbed canonical energy (associated to the stationary Killing field \(t^\mu\)) and angular momenta (associated to the axial Killing fields \(\phi^\mu_{(i)}\); here the index \((i)\) is used to denote the multiple axial Killing fields in greater than \(4\)-spacetime dimensions) defined at spatial infinity (\cref{eq:E-J-can}). The form of the perturbed canonical energy and angular momenta at infinity depends on the theory under consideration and also the asymptotic fall-off conditions on the fields, and they contain contributions from both the gravitational dynamical fields and other matter dynamical fields in the theory. For instance, in Einstein-Yang-Mills theory, \(\delta E_{can}\) contains both the perturbed ADM mass and a ``potential times perturbed charge" term from the Yang-Mills gauge field at infinity (see \cref{sec:yang-mills}). Similarly, \(\delta J_{can}\) contains both the perturbed ADM angular momentum and the perturbed angular momentum of the Yang-Mills fields.

	Note that for Einstein-Yang-Mills theory, Sudarsky and Wald \cite{SW1} get a vanishing Yang-Mills potential term at the horizon because of their assumption that there exists a smooth choice of gauge such that the gauge-fixed Yang-Mills fields are stationary \(\Lie_{K} A_\mu^I = 0\). We will argue that in general such a gauge choice cannot be made, and the ``potential times perturbed charge" at the horizon can be set to vanish only in special situations (at the cost of changing the contributions to perturbed canonical energy and angular momenta at infinity; see \cref{rem:YM-potential-amb}). The existence of this non-vanishing term at the horizon was also pointed out in \cite{Gao-YM}, though they could not write the term in terms of potentials and perturbed charges for non-abelian Yang-Mills fields.

	We also show that the ambiguities in defining the Noether charge for a Lagrangian do not affect the first law and the perturbed entropy. We also discuss the ambiguities in defining a total entropy for a stationary axisymmetric black hole. We argue that a second law of black mechanics could fix at least some of these ambiguities in the total entropy. Since we do not know of a general derivation of the second law for arbitrary theories of gravity (except in the case of General Relativity), we do not make an attempt to define the total entropy or a notion of dynamical black hole entropy in this paper.\\

	Even though we use a first-order coframe formulation for gravity, the general form of the first law described above is applicable to any Lagrangian theory for gravity where some fields with internal gauge freedom are considered as dynamical fields instead of the metric \cite{Peldan, Plebanski, CJDM} including, higher-derivative theories of gravity \cite{Sotiriou-thesis} with Lagrangians depending on torsion, curvature and finitely many of their derivatives. One can also include ``non-metricity", metric-affine theories of gravity \cite{Hehl, BH} by a simple extension of the formalism. On the matter side, we can include all the charged matter fields in the Standard Model of particle physics. We also expect that our results can be generalised to include supersymmetric theories following, for instance, \cite{HM}.

	Despite this generality, there remain some potentially physically interesting theories that are not covered by our formalism. These include higher \(p\)-form gauge theories in the presence of magnetic charges (see \cite{Compere, Rogatko, Keir} for work in this direction) and Chern-Simons Lagrangians which are only gauge-invariant up to a total derivative term. The entropy contribution of gravitational Chern-Simons Lagrangians was computed from a spacetime point of view in \cite{Tach-CS, BCDPS-CS} (using a ``modified Lie derivative"), and in \cite{ALNR} (using a modification of the symplectic current). We shall defer the analysis of Chern-Simons theories from a bundle point of view to a forthcoming paper \cite{KP-CS}.

\hr

	The remainder of this work is organised as follows. We describe the principal bundle formulation of dynamical fields and gauge-invariant Lagrangians in \cref{sec:fields-L-bundle} and give a general form for the symplectic potential, symplectic current, and Noether charge for any bundle automorphism (i.e. combined diffeomorphisms and gauge transformations) for such theories. We define the horizon potentials and charges, and derive a generalised zeroth law for bifurcate Killing horizons in \cref{sec:zeroth-law}. In \cref{sec:first-law} we give a formulation of the first law of black hole mechanics for the theories under consideration. In \cref{sec:examples} we use this formalism to derive a first law for General Relativity in a first-order tetrad formulation, Einstein-Yang-Mills theory and Einstein-Dirac theory. The appendix \cref{sec:math} collects some technical definitions and formulae along new results which will be used in the main arguments of the paper.

\section*{Notation}

	We will use an abstract index notation for vector spaces and tensor fields whenever convenient. Tensor fields on spacetime (or some base space for a principal bundle) will be denoted by indices \(\mu,\nu,\lambda,\ldots\) from the middle of the lower case Greek alphabet, e.g. \(X^\mu\) is a vector field and \(\sigma_\mu\) is a covector field on spacetime. Similarly, lower case Latin indices \(m,n,l,\ldots\) denote tensors on a principal bundle, e.g. \(X^m\) is a vector field and \(\sigma_m\) is a covector field on a principal bundle. 

	It will be convenient to often use an index-free notation for differential forms and vector fields. We will use the factor and sign conventions of Wald \cite{Wald-book} when translating differential forms to and from an index notation and use the symbol \(\equiv\) to denote such a translation. When using an index-free notation we denote differential forms by a bold-face symbol, for instance, a differential \(k\)-form on a principal bundle is denoted by \( \df\sigma \equiv \sigma_{m_1 \ldots m_k} = \sigma_{[m_1 \ldots m_k]} \) and for a vector field \(X^m\), we denote the  \emph{interior product} by \(X \cdot \df\sigma \equiv X^{m_1}\sigma_{m_1 \ldots m_k}\) and the \emph{Lie derivative} by \(\Lie_X \df\sigma = X \cdot d\df\sigma + d (X\cdot \df\sigma)\). When using an index-free notation (and for scalars which have no indices) we will also use an underline to distinguish between functions, differential forms and vector fields on the base space from those on the principal bundle i.e. \(\dfM \varphi\) is a function, and \(\dfM{\df \sigma} \equiv \sigma_{[\mu_1\ldots \mu_k]}\) a \(k\)-form, respectively on \(M\), and \(\dfM X \cdot \dfM{\df \sigma}\) is the interior product, on the base space.

	Upper case indices \(I,J,K,\ldots\) from middle of the Latin alphabet will denote elements of a finite-dimensional Lie algebra \(\mf g\) e.g. \(X^I\) is an element of a Lie algebra and \({c^I}_{JK} = {c^I}_{[JK]}\) denotes the \emph{structure constants}. We write the Lie bracket on \(\mf g\) as \([X,Y]^K = {c^K}_{IJ}X^I Y^J\). The \emph{Killing form} on \(\mf g \) is a bilinear, symmetric form defined by
	\be\label{eq:killing-form}
		k_{IJ} \defn {c^L}_{IK}{c^K}_{JL}
	\ee
	The Killing form is invariant under the adjoint action of the group \(G\) on its Lie algebra \(\mf g\). It is non-degenerate if and only if \(\mf g\) is \emph{semisimple}, and hence defines a metric on the Lie algebra. Further, when the group \(G\) is \emph{compact} the Killing form is \emph{negative} definite. Throughout the paper we stick to the semisimple case and use \(k_{IJ}\) and its inverse \(k^{IJ}\) to raise and lower the abstract indices on elements of \(\mf g\).

	Upper case letters \(A,B,\ldots\) from the beginning of the Latin alphabet denote elements of a vector space \(\bb V\) with some representation \(R\) of a finite-dimensional group \(G\). The action of any \(g \in G\) on any element \(\varphi^A \in \bb V\) under the representation \(R\) is denoted by \(R(g)\varphi\); omitting the indices for simplicity. The corresponding action \(r\) of a Lie algebra element \(X^I \in \mf{g}\) is denoted (using the abstract index notation) by \(X^I {{r_I}^A}_B \varphi^B\). We shall also use \(\alpha,\beta,\ldots\) as indices to denote some collection of fields, each of which can be tensor fields valued in different vector spaces (for example in \cref{eq:chi-defn,eq:psi-defn}).

\section{Dynamical fields and Lagrangian theories on a principal bundle}\label{sec:fields-L-bundle}

	Fields with internal gauge freedom under the action of a group \(G\) (which we assume to be semisimple) are usually written as tensor fields on spacetime \(M\) valued in some vector space which transform under a representation of \(G\). As noted in the Introduction, in general such fields can only be represented locally as smooth tensor fields. Since we are interested in the first law of black hole mechanics which is a global equality relating quantities defined at the horizon to those defined at spatial infinity, it will be very convenient to have globally well-defined smooth dynamical fields to describe the physical theory. Such fields with internal gauge transformations under some group \(G\) can be defined globally on a \emph{\(G\)-principal bundle} \(P\). For details of principal bundle formalism we refer to the classic treatment of \cite{K-conn, KN-book1, KN-book2, CDD-book, CD-book}.\footnote{Note that these references may use different conventions for numerical factors and signs for differential forms when converting to and from an index notation. Throughout this paper we use the conventions of \cite{Wald-book}.} We briefly recall the essential concepts needed below, while some new technical results are collected in \cref{sec:math}.

	Let \(\pi : P \to M\) be a \(G\)-principal bundle over \(M\). Denote the space of \emph{vertical vector fields} by \(VP\) containing vector fields \(X^m\) whose projection vanishes i.e. \((\pi_*)^\mu_m X^m = 0\). The space of \emph{horizontal} \(k\)-forms is denoted \(\Omega^k_{hor}P\) containing differential forms \(\df\sigma\) such that \(X \cdot \df\sigma = 0\) for all vertical vector fields \(X^m\). Recall that differential forms on the base space that are \emph{invariant} under internal gauge transformations are isomorphic to horizontal differential forms on the bundle i.e. \(\pi^*\Omega^kM \cong \Omega^k_{hor}P\) (see Example 5.1 in \S~II.5 \cite{KN-book1}). Similarly, the space of horizontal \(k\)-forms which are valued in a vector space \(\bb V\) transforming under a representation \(R\) of the group \(G\) is denoted \(\Omega^k_{hor}P(\bb V, R)\); these correspond to gauge-covariant differential forms on spacetime.

 Since we are interested in theories with gravity described by some orthonormal coframes, we choose \(P\) to have the structure\footnote{Most of our results generalise straightforwardly to the more general case where the Lorentz bundle \(P_O\) is simply a subbundle of principal bundle \(P\).}
	\be\label{eq:P-split}
		P = P_O \oplus P'
	\ee
	where \(P_O\) is a Lorentz bundle, and \(P'\) is a principal bundle with structure group \(G'\) corresponding to other internal gauge transformations of the matter fields; the structure group of \(P\) is then \(G = O(d-1, 1) \times G'\). As described in \cref{sec:bundles}, the coframes \(\df e^a\) are horizontal forms on \(P\), while the frames \(E_a^m\) are represented by vector fields.  The gauge fields of the theory are represented by a connection \(\df A^I\) on \(P\) which is a \(1\)-form valued in the Lie algebra \(\mf g\) transforming in the adjoint representation \(\rm{Ad}\). For the bundle \(P\) in \cref{eq:P-split} it is of the form 
	\be\label{eq:conn-split}
	\df A^I = \begin{pmatrix}\df \omega^a{}_b, \df A'^{I'}\end{pmatrix} \in \Omega^1P(\mf g, \rm{Ad})
	\ee
	where \(\df \omega^a{}_b\) is a Lorentz \(SO(d-1,1)\)-connection on \(P_O\) and \(\df A'^{I'}\) is a \(G'\)-connection on \(P'\).\\

	Now we describe our strategy to write charged fields with internal gauge transformations as tensor fields on the bundle \(P\) instead of the spacetime \(M\). First let us consider the case of a charged scalar field i.e. a (local) function \(\dfM \varphi^A\) on \(M\) valued in some vector space \(\bb V\) which has a representation \(R\) of the internal gauge group \(G\). The field \(\dfM \varphi^A\) is represented on the principal bundle \(P\) as an \emph{equivariant} function valued in \(\bb V\) i.e. by \(\varphi^A \in \Omega^0P(\bb V, R) \). The main example of such a charged scalar field we'll consider is the Dirac spinor field in \cref{sec:dirac} in which case the group \(G\) is the spin group \(Spin^0(3,1)\) and the corresponding bundle is a spin bundle \(P_{Spin}\) (described in \cref{sec:bundles}). Another example of such a field is the Higgs field of the Standard Model where the group \(G\) is taken to be \(SU(2) \times U(1)\).\\

	To write more general charged tensor fields on \(P\) it will be convenient to express them in terms of their frame components as follows. Let \(\sigma_\mu^A\) be a charged covector field and and \(\eta^\mu_A\) a charged vector field on \(M\) valued in some vector space \(\bb V\) with some internal gauge transformation. Locally choosing a set of frames and coframes, we write the frame components as
	\be\label{eq:sigma-eta-frame-comp}
	\sigma_a^A \defn \sigma_\mu^A E^\mu_a \eqsp \eta^a_A \defn \eta^\mu_A e_\mu^a
	\ee
	Now we can view the frame components \(\sigma^A_a\) and \(\eta_A^a\) as scalar fields valued in \(\bb R^d \otimes \bb V\) with internal gauge transformations under both \(O(d-1,1)\) and \(G'\). Then, we can consider the frame components as globally smooth functions valued in \(\bb R^d \otimes \bb V\) on the bundle \(P\), in the same manner as the charged scalar field discussed above. Similarly, we can write any charged tensor field --- with arbitrary tensor structure and with internal gauge transformations under the full structure group \(G = O(d-1,1) \times G'\) --- on \(P\) in terms of its frame components, the frames and coframes. Henceforth we will always represent charged tensor fields defined on spacetime \(M\) by their frame components on the bundle \(P\) written as charged scalars \(\varphi^A\) where now the \(A\) index includes the frame component indices.

	Using the covariant exterior derivative \(D\) associated to the connection \(\df A^I\) in \cref{eq:conn-split} we can similarly represent the covariant derivatives of any charged tensor field in terms of their frame components on the principal bundle. We write the frame components of the \(k\)-derivatives of \(\varphi^A\) on the bundle \(P\) using the shorthand \( \varphi^A_{a_1\ldots a_k} \) where
	\[
		\varphi^A_{a_1\ldots a_k} \defn E_{a_k}\cdot D (\ldots E_{a_1}\cdot D\varphi^A) = E_{a_k}\cdot D \varphi^A_{a_1\ldots a_{k-1}}
	\]
	Thus we can describe all the dynamical fields with internal gauge transformations and their covariant derivatives as globally smooth tensor fields on the bundle \(P\) without making any choice of gauge.\\

	The other problem that arises if the dynamical fields have some internal gauge freedom is that we can only define a notion of ``diffeomorphisms up to a gauge transformation", and consequently there is only notion of ``stationarity up to internal gauge transformations". As discussed in the Introduction \cref{sec:intro}, if one uses the ordinary Lie derivative on spacetime (ignoring the internal gauge transformations), the result is not gauge-invariant. Also, the various attempts at defining ``covariant Lie derivatives" (\crefrange{eq:YM-gauge-Lie}{eq:LK-Lie}) do not implement the Lie algebra of diffeomorphisms of spacetime (see \cref{eq:Lie-comm-gauge}).

	Since we have defined the dynamical fields on a principal bundle \(P\), the source of this problem becomes more apparent. For theories with dynamical tensor fields defined on spacetime \(M\), the group of transformations is the group of diffeomorphisms of spacetime i.e. {\rm Diff}(M). Similarly, for theories with dynamical fields defined on the bundle \(P\) the group of transformations consists of automorphisms of the principal bundle (see \S~I.5 \cite{KN-book1}). The automorphism group of \(P\) has the semi-direct product structure \({\rm Aut}(P) \cong {\rm Diff}(M) \ltimes {\rm Aut}_V(P)\). Here \({\rm Aut}_V(P)\) is the normal subgroup of vertical automorphisms that do not move the points in the base space \(M\) i.e. these correspond to internal gauge transformations. Since, \({\rm Aut}_V(P)\) is a normal subgroup, the action of internal gauge transformations leaving the spacetime points fixed is well-defined. However, without picking a gauge choice there is only a notion of ``diffeomorphisms up to internal gauge transformations" and any attempt to define an action of just diffeomorphisms of \(M\) on charged fields is doomed to fail. In fact from \cref{eq:Lie-comm-gauge} we see that even when one defines some ``gauge covariant Lie derivative" one has to consider diffeomorphisms and internal gauge transformations simultaneously. Thus, again we are lead to work directly with fields defined on the principal bundle with \({\rm Aut}(P)\) acting as the full group of transformations. The corresponding Lie algebra of infinitesimal automorphisms \(\mf{aut}(P)\) consists of vector fields on the bundle which act on charged fields by the usual Lie derivative. We then define stationary (axisymmetric) charged fields as fields that are preserved under those automorphisms of the bundle \(P\) which project to stationary (axisymmetric) diffeomorphisms of spacetime \(M\) (see \cref{def:stationary-axisymm}). Viewed from the base spacetime this gives the appropriate notion of dynamical fields being stationary (axisymmetric) up to gauge. This point of view has the further advantage that we can treat both diffeomorphisms and gauge transformations simultaneously using standard tools of differential calculus on the bundle.

	Even though in general there is no unique way to associate a given diffeomorphism of spacetime to an automorphism of the bundle, if we require that the automorphism preserves a given connection on the bundle, we can prove that the non-uniqueness is given by a \emph{global symmetry} (if any exist) which keeps the chosen connection fixed at every point (see \cref{lem:aut-conn-unique} and \cref{rem:global-symm}). As we will show, in Einstein-Yang-Mills theory, when such a global symmetry of the solution Yang-Mills connection does not exist we \emph{cannot} set the Yang-Mills potential at the horizon to vanish and we get a new non-vanishing ``potential times perturbed charge" term at the horizon for the first law, generalising the results of \cite{SW1, SW2, Gao-YM} on the first law for Einstein-Yang-Mills theory.

	Similarly, if an automorphism of the Lorentz bundle \(P_O\) is required to preserve the coframes, then it is \emph{uniquely} determined by the corresponding isometry of the spacetime metric (see \cref{lem:aut-e-unique}). This uniqueness essentially implies that for a Killing field of the spacetime metric, the Lie derivative on the bundle coincides with the Lorentz-Lie derivative \cref{eq:Lor-Lie} on coframes and the Lichnerowicz-Kosmann-Lie derivative \cref{eq:LK-Lie} on spinors (see \cref{eq:get-Lor-Lie,eq:get-LK-Lie}). Thus, even though our Noether charges for arbitrary automorphisms differ from those derived by \cite{JM} for a coframe formulation of gravity, we get the same first law for stationary spacetimes for the first-order formulation of gravity.\\

	Now that we have defined the dynamical fields (and the Lagrangian of the theory; see the next \cref{sec:L-bundle}) on the bundle \(P\) instead of the spacetime \(M\), we can derive a first law of black hole mechanics. Note that we do not make any choice of gauge and consider the full group \({\rm Aut}(P)\) instead of just diffeomorphisms of \(M\). Most of the computations proceed in direct analogy to the computations of \cite{LW,W-noether-entropy,IW-noether-entropy} except they are carried out on the bundle. The only additional task is to check that the relevant quantities are infact gauge-invariant (or covariant), which is easily done by verifying that the computations yield horizontal forms on the bundle.

\subsection{The form of the gauge-invariant Lagrangian}\label{sec:L-bundle}

	On spacetime the Lagrangian is a  \(d\)-form \(\dfM{\df  L} \in \Omega^dM\) and we further assume that the Lagrangian is invariant under internal gauge transformations.\footnote{For instance, in this paper we do not consider theories with a Chern-Simons Lagrangian deferring their analysis to future work \cite{KP-CS}.} Thus we can pullback the Lagrangian \(\dfM{\df L}\) from the spacetime \(M\) to the bundle \(P\) that is, we consider the Lagrangian of the theory as a real horizontal \(d\)-form on the \(P\) given by \(\df L \in \Omega^d_{hor}P\).

 We will take the Lagrangian to depend on the frames \(E^m_a\), the coframes \(\df e^a\), the connection \(\df A^I\) \cref{eq:conn-split}, the frame components \(\varphi^A\) of charged tensor fields, and their finitely many covariant derivatives \(\varphi^A_{a_1\ldots a_k}\) (written as functions on \(P\); see the discussion above). We also allow dependence on the curvature \(\df F^I\) and the torsion \(\df T^a\), and finitely many of their covariant derivatives.

	Any antisymmetrisation in the derivatives  of the tensor fields \(\varphi^A\) can be converted to terms with lower order derivatives and torsion and curvature terms using
	\be
		2 \varphi^A_{[ab]} = - T^c{}_{ab} \varphi^A_c + F^I{}_{ab} {{r_I}^A}_{B} \varphi^B
	\ee
	where \(T^c{}_{ab} = E_b \cdot E_a \cdot \df T^c\) and \(F^I{}_{ab} = E_b \cdot E_a \cdot \df F^I\) are the frame components of the torsion and curvature. Using this on higher order antisymmetrised derivatives we can write all derivative terms in terms of completely symmetrised derivatives and derivatives of torsion and curvature. Then in a similar manner we can eliminate any antisymmetrised derivatives of the torsion and curvature. Finally, using \cref{eq:torsion-bianchi} we eliminate any dependence of the Lagrangian on \(D \df T^a\) in favour of the Lorentz curvature \(\df R^a{}_b\) and the coframes \(\df e^a\). For later convenience we introduce the shorthand
\be\label{eq:chi-defn}
	\chi^\alpha \defn \{\varphi^A, T^c{}_{ab}, F^I{}_{ab}\}
\ee
and the frame components of their completely \emph{symmetrised} derivatives by \(\chi^\alpha_{a_1\ldots a_i}\).\\

	Thus the dependence of the Lagrangian on the dynamical fields can be written as\footnote{Note that we simply assume that the Lagrangian is independent of any background fields and do not attempt to prove a ``Thomas replacement theorem" as done in Lemma 2.1 \cite{IW-noether-entropy}.}
	\be\label{eq:L}
	\df L(E^m_a, \df e^a, \df A^I, \{\chi^\alpha_{a_1\ldots a_i}\}) \in \Omega^d_{hor}P
	\ee
	where \(0\leq i \leq k\) counts the number of completely symmetrised derivatives of the corresponding fields in \cref{eq:chi-defn}. As discussed above the frames \(E^m_a\) on the bundle are only defined up to vertical vector fields, so we also demand that the Lagrangian depend on the frames so that \(\df L[E^m_a] = \df L[E'^m_a]\) whenever \(E'^m_a - E^m_a\) is a vertical vector field. The full set of \emph{dynamical} fields of the theory then includes the coframes \(\df e^a\), the connection \(\df A^I\) and the frame components of the charged tensor fields \(\varphi^A\) which we collectively denote as a differential form on \(P\) valued in a collective vector space \(\bb V\)
	\be\label{eq:psi-defn}
	\df\psi^\alpha \defn \{ \df e^a, \df A^I, \varphi^A \} \in \Omega^{\degf{\alpha} }P(\bb V)
	\ee
	Here and henceforth, we use the notation \(\degf{\alpha}\) to denote the degree of the differential form corresponding to the dynamical field \(\df\psi^\alpha\) with an \(\alpha\) index i.e. 
	\be\label{eq:deg-not}
	\degf{\alpha} = \{ \degf{a}, \degf{I}, \degf{A} \} = \{ 1, 1, 0 \}
	\ee

	The Lagrangian is further required to be a \emph{local and covariant} functional of the fields in the sense of \cref{def:loc-cov-func} i.e. for any automorphism of the bundle \(f \in {\rm Aut}(P)\) we have
	\be
	f^* \df L[\df\psi] = \df L[f^*\df\psi] \quad \forall f \in {\rm Aut}(P)
	\ee
	where it is implicit that on the right-hand-side that \(f\) also acts on the derivatives of \(\df\psi\). If \(X^m \in \mf{aut}(P)\) is the vector field generating the automorphism \(f\) then the above equation implies that
	\be
	\Lie_X \df L[\df\psi] = \df L[\Lie_X\df\psi] \quad \forall X^m \in \mf{aut}(P)
	\ee
	Note that since we assume that the Lagrangian is gauge-invariant, we have
	\be\begin{split}
	\Lie_X \df L[\df\psi] & = 0 \quad \forall X^m \in \mf{aut}_V(P)
	\end{split}\ee

\subsection{Equations of motion, the symplectic potential and symplectic current}\label{sec:symp-pot-current}

	With the above described Lagrangian the equations of motion of the theory are obtained by a variation of the Lagrangian with respect to the dynamical fields \cref{eq:psi-defn}. To consider such variations, we take any smooth \(1\)-parameter family of dynamical fields \(\df\psi^\alpha(\lambda)\) with \(\df\psi^\alpha(0) = \df\psi^\alpha\) corresponding to the background dynamical fields of interest. Define the first variation or perturbation about \(\df\psi^\alpha\) by
	\be\label{eq:var-defn}
	\delta \df\psi^\alpha \defn \left.\frac{d}{d\lambda}\df\psi^\alpha(\lambda)\right\rvert_{\lambda = 0}
	\ee
	We also use the symbol \(\delta\) to denote variations of any functional \(\mc F\) of the dynamical fields defined in the same way i.e.
	\be\label{eq:var-func}
		\delta \mc F[\psi] \defn \left.\frac{d}{d\lambda}\mc F [\psi(\lambda)]\right\rvert_{\lambda = 0}
	\ee\\

	Since the difference of two connections is horizontal and all the other dynamical fields are already horizontal the perturbations of the dynamical fields \cref{eq:psi-defn} given by \(\delta \df\psi^\alpha \in \Omega_{hor}^{\degf{\alpha}} P(\mathbb V) \) are all horizontal forms on \(P\). Further, since \(E_a \cdot \df e^b = \delta^b_a\) holds at each \(\lambda\) of the \(1\)-parameter family of frames and coframes, we have  \( \delta E_a \cdot \df e^b = - E_a \cdot \delta \df e^b \). Since the frames are considered equivalent if their difference is vertical we have
	\be\label{eq:var-frame}
		\delta E^m_a = - (E_a \cdot \delta \df e^b) E^m_b = - E^m_b E^n_a \delta e_n^b
	\ee
	and we can convert all variations of the Lagrangian of the form \cref{eq:L} obtained from frame variations to variations of the dynamical coframe fields as
	\be
		\frac{\delta \df L}{\delta E^m_a} \delta E^m_a = \lb( - \frac{\delta \df L}{\delta E^n_b} E^n_a E^m_b \rb) \delta e_m^a
	\ee\\

	The variation of the Lagrangian can be written in the form (we will prove this in \cref{lem:hor-theta}):
	\be\label{eq:var-L-1}
		\delta \df L = \tilde{\df{\mc E}}^{m_1 \ldots m_{\degf{\alpha}}}_\alpha  \delta \psi_{m_1 \ldots m_{\degf{\alpha}}}^\alpha + d\df\theta(\psi;\delta\psi)
	\ee
	where the \emph{equations of motion} \(\tilde{\df{\mc E}}_\alpha : \Omega^{\degf{\alpha}}_{hor} P(\bb V) \to \Omega^d_{hor} P(\bb V^*)\) are the following functional derivative
	\be\label{eq:eom-maps}
		\tilde{\df{\mc E}}^{m_1 \ldots m_{\degf{\alpha}}}_\alpha = \frac{\delta\df L}{\delta \psi_{m_1 \ldots m_{\degf{\alpha}}}^\alpha}
	\ee
	The \emph{symplectic potential} \(\df \theta\) denotes the ``boundary term" in the variation and depends locally and covariantly on the background \(\df\psi\) and linearly on the perturbation \(\delta\df\psi\) and its derivatives.

	For subsequent computations it will be very convenient to express the equations of motion purely as differential forms rather than linear maps valued in differential forms as in \cref{eq:eom-maps}. Defining the equations of motion \(\df{\mc E}_\alpha\) corresponding to the variation of each dynamical field \(\df\psi^\alpha \) (\cref{eq:psi-defn}), with \(k \in \degf{\alpha} = \{ 1,1,0 \}\) (\cref{eq:deg-not}) by
	\be\label{eq:eom-form}
		\df{\mc E}_\alpha \equiv ({\mc E}_\alpha)_{m_1\ldots m_{d-k}} = \frac{(d-k)!k!}{d!} \frac{\delta L_{m_1\ldots m_{d-k}l_1\ldots l_k}}{\delta \psi_{n_1 \ldots n_k}^\alpha} \delta^{l_1}_{n_1}\cdots \delta^{l_k}_{n_k}  \in \Omega_{hor}^{d-k} P(\mathbb V^*)
	\ee
	a straightforward computation shows that the variational principle can be written in the more convenient form
	\be\label{eq:var-L}
		\delta \df L = \df{\mc E}_\alpha(\psi) \wedge \delta \df\psi^\alpha + d\df\theta(\psi;\delta\psi)
	\ee
	This rewriting of the variational principle (as opposed to \cref{eq:var-L-1}) will simplify a lot of the later computations as compared to similar ones in \cite{LW, IW-noether-entropy}. That the equations of motion \cref{eq:eom-form} are horizontal forms expresses the well-known fact that gauge-invariant Lagrangians give gauge-covariant equations of motion. The dynamical fields \(\df\psi^\alpha \) which satisfy the equations of motion \(\df{\mc E}_\alpha(\psi) = 0\) form the subspace of \emph{solutions}. Given a solution \(\df\psi^\alpha\), any perturbation \(\delta\df\psi^\alpha\) is called a \emph{linearised solution} if it satisfies the \emph{linearised equations of motion} \(\delta\df{\mc E}_\alpha = 0\) at \(\df\psi^\alpha\).

	The variational principle \cref{eq:var-L} implies that \(d\df\theta\) is a horizontal form but we can show \(\df\theta\) itself can be chosen to be horizontal (i.e. gauge-invariant) in the following lemma which is an extension of Lemma 3.1 \cite{IW-noether-entropy} to the bundle.

	\begin{lemma}[horizontal symplectic potential]\label{lem:hor-theta}
	For any Lagrangian of the form specified in \cref{eq:L}, the symplectic potential \(\df\theta(\psi;\delta\psi)\) can be chosen to be a horizontal form on \(P\) of the form
	\be
	\df\theta = (-)^{d-2} \df Z_I \wedge \delta \df A^I + (-)^{d-2} \df Z_a \wedge \delta \df e^a + \df\theta' \in \Omega^{d-1}_{hor}P
	\ee
	with
	\be\label{eq:theta-add}
	\df\theta' =  - \sum_{i=1}^k \lb( E_{a_i} \cdot \df Z_\alpha^{a_1\ldots a_i} \rb) \delta \chi^\alpha_{a_1\ldots a_{i-1}}
	\ee
	Here
	\begin{subequations}\label{eq:Z-form}\begin{align}
		\df Z_I & \equiv (Z_I)_{m_1\ldots m_{d-2}} = \frac{(d-2)!2!}{d!} \frac{\delta L_{m_1\ldots m_{d-2}l_1l_2}}{\delta F_{n_1n_2}^I} \delta^{l_1}_{n_1} \delta^{l_2}_{n_2} \in \Omega^{d-2}_{hor}P(\mf g^*) \label{eq:ZI}\\
		\df Z_a & \equiv (Z_a)_{m_1\ldots m_{d-2}} = \frac{(d-2)!2!}{d!} \frac{\delta L_{m_1\ldots m_{d-2}l_1l_2}}{\delta T_{n_1n_2}^a} \delta^{l_1}_{n_1} \delta^{l_2}_{n_2} \in \Omega^{d-2}_{hor}P({\bb R^d}^*) \label{eq:Za}
	\end{align}\end{subequations}
	would be the equations of motion obtained if the curvature \(\df F^I\) and torsion \(\df T^a\), respectively, are viewed as an independent fields. Similarly 
	\be
		\df Z_\alpha^{a_1\ldots a_i} = \frac{\delta \df L}{\delta \chi^\alpha_{a_1\ldots a_i}}
	\ee
would be the equations of motion if the all the derivatives up to the \(i\)-th derivative of \(\chi^\alpha\) \cref{eq:chi-defn} (but not higher derivatives) are viewed as independent fields.
	\begin{proof}
	The proof proceeds by varying the Lagrangian \cref{eq:L} considering all the fields and their derivatives as independent and then ``integrating by parts" the variations due to the derivatives. Write the variation of the Lagrangian as
	\[
		\delta \df L = \sum_{i=0}^k \df U_\alpha^{a_1\ldots a_i}  \delta \chi^\alpha_{a_1\ldots a_i} + [\cdots]
	\]
	where here (and throughout the rest of this proof) \([\cdots]\) denotes terms proportional to \(\delta \df e^a\) and \(\delta \df A^I\) and
	\be\label{eq:U-defn}
		\df U_\alpha^{a_1\ldots a_i} \defn \frac{\partial \df L}{\partial \chi^\alpha_{a_1\ldots a_i}}
	\ee
	is a horizontal \(d\)-form valued in the appropriate representation of the structure group and we fix the index permutation symmetries of \(\df U\) to be the same as the corresponding field \(\chi\). Note that we have used \(\partial\) in \cref{eq:U-defn} to emphasise that we have not performed any ``integration by parts" yet. To get the form of the variational principle we have to rewrite the terms obtained by a variation of the derivatives of \(\chi^\alpha\) in terms of variations of \(\chi^\alpha\) by ``integrating by parts". Consider the variation due to the \(i\)-th derivatives as
	\[\begin{split}
		\df U_\alpha^{a_1\ldots a_i} \delta \chi^\alpha_{a_1\ldots a_i} & = \df U_\alpha^{a_1\ldots a_i} \delta \lb( E_{a_i}\cdot D\chi^\alpha_{a_1\ldots a_{i-1}} \rb) \\
		& =  \df U_\alpha^{a_1\ldots a_i} \lb[ -(E_{a_i}\cdot\delta \df e^b) E_b\cdot D\chi^\alpha_{a_1\ldots a_{i-1}} + E_{a_i}\cdot \delta D\chi^\alpha_{a_1\ldots a_{i-1}} \rb] \\
		& = (-)^{d+1}E_{a_i} \cdot \df U_\alpha^{a_1\ldots a_i} \wedge D\delta\chi^\alpha_{a_1\ldots a_{i-1}} + [\cdots] \\
		& = \df Y_\alpha^{a_1\ldots a_{i-1}} \delta\chi^\alpha_{a_1\ldots a_{i-1}} + d\df \theta^{(i)} + [\cdots]
	\end{split}\]
	where the second line uses \cref{eq:var-frame}. The term \(Y\) then contributes to the variation with respect to the \((i-1)\)-th derivative term. Thus define recursively, for any \(0 \leq i \leq k\), the term obtained by a variation of the Lagrangian considering the derivatives up to the \(i\)-th derivative of \(\chi^\alpha\), but not higher derivatives, as independent
	\[
		\df Z_\alpha^{a_1\ldots a_i} \defn
		\begin{cases}
		\df U_\alpha^{a_1\ldots a_i} &\quad\text{for}\quad i = k \\
		\df U_\alpha^{a_1\ldots a_i} + D\lb( E_{a_{i+1}} \cdot \df Z_\alpha^{a_1\ldots a_{i+1}} \rb) &\quad\text{for}\quad 0 \leq i < k
		\end{cases}
	\]
	Using this the variation due to the \(i\)-th derivative term has the terms
	\[\begin{split}
		\df Y_\alpha^{a_1\ldots a_i} &= D\lb( E_{a_{i+1}} \cdot \df Z_\alpha^{a_1\ldots a_{i+1}} \rb) \\
		\df\theta^{(i)} &= - \lb( E_{a_i} \cdot \df Z_\alpha^{a_1\ldots a_i} \rb) \delta \chi^\alpha_{a_1\ldots a_{i-1}}
	\end{split}\]

	Iterating the above computation for each derivative order we can write
	\[
		\delta \df L = \df Z_\alpha \delta\chi^\alpha + \sum_{i=1}^k d\df\theta^{(i)} + [\cdots]
	\]
	The second term above gives us the term in \(\df\theta'\) in \cref{eq:theta-add}. From the collective notation \cref{eq:chi-defn}, the first term has variations of the charged tensor fields \(\varphi^A\) (which contribute to the equation of motion) as well as those of the curvature and torsion. We then convert the terms obtained from the variations of the curvature and torsion to variations of the connection and coframes.
	\be\begin{split}
		\df Z_I^{ab} \delta F^I_{ab} = \df Z_I^{ab} \delta (E_b \cdot E_a \cdot \df F^I) & = (-)^{d-2} d( \df Z_I \wedge \delta \df A^I) + [\cdots] \\
		\df Z_c^{ab} \delta T^c_{ab} = \df Z_c^{ab} \delta (E_b \cdot E_a \cdot \df T^c)  & = (-)^{d-2} d( \df Z_a \wedge \delta \df e^a) + [\cdots]
	\end{split}\ee
	where \(\df Z_I = E_b \cdot E_a \cdot \df Z_I^{ab}\) and \(\df Z_a = E_c \cdot E_b \cdot \df Z_a^{bc}\), and are explicitly given by \cref{eq:Z-form}.

	Thus, in the total variation of the Lagrangian we can collect all terms proportional to \(\delta \df e^a\), \(\delta \df A^I\) and \(\delta\varphi^A\) into the respective equations of motion and an exact form
	\[
		\delta \df L = \df{\mc E}_\alpha \wedge \delta\df\psi^\alpha + d\df\theta
	\]
	with \(\df\theta\) given by the claim of the lemma.
	\end{proof}
	\end{lemma}

	The above algorithm for choosing a horizontal \(\df\theta\) has some ambiguities which we enumerate next.
	\begin{enumerate}
	\item For some \(\df\mu(E^m_a, \df e^a, \df A^I, \{\chi^\alpha_{a_1\ldots a_i}\}) \in \Omega^{d-1}P\), we can add a exact form \(d\df\mu \in \Omega^d_{hor}P\) to the Lagrangian without changing the equations of motion (i.e. the dynamical content of the theory) as
	\be\label{eq:mu-amb}
		\df L \mapsto \df L + d\df \mu
	\ee
	Since we restrict to horizontal Lagrangians, we only consider \(d\df\mu\) that are horizontal forms, but we do not demand that \(\df\mu\) itself be horizontal i.e. gauge-invariant in contrast to \cite{IW-noether-entropy}. For instance, \(d\df\mu\) could be the integrand of a topological invariant of the bundle (for example, the Euler density), in which case \(\df\mu\) itself would not be horizontal. This shifts the symplectic potential as
	\[
		\df\theta(\delta \psi) \mapsto \df\theta(\delta \psi) + \delta \df\mu
	\]
	Note that we can apply \cref{lem:hor-theta} to the Lagrangian \(\df L + d\df\mu\) and conclude that \(\delta\df\mu\) is horizontal i.e. invariant under internal gauge transformations, even if \(\df\mu\) is not.
	\item Given a choice of Lagrangian, the variational principle \cref{eq:var-L} only determines the symplectic potential up to the addition of a local, covariant and horizontal \((d-1)\)-form \(\df\lambda'(\psi;\delta\psi)\) which is linear in the perturbation \(\delta\df\psi\) and \(d\df\lambda' = 0 \). Using \cref{lem:W-lemma} (with \(\df\psi^\alpha\) as the ``background field" and \(\delta\df\psi^\alpha\) as the ``dynamical field") we get, \(\df\lambda' = d\df\lambda\) for some local and covariant horizontal form \(\df\lambda(\psi;\delta\psi) \in \Omega^{d-2}_{hor}P\). Thus, this additional ambiguity in the symplectic current is 
	\be\label{eq:lambda-amb}
		\df\theta(\delta \psi) \mapsto \df\theta(\delta \psi) + d\df\lambda(\delta \psi)
	\ee
	\end{enumerate}

	Using the symplectic potential we define the \emph{symplectic current} as an antisymmetric bilinear map on perturbations (see \cite{LW,IW-noether-entropy})
	\be\label{eq:omega-defn}
		\df\omega(\psi;\delta_1\psi, \delta_2\psi) \defn \delta_1\df\theta(\delta_2\psi) - \delta_2\df\theta(\delta_1\psi) - \df\theta([\delta_1,\delta_2]\psi) \in \Omega^{d-1}_{hor} P
	\ee
	In the above definition we have considered the perturbations \(\delta\psi\) as vector fields on the space of field configurations evaluated at the background given by \(\psi\). The commutator \( [\delta_1,\delta_2]\psi \defn \delta_1\delta_2 \psi - \delta_2\delta_1 \psi \) depends on how one chooses to extend these vector fields away from the background \(\psi\) in configuration space, even though the symplectic current \(\df\omega\) at \(\psi\) is independent of this choice. If the variations \(\delta_1\) and \(\delta_2\) are extended to correspond to ``independent" one-parameter families of dynamical fields then the commutator vanishes, which suffices for most situations. On the other hand, if the variations are extended to correspond to the same infinitesimal action of bundle automorphisms the commutator is non-vanishing in general (this is useful, for instance, when considering Einstein-fluid systems; see \cite{GSW}).

	Note that the symplectic current \cref{eq:omega-defn} is a horizontal form and hence gauge-invariant, in the sense that it is invariant under any vertical automorphism \(f \in {\rm Aut}_V(P)\) of the bundle. However, it does not vanish if we substitute one of the perturbations, say \(\delta_2\df\psi\), by a perturbation \(\Lie_X\df\psi\) generated by an infinitesimal vertical automorphism \(X^m \in \mf{aut}_V(P)\) (see \cref{lem:omega-Q}).

By taking a second variation of the Lagrangian the symplectic current can be shown to be a closed form on solutions (see \cite{LW, SeiW}) that is
	\begin{lemma}[closed symplectic current]\label{lem:closed-omega}
The symplectic current is closed when restricted to solutions and linearised solutions.
	\begin{proof}
	Consider a second variation of the Lagrangian \cref{eq:L} 
	\[
	\delta_1\delta_2\df L = \delta_1\df{\mc E}_\alpha \wedge \delta_2 \df\psi^\alpha + \df{\mc E}_\alpha \wedge \delta_1\delta_2 \df\psi^\alpha + d\delta_1\df\theta(\delta_2\psi)
	\]
	Using the identity \((\delta_1\delta_2 - \delta_2\delta_1 - [\delta_1,\delta_2])\df L = 0\) we have:
	\be
		0 = \delta_1\df{\mc E}_\alpha \wedge \delta_2 \df\psi^\alpha - \delta_2\df{\mc E}_\alpha \wedge \delta_1 \df\psi^\alpha - \df{\mc E}_\alpha \wedge [\delta_1,\delta_2] \df\psi^\alpha + d\df\omega
	\ee
	Restricting the above on solutions \(\df\psi\) and linearised solutions \(\delta\df\psi\) we have \( d\df\omega = 0 \).
	\end{proof}
	\end{lemma}

	Since the sympectic current \(\df\omega\) is a horizontal form on \(P\) there is a corresponding gauge-invariant form on spacetime which we denote by \(\dfM{\df\omega}\). Given a Cauchy surface \(\Sigma\), the symplectic current defines a \emph{symplectic form} \(W_\Sigma\) on perturbations as
	\be\label{eq:W-defn}
		W_\Sigma(\psi;\delta_1\psi, \delta_2\psi) \defn \int_\Sigma \dfM{\df\omega}(\psi;\delta_1\psi, \delta_2\psi)
	\ee

	From \cref{lem:closed-omega} we can conclude that the symplectic form is conserved on linearised solutions i.e. if \(\Sigma_t\) is a time-evolved Cauchy surface then \(W_\Sigma(\psi;\delta_1\psi, \delta_2\psi) = W_{\Sigma_t}(\psi;\delta_1\psi, \delta_2\psi)\), whenever \(\psi\) is a solution, \(\delta_1\psi\) and \(\delta_2\psi\) are linearised solutions with boundary conditions such that there is no symplectic flux at infinity.

	The \(\df\mu\)-ambiguity in the Lagrangian \cref{eq:mu-amb} does not affect the symplectic current and from the \(\df\lambda\)-ambiguity \cref{eq:lambda-amb} we have
	\[
		\df\omega(\delta_1\psi, \delta_2 \psi) \mapsto \df\omega(\delta_1\psi, \delta_2 \psi) + d\lb[ \delta_1\df\lambda(\delta_2\psi) - \delta_2\df\lambda(\delta_1 \psi) - \df\lambda([\delta_1, \delta_2] \psi) \rb]
	\]
	which adds a boundary term to the symplectic form \(W_\Sigma\)
	\be
		W_\Sigma \mapsto W_\Sigma + \int_{\partial\Sigma} \lb[ \delta_1\dfM{\df\lambda}(\delta_2\psi) - \delta_2\dfM{\df\lambda}(\delta_1\psi) - \dfM{\df\lambda}([\delta_1, \delta_2] \psi)  \rb]
	\ee
	where \(\dfM{\df\lambda}\) is the unique gauge-invariant differential form on \(M\) corresponding to \(\df\lambda\).\\

	Following \cite{IW-noether-entropy}, we will use the symplectic form to derive the first law of black hole mechanics and show that the above ambiguities do not effect the first law. At this point one can generalise the entire analysis of \cite{LW} to construct the phase space and Poisson brackets for such theories which is certainly of independent interest. Since we are primarily interested in the first law of black hole mechanics, we turn next to the definition of the Noether charge for any bundle automorphism. The first law then follows from the relation between the symplectic form defined above and the Noether charge (see \cref{lem:omega-Q}).

\subsection{Noether current, Noether charge and boundary Hamiltonians}\label{sec:noether-current-charge}

	As is well known, Noether's theorem associates gauge symmetries of a Lagrangian theory to conserved currents and charges (see \cite{LW} for instance). The Lagrangians we are considering are both covariant under diffeomorphisms of the base spacetime \(M\) as well as invariant under internal gauge transformations. Though, as discussed earlier, there is no natural group action of the diffeomorphisms of \(M\) on the dynamical fields with non-trivial internal gauge transformations and we only have a notion of ``diffeomorphism up to internal gauge". Thus, we cannot separately define Noether currents associated to only diffeomorphisms and have to consider the full gauge group of the theory given by the group of automorphisms \({\rm Aut}(P)\) of the principal bundle \(P\). Thus, we will define Noether currents associated to any automorphism in \({\rm Aut}(P)\) by adapting the procedure used in \cite{LW} to work directly on the principal bundle instead of the base spacetime.

	We denote the variation obtained by the bundle automorphism generated by a vector field \(X^m \in \mf{aut}(P)\) as \(\delta_X \df\phi \defn \Lie_X\df\phi \). Since we have assumed that the Lagrangian is covariant under such automorphisms, to each automorphism we can associate a Noether current as follows. For the gauge-invariant Lagrangians under consideration \(d\df L \in \Omega^{d+1}_{hor}P\), and hence \(d\df L = 0\). Then for the variation of the Lagrangian under an automorphism we have
	\be
		\delta_X \df L = \Lie_X \df L = X \cdot d\df L + d(X\cdot \df L) = d(X \cdot \df L)
	\ee
	The \emph{Noether current} corresponding to any \(X^m \in \mf{aut}(P) \) is defined by (see \cite{LW}):
	\be\label{eq:noether-current-defn}
		\df J_X \defn \df\theta(\delta_X \psi) - X \cdot \df L
	\ee
	Here we note that if \(X^m \in \mf{aut}_V(P)\) generates vertical automorphisms of the bundle i.e. internal gauge transformations we have \(X \cdot \df L = 0\).

	We can define the Noether charge associated to the Noether current \(\df J_X\) as follows. Consider the following computation:
	\be\label{eq:dJ}\begin{split}
		d\df J_X & = d\df\theta(\delta_X \psi) - d(X \cdot \df L) = \delta_X \df L - \df{\mc E}_\alpha \wedge \delta_X \df\psi^\alpha - \delta_X \df L \\
			& = - \df{\mc E}_\alpha \wedge \Lie_X \df\psi^\alpha \\
	\end{split}\ee

	By adapting the procedure in \cite{SeiW} to work on the bundle we can define a Noether charge without using the equations of motion (``off-shell"). To do this, we first define the following linear maps from infinitesimal automorphisms of the bundle to horizontal forms, which are generalised versions of the \emph{constraints} and \emph{Bianchi identities} (see \cite{LW, SeiW}).

	The \emph{constraints} are linear maps \( \df{\mc C}:\mf{aut}(P) \to \Omega^{d-1}_{hor}P \) given by
	\be\label{eq:constraint-defn}\begin{split}
	\df{\mc C}(X) & \defn (-)^{d-\degf{\alpha}+1} \df{\mc E}_\alpha \wedge X \cdot \df\psi^\alpha \\
		&~ = (-)^d \lb[ \df{\mc E}_a (X \cdot \df e^a) + \df{\mc E}_I  (X \cdot \df A^I) \rb]
	\end{split}\ee
	and in the second line we have used \cref{eq:psi-defn,eq:deg-not}. Since \(\df{\mc C}(X)\) is a horizontal form on the bundle we can consider the corresponding gauge-invariant \((d-1)\)-form \(\dfM{\df{\mc C}}(X)\) on spacetime. The pullback of \(\dfM{\df{\mc C}}(X)\) to any Cauchy surface then are the constraint equations that hold for any initial data for dynamical fields which correspond to a solution to the equations of motion. Note that none of the charged tensor fields \(\varphi^A\) and their equations of motion contribute to the constraints, since we consider the frame components of tensor fields as the dynamical fields. Further, if \(X^m \in \mf{aut}_V(P)\) then from \cref{eq:psi-defn} only the connection \(\df A^I\) and its equation of motion \(\df{\mc E}_I\) contribute to the constraints corresponding to gauge transformations.

	The Bianchi identities \(\df{\mc B}:\mf{aut}(P) \to \Omega^d_{hor}P \) are linear maps given explicitly by
	\be\label{eq:bianchi-defn}\begin{split}
	\df{\mc B}(X) & \defn - \df{\mc E}_\alpha \wedge X\cdot d\df\psi^\alpha +(-)^{d-\degf{\alpha}} d\df{\mc E}_\alpha \wedge X\cdot \df\psi^\alpha \\
		&~ = - \df{\mc E}_a \wedge (X \cdot \df T^a) + (-)^{d-1} D\df{\mc E}_a (X \cdot \df e^a) + \df{\mc E}_a \wedge \df e^b (X \cdot {\df \omega}^a{}_b) \\
		&~ \quad - \df{\mc E}_I \wedge (X \cdot \df F^I) + (-)^{d-1} D \df{\mc E}_I (X \cdot \df A^I) \\
		&~ \quad - \df{\mc E}_A (X \cdot D \varphi^A) + \df{\mc E}_A \varphi^B (X \cdot {\df A^I}){r_I}^A{}_B \\
	\end{split}\ee
	where in the second equality we have used \cref{eq:psi-defn} and written all terms as manifestly horizontal forms. Note that the second line only depends on the gravitational connection \({\df \omega}^a{}_b\) while the rest depend on the full connection \(\df A^I\) on \(P\) (see \cref{eq:conn-split}). These Bianchi identities above are a generalisation of the ones for diffeomorphism covariant theories given in \cite{LW, SeiW} to include all automorphisms of the bundle i.e. both gauge transformations as well as diffeomorphisms.

 Using \cref{eq:constraint-defn,eq:bianchi-defn}, we can rewrite \cref{eq:dJ} in the following form
	\be\label{eq:J-C-B}
	d\left[ \df J_X - \df{\mc C}(X) \right] = \df{\mc B}(X)
	\ee
	Using the arguments in \S~IV. \cite{SeiW}, we can show that the Bianchi identities \(\df{\mc B}(X)\) vanish identically on all dynamical fields even those that do not satisfy the equations of motion.
	\begin{prop}[Bianchi identities]\label{prop:bianchi}
	The Bianchi identities \(\df{\mc B}(X)\) vanish for any dynamical field \(\df\psi^\alpha\) for all \(X^m \in \mf{aut}(P)\).
	\begin{proof}
	Since \(\df{\mc B}(X)\) is a horizontal form denote the corresponding gauge-invariant form on spacetime by \(\dfM{\df{\mc B}}(X) \in \Omega^dM\). Then we can show that \(\dfM{\df{\mc B}}(X) = 0\) using same argument as in \S~IV. \cite{SeiW}. Thus \(\df{\mc B}(X) = \pi^*\dfM{\df{\mc B}}(X) = 0\) for any \(\df\psi\) and all \(X \in \mf{aut}(P)\). 
  	\end{proof}
	\end{prop}

	Using the above we can define the Noether charge without using any equations of motion.
	\begin{lemma}[Noether charge]\label{lem:noether-charge}
	For any infinitesimal automorphism \(X^m \in \mf{aut}(P)\) there exists a horizontal \((d-2)\)-form \(\df Q_X \in \Omega^{d-2}_{hor}P\) called the Noether charge, such that the Noether current \(\df J_X\) can be written in the form
	\be\label{eq:J-C-Q}
		\df J_X = d\df Q_X + \df{\mc C}(X)
	\ee
	\begin{proof}
	Since \(\df{\mc B}(X) = 0\), from \cref{eq:J-C-B} we have \(d\left[ \df J_X - \df{\mc C}(X) \right] = 0\). Then using \cref{lem:W-lemma} (with \(X^m\) as the ``dynamical field" and \(\df\psi^\alpha\) as a ``background field") we conclude that there exists a \(\df Q_X \in \Omega^{d-2}_{hor}P\) (which depends linearly on \(X^m\) and finitely many of it derivatives) such \cref{eq:J-C-Q} holds.
	\end{proof}
	\end{lemma}

	\cref{lem:noether-charge} shows that the Noether charge exists but for any theory based on a Lagrangian of the form \cref{eq:L} we can obtain an explicit useful expression for the Noether charge (this generalises Prop.~4.1 \cite{IW-noether-entropy}) as follows

	\begin{lemma}[Form of the Noether charge]\label{lem:noether-charge-form}
	The Noether charge \(\df Q_X\) for \(X^m \in \mf{aut}(P)\) can be chosen to be of the form
	\be\label{eq:noether-charge-form}
		\df Q_X = \df Z_I (X \cdot \df A^I) + \df Z_a (X\cdot \df e^a) \in \Omega^{d-2}_{hor}P
	\ee
	where \(\df Z_I\) and \(\df Z_a\) are as in \cref{lem:hor-theta} and can be computed directly from the Lagrangian using \cref{eq:Z-form}.
	\begin{proof}
	To get an explicit form for the Noether charge we start with \cref{eq:noether-current-defn} and use the form of the symplectic current given by \cref{lem:hor-theta} to compute
	\[
		\df\theta(\delta_X\psi) = (-)^{d-2} \df Z_I \wedge \Lie_X \df A^I + (-)^{d-2}\df Z_a \wedge \Lie_X \df e^a + \df\theta'(\delta_X \psi)
	\]
	Using \cref{eq:Lie-cov-D} for the Lie derivatives of the connection and coframes, we can write the first two terms as
	\[\begin{split}
		& (-)^{d-2} \df Z_I \wedge \lb[X\cdot \df F^I + D(X \cdot \df A^I)\rb] + (-)^{d-2} \df Z_a \wedge \lb[ X \cdot \df T^a + D(X\cdot \df e^a) - (X\cdot {\df A^a}_b)\df e^b \rb] \\
	= &~ d\lb[ \df Z_I (X \cdot \df A^I) + \df Z_a (X\cdot \df e^a) \rb] + [\cdots]
	\end{split}\]
	where thoughout this proof the \([\cdots]\) represents a local, covariant, and horizontal \(d-1\) form which is linear in \(X^m\) and independent of its derivatives. A similar computation for the \(\df\theta'\) term only gives \([\cdots]\)-type terms since the \(\chi^\alpha_{a_1\ldots a_i}\) are all functions i.e. \(0\)-forms. Thus, using \cref{eq:noether-current-defn} the Noether current can be written as
	\[
		\df J_X = d\lb[ \df Z_I (X \cdot \df A^I) + \df Z_a (X\cdot \df e^a) \rb] + [\cdots] 
	\]
	where we have again absorbed \(X \cdot \df L\) into the \([\cdots]\)-term. Adding the constraints \(\df{\mc C}(X)\) in \cref{eq:constraint-defn}, which are also of \([\cdots]\)-type, to both sides we get
	\[
	\df J_X - \df{\mc C}(X) - d\lb[ \df Z_I (X \cdot \df A^I) + \df Z_a (X\cdot \df e^a) \rb] = [\cdots]
	\]
	Since \(d\left[ \df J_X - \df{\mc C}(X) \right] = 0 \) from \cref{eq:J-C-B} and \cref{prop:bianchi}, we get that the right-hand-side is a closed horizontal \((d-1)\)-form that does not depend on derivatives of \(X^m\). Using \cref{lem:W-lemma} (with \(X^m\) as the ``dynamical field" and \(\df\psi^\alpha\) as a ``background field") we conclude that right-hand-side vanishes and we get
	\[
		\df J_X =  \df{\mc C}(X) + d\lb[ \df Z_I (X \cdot \df A^I) + \df Z_a (X\cdot \df e^a) \rb]
	\]
	Thus the Noether charge can be chosen to be of the form in \cref{eq:noether-charge-form}.
	\end{proof}
	\end{lemma}

	Note here that only the coframes and connection contribute explicitly to the form of the Noether charge since we have converted all other tensor fields and their derivatives into functions (using the coframe and frames). Consequently the form of the Noether charge given by \cref{lem:noether-charge-form} is much simpler than the corresponding one in Prop.~4.1 \cite{IW-noether-entropy}. Further the expression \cref{eq:noether-charge-form} for the Noether charge is completely specified by the dependence of the Lagrangian on the curvature and torsion.

	The ambiguities \cref{eq:mu-amb,eq:lambda-amb} in the Lagrangian and the symplectic potential lead to the following ambiguities in the Noether current and Noether charge
	\be\label{eq:J-Q-amb}\begin{split}
		\df J_X & \mapsto \df J_X + d(X \cdot \df\mu) + d\df\lambda(\delta_X\psi) \\
		\df Q_X &\mapsto \df Q_X + X \cdot \df\mu + \df\lambda(\delta_X\psi) + d\df\rho
	\end{split}\ee
	where \(\df\rho\) is an extra ambiguity in the Noether charge, since the charge is defined by the Noether current only up to a closed and hence exact form (see \cref{lem:W-lemma}). We note that the form of the Noether charge given by \cref{lem:noether-charge-form} is not unambiguous. If the terms \(\df\mu\) and \(\df\lambda\) have suitable dependence on the curvature and torsion they can contribute non-trivially to the Noether charge.\\

	The utility of the above formalism in deriving a first law stems from the following relation between the symplectic current and the Noether charge. The proof follows by a simple computation on the bundle \(P\), in exact parallel to the ones on spacetime in \cite{LW, IW-noether-entropy, SeiW, HW}.
	\begin{lemma}\label{lem:omega-Q}
	For any perturbation \(\delta\df\psi\) and \(X^m \in \mf{aut}(P)\), the symplectic current \(\df\omega(\delta\psi, \Lie_X\psi)\) is related to the Noether charge \(\df Q_X\) by
	\be\label{eq:omega-Q}
	\df\omega(\delta\psi, \Lie_X\psi) = d \left[ \delta \df Q_X -  X \cdot \df\theta(\delta\psi) \right] + \delta \df{\mc C}(X) + X\cdot (\df{\mc E}_\alpha \wedge \delta\df\psi^\alpha )
	\ee
	\begin{proof}
	Consider a variation of \cref{eq:noether-current-defn} with a given fixed \(X^m\)
	\be\begin{split}
		\delta \df J_X & = \delta\df\theta(\delta_X\psi) - X \cdot \delta \df L \\
				& = \delta\df\theta(\delta_X\psi) - X \cdot d\df\theta(\delta\psi) - X\cdot (\df{\mc E}_\alpha \wedge \delta\df\psi^\alpha ) \\
				& = \delta\df\theta(\Lie_X\psi) - \Lie_X \df\theta(\delta\psi) + d(X \cdot \df\theta(\delta\psi) ) - X\cdot (\df{\mc E}_\alpha \wedge \delta\df\psi^\alpha )
	\end{split}\ee
	The first two terms on the right-hand-side can be rewritten in terms of the symplectic current using \cref{eq:omega-defn} to get
	\be\begin{split}
		\df\omega(\delta\psi, \Lie_X\psi) & = \delta \df J_X -  d(X \cdot \df\theta(\delta\psi) ) + X\cdot (\df{\mc E}_\alpha \wedge \delta\df\psi^\alpha ) \\
			& =  d \left[ \delta \df Q_X -  X \cdot \df \theta(\delta\psi) \right] + \delta \df{\mc C}(X) + X\cdot (\df{\mc E}_\alpha \wedge \delta\df\psi^\alpha )
	\end{split}\ee
	\end{proof}
	\end{lemma}

	From \cref{lem:omega-Q} we see that for linearised solutions \(\delta\df\psi^\alpha\) we have
	\be
		\df\omega(\delta\psi, \Lie_X\psi) = d \left[\delta \df Q_X -  X \cdot \df\theta(\delta\psi) \right]
	\ee
	and the corresponding symplectic form \cref{eq:W-defn} on a Cauchy surface \(\Sigma\) is an integral of a boundary term on \(\partial\Sigma\).
	\be\label{eq:W-boundary}
		W_\Sigma(\psi;\delta\psi,\Lie_X\psi) = \int_{\partial\Sigma} \delta \dfM{\df Q}_X  - \dfM{X} \cdot \dfM{\df\theta}(\delta\psi)
	\ee
	Here we have used the fact that both \(\df Q_X\) and \(\df\theta\) are horizontal (i.e. gauge-invariant) forms on \(P\) and thus can be represented as gauge-invariant forms \(\dfM{\df Q}_X\) and \(\dfM{\df\theta}\) on spacetime \(M\). Further, since \(\df\theta\) is horizontal, only the projection \(\dfM X \equiv X^\mu = (\pi_*)^m_\mu X^\mu\) contributes in the second term, but the Noether charge \(\dfM{\df Q}_X\) depends on the full vector field \(X^m \in \mf{aut}(P)\).

	As discussed in \cite{IW-noether-entropy}, a boundary Hamiltonian for the dynamics generated by \(X^m\) exists if and only if there is a function \(H_X\) on the space of solutions such that its variation is given by
	\be\label{eq:var-H}
		\delta H_X = W_\Sigma(\psi;\delta\psi,\Lie_X\psi) = \int_{\partial\Sigma} \delta \dfM{\df Q}_X - \dfM X \cdot \dfM{\df\theta}(\delta\psi)
	\ee
	which is equivalent to the existence of \(\dfM{\df\Theta}(\psi) \in \Omega^{d-1}M\) so that
	\be\label{eq:Theta}
		\int_{\partial\Sigma} \dfM X \cdot \dfM{\df\theta}(\delta\psi) = \delta \int_{\partial\Sigma} \dfM X \cdot \dfM{\df\Theta}(\psi)
	\ee
	Note here that \(\dfM{\df\Theta}\) need not be covariant or gauge-invariant in its dependence on the dynamical fields. Thus the boundary Hamiltonian becomes
	\be\label{eq:hamiltonian}
		H_X = \int_{\partial\Sigma}\dfM{\df Q}_X - \dfM X \cdot \dfM{\df\Theta}(\psi)
	\ee

	The existence of the Hamiltonian \(H_X\) is intimately related to the boundary conditions at \(\partial\Sigma\) on the dynamical fields \(\df\psi\), the perturbation \(\delta\df\psi\), and the vector field \(X^m\); for general field configurations and arbitrary perturbations the Hamiltonian might not exist or may not be unique. Imposing boundary conditions so that the symplectic form \(W_\Sigma (\psi; \delta\psi, \Lie_X\psi)\) is finite ensures that the perturbed Hamiltonian \(\delta H_X\) is also well-defined. But, even if we choose boundary conditions so that \(\delta H_X\) is well-defined (it is manifestly covariant and gauge-invariant as it is defined in terms of horizontal forms on \(P\)) there still might not exist a unique, or covariant, or gauge-invariant Hamiltonian \(H_X\).

\section{Horizon potentials and charges, and the zeroth law for bifurcate Killing horizons}\label{sec:zeroth-law}

	We describe next the spacetimes for which we will derive a zeroth law for bifurcate Killing horizons and a first law of black hole mechanics.

	To formulate the zeroth law for bifucate Killing horizons, we will consider dynamical fields \(\df\psi^\alpha\) which determine a \(d\)-dimensional spacetime \(M\) with a bifurcate Killing horizon \( \ms H \defn  \ms H^+ \union \ms H^- \) and let the bifurcation surface be \(  B \defn \ms H^+ \inter \ms H^- \). The \emph{horizon Killing field} \(K^\mu\) is null on \(\ms H\) and vanishes on \(B\). We denote the corresponding infinitesimal automorphism on the bundle by \(K^m\), which preserves the background dynamical fields i.e. \(\Lie_K\df\psi^\alpha = 0\). The possible ambiguity in the choice of such \(K^m\) is given in \cref{lem:aut-conn-unique}, \cref{rem:aut-field-unique} and \cref{lem:aut-e-unique}. For the zeroth law, we do not demand that the spacetime described above be determined by solutions to the equations of motion \(\df{\mc E}_\alpha = 0\) (\cref{eq:eom-form}) nor do we require any asymptotic conditions.\\

	\begin{figure}[h!]
	\centering
	\includegraphics[width=0.5\textwidth]{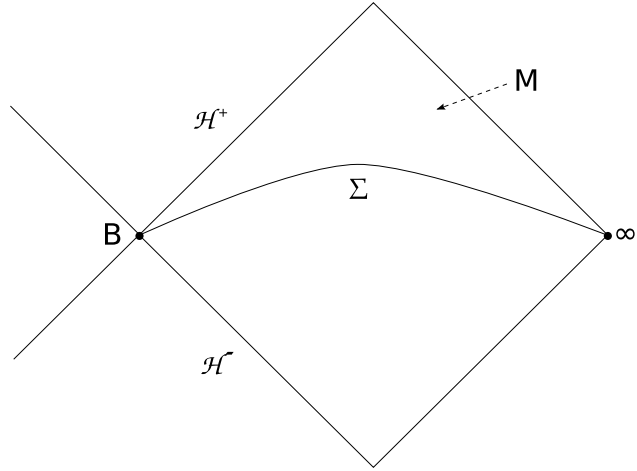}
	\caption{Carter-Penrose diagram of the black hole exterior spacetime \((M,g_{\mu\nu})\).}\label{fig:spacetime}
	\end{figure}

	For the first law of black hole mechanics we will consider stationary and axisymmetric dynamical fields \(\df\psi^\alpha\) (\cref{eq:psi-defn}) which determine a \(d\)-dimensional, \emph{asymptotically flat}, stationary and axisymmetric black hole spacetime with Lorentzian metric \(g_{\mu\nu}\) with a bifurcate Killing horizon as described above. For the bundle \(P\) (\cref{eq:P-split}) on which we have formulated the theory, we choose as the base space \(M\), the exterior (including the horizon) of the black hole (see \cref{fig:spacetime}). The metric \(g_{\mu\nu}\) on \(M\) is determined by the coframes \(\df e^a\) in the standard manner. Thus, we can now identify the abstract Lorentz bundle \(P_O\) in \cref{eq:P-split} with the bundle \(F_OM\) of orthonormal frames determined by \(g_{\mu\nu}\).

	Since, the dynamical fields \(\df\psi^\alpha\) are defined on the bundle \(P\) we define stationarity and axisymmetry of \(\df\psi^\alpha\) using \cref{def:stationary-axisymm} which we summarise as follows. Let \(t^\mu\) denote the time translation Killing field, i.e. the Killing field that is timelike near infinity, and \(\phi^\mu_{(i)}\) the axial Killing fields (we use the index \({(i)}\) to account for more than one axial Killing fields at infinity in greater than \(4\) dimensions). Then, there exist infinitesimal automorphisms, \(t^m,\phi^m_{(i)} \in \mf{aut}(P;\df\psi)\) which preserve the dynamical fields i.e. \(\Lie_t\df\psi^\alpha = 0 = \Lie_{\phi_{(i)}}\df\psi^\alpha\) on the bundle \(P\), which project to the corresponding stationary and axial Killing fields \(t^\mu\) and \(\phi^\mu_{(i)}\) respectively. The ambiguity in the choice of such \(t^m\) and \(\phi^m_{(i)}\) is given in \cref{lem:aut-conn-unique} and \cref{lem:aut-e-unique}. In this case, the horizon Killing field is \(K^\mu = t^\mu + \Omega_{\ms H}^{(i)} \phi^\mu_{(i)}\) (where \(\Omega_{\ms H}^{(i)}\) are constants representing the \emph{horizon angular velocities}) is null on \(\ms H\) and vanishes on \(B\). We denote the corresponding infinitesimal automorphism on the bundle by \(K^m = t^m + \Omega_{\ms H}^{(i)} \phi^m_{(i)}\). 

	We will use the notion of asymptotic flatness defined as follows.\footnote{One could define asymptotic flatness for principal bundles more rigorously, generalising those in Ch.~11 \cite{Wald-book}, however for our purposes it will suffice to specify the fall-off of the fields in terms of some asymptotic radial coordinate.} There exist asymptotically Minkowskian coordinates \((t, x^i)\) near spatial infinity, with \(r \defn \sqrt{\sum_i(x^i)^2}\) being the asymptotic radial coordinate and \(\eta_{\mu\nu} \equiv {\rm diag}(-1,1,\ldots,1)\) be the asymptotic flat metric in these coordinates. We require the asymptotic fall-off of the metric to be
	\be\label{eq:g-falloff}
		g_{\mu\nu} = \eta_{\mu\nu} + O(1/r^{d-3})
	\ee
	with each derivative of the metric falling-off faster by a factor of \(1/r\). To prescribe the fall-off conditions for the dynamical fields defined on the bundle, we lift the asymptotic radial coordinate \(r\) (viewed as a function on \(M\) near infinity) to the bundle near spatial infinity. Then, the fall-off conditions on dynamical fields on the bundle are prescribed as their behaviour in \(1/r\); the precise fall-off conditions are chosen depending on the equations of motion of the theory under consideration so that the solution metric behaves as in \cref{eq:g-falloff}. We expect our results can be generalised to spacetimes with different asymptotics but we stick to the asymptotically flat case.

	In the formulation of the first law we will also consider an asymptotically flat Cauchy surface \(\Sigma\) which smoothly terminates at the bifurcation surface \(B\). We further assume that the embeddings of \(\Sigma\) and \(B\) in \(M\) are regular i.e. they admit smooth no-where vanishing normals.\\

\hr

	Next, we show that the black hole spacetime described above satisfies a generalisation of \emph{the zeroth law for bifurcate Killing horizons} in the sense that, we can define certain potentials which are constant on the bifurcate Killing horizon \(\ms H\). The term ``zeroth law" for this result is justified by \cref{thm:temp-entropy} (see \cref{rem:zeroth-law}), where we show that the horizon potential contributed by the gravitational Lorentz connection can be identified with the surface gravity of the black hole.

	To prove the zeroth law we first show that the Lie-algebra-valued function \(K \cdot \df A^I\) is covariantly constant on the bundle over the bifurcate Killing horizon.
	\begin{prop}\label{prop:K-A-H}
	Let the principal bundle \(P\) restricted to the bifurcate Killing horizon \(\ms H\) be \(P_{\ms H}\). Then, the pullback of \(D(K\cdot \df A^I)\) to \(P_{\ms H}\) vanishes i.e.
	\be
	\lb. D(K\cdot \df A^I) \rb\vert_{P_{\ms H}} = 0
	\ee
	\begin{proof}
	First let \(P_B\) be the restriction of the principal bundle to the bifurcation surface \(B\). Since the horizon Killing field satisfies \(K^\mu\vert_{B} = 0\), the corresponding vector field \(K^m\) is vertical on \(P_B\). As a result we have \(\lb. K \cdot \df F^I \rb\vert_{P_B} = 0\), since \(\df F^I\) is a horizontal form.

	Now, consider any cross-section \(B'\) of the bifurcate Killing horizon \(\ms H\). Along the flow generated by the vector field \(K^\mu\), \(B'\) limits to the bifurcation surface \(B\) (see \S\! 2 \cite{KayWald}). Similarly, the principal bundle \(P_{B'}\) limits to \(P_B\) under the flow along the integral curves of the corresponding vector field \(K^m\) on the bundle \(P_{\ms H}\). Clearly, the pullback of \(K \cdot \df F^I\) to the integral curves of \(K^m\) vanishes, and so consider then \(\lb. K \cdot \df F^I \rb\vert_{P_{B'}}\). Taking the limit of \(\lb. K \cdot \df F^I \rb\vert_{P_{B'}}\) along the flow of \(K^m\) as \(P_{B'} \to P_B\), we have \(\lb. K \cdot \df F^I \rb\vert_{P_{B'}} \to 0\). Since, \(K^m\) is an infinitesimal automorphism which preserves the connection \(\df A^I\), \(\Lie_K (\lb. K \cdot \df F^I) \rb\vert_{P_{\ms H}} = 0\) and since the curvature and \(K^m\) are smooth on the bundle \(P_{\ms H}\) we must have \(\lb. K \cdot \df F^I \rb\vert_{P_{B'}} = 0\), and thus
	\be
		\lb. K \cdot \df F^I \rb\vert_{P_{\ms H}} = 0
	\ee
	Thus, using \cref{eq:Lie-cov-D} for the Lie derivative along \(K^m\) of the connection we have
	\be\label{eq:Lie-A-H}\begin{split}
		0 & = \lb. \Lie_K \df A^I \rb\vert_{P_{\ms H}} = \lb. K \cdot \df F^I \rb\vert_{P_{\ms H}} + \lb. D(K \cdot \df A^I) \rb\vert_{P_{\ms H}} \\
		& = \lb. D(K \cdot \df A^I) \rb\vert_{P_{\ms H}}
	\end{split}\ee
	\end{proof}
	\end{prop}

	To define the horizon potentials we will expand the covariantly constant Lie-algebra-valued function \(K \cdot \df A^I\) on \(P_{\ms H}\) in a suitable choice of basis of the Lie algebra \(\mf g\). To motivate the construction first consider the case where \(\mf g = \mf{su}(2)\) (which can be thought of as the Lie algebra of \(3\)-dimensional Euclidean rotations) and the basis of \(\mf{su}(2)\) given by the Pauli matrices \(\{\sigma^I_x, \sigma^I_y, \sigma^I_z\}\) where \(\sigma^I_z\) is diagonal. Now, we note that any given element of \(\mf{su}(2)\) can be aligned with \(\sigma^I_z\) by the adjoint action of some element in the group \(SU(2)\) (i.e. any direction in \(3\)-dimensional Euclidean space can be rotated to align with the \(Z\)-axis). In particular, we align \(K \cdot \df A^I\) with \(\sigma^I_z\) and define the corresponding ``potential" \(\ms V\) by the expansion \(K \cdot \df A^I = \ms V \sigma^I_z\).

	The above construction for \(\mf{su}(2)\) can be generalised to any semisimple Lie algebra \(\mf g\) by picking a maximal abelian subalgebra of diagonalisable elements called a \emph{Cartan subalgebra} (for \(\mf{su}(2)\) the Cartan subalgebra can be chosen to be spanned by \(\sigma^I_z\) as done above) and the corresponding \emph{Weyl-Chevalley basis} of \(\mf g\); the relevant properties of this construction are recalled in \cref{sec:math}. In the following we assume that a Cartan subalgebra \(\mf h\) of \(\mf g\) has been picked and denote the basis of \(\mf h\) by \(h^I_\Lambda\) where the index \(\Lambda\) enumerates the chosen basis.

	Using the properties of a Cartan subalgebra of \(\mf g\) (see \cref{sec:math}) we define the horizon potentials \(\ms V^\Lambda\) at a point of the bifurcate Killing horizon \(\ms H\) as follows.\footnote{Our strategy to define the horizon potentials in terms of the Cartan subalgebra parallels the one used in \S~V. \cite{CK} to define global charges in Yang-Mills theory.} 

	\begin{prop}[Horizon potentials at a point]\label{prop:potential-defn}
	Let \(\mf h\) be some fixed choice of Cartan subalgebra of \(\mf g\) and let \(h^I_\Lambda\) given by the simple coroots be a choice of basis of \(\mf h\) (see \cref{pr:WC-basis}). For any point \(x \in \ms H\) on the horizon there exists a point \(u \in P_{\ms H}\) such that \(\pi(u) = x\) and \(K \cdot \df A^I(u) \in \mf h\). Thus, in the chosen basis of \(\mf h\) we can write
	\be\label{eq:potential-defn}
		K\cdot \df A^I (u) = \ms V^\Lambda h^I_\Lambda 
	\ee
	The set of coefficients \(\ms V^\Lambda\) is determined up to the action of the Weyl group of \(\mf g\) (see \cref{pr:weyl-group}), irrespective of the chosen point \(u \in \pi^{-1}(x)\) and the chosen simple coroots \(h^I_\Lambda\). We call the coefficients \(\ms V^\Lambda\) in the above expansion, the horizon potentials.
	\begin{proof}
	For any point \(x \in \ms H\), the fibre of the principal bundle \(P_{\ms H}\) over \(x\) is \(\pi^{-1}(x) \cong G\). Note that \(K \cdot \df A^I\) is a \(\mf g\)-valued function with the adjoint representation of \(G\). Using \cref{pr:CSA-map-into}, there exists some point \(u \in \pi^{-1}(x)\) in the fibre where \(K\cdot \df A^I(u) \in \mf h\). Then, using the chosen basis \(h^I_\Lambda\) of \(\mf h\) we can define the coefficients \(\ms V^\Lambda\) as in \cref{eq:potential-defn} at the point \(u \in P_{\ms H}\).

	Suppose there exists another point \(\tilde u \in \pi^{-1}(x)\) such that \(K \cdot \df A^I(\tilde u) \in \mf h\) and hence \cref{eq:potential-defn} holds with some other set of coefficients \(\tilde{\ms V}^\Lambda\). From \cref{pr:CSA-map-into}, the new set of potentials \(\tilde{\ms V}^\Lambda\) are related to the original set \(\ms V^\Lambda\) by the action of some element of the Weyl group on \(\mf h\). 
	Similarly by \cref{pr:weyl-group}, the possible choice of simple coroots are given by the action of the Weyl group on the original choice \(h^I_\Lambda\).

	Thus, the potentials \(\ms V^\Lambda\) are well-defined at any point on the horizon up to the action of the Weyl group of \(\mf g\).
	\end{proof}
	\end{prop}

	By parallel-transporting the covariant constant \(K \cdot \df A^I\) (from \cref{prop:K-A-H}) and using \cref{prop:potential-defn} as the definition of the horizon potentials we show that the potentials can be consistently chosen to be constant over the entire horizon. This gives us the following generalised zeroth law for bifurcate Killing horizons. 

	\begin{thm}[The zeroth law for bifurcate Killing horizons]\label{thm:zeroth-law}
	The horizon potentials \(\ms V^\Lambda\) given by \cref{prop:potential-defn} can be chosen to be constant on the horizon 
	\be\label{eq:zeroth-law}
		\lb. d\ms V^\Lambda \rb\vert_{\ms H} = 0
	\ee
	\begin{proof}
	Let \(x, x' \in \ms H\) be any two points on the horizon connected by a path \(\gamma\), and let \cref{eq:potential-defn} hold at some choice of \(u \in \pi^{-1}(x)\) as discussed in \cref{prop:potential-defn}. Let \(\Gamma\) be the unique path in \(P_{\ms H}\) starting at \(u\) which is horizontal with respect to the given connection \(\df A^I\) and projects to the path \(\gamma\) (see Prop. 3.1 \S~II.3 \cite{KN-book1}), and let \(u' \in \pi^{-1}(x')\) be the endpoint of \(\Gamma\). From \cref{prop:K-A-H} we have \(\lb. D(K \cdot \df A^I) \rb\vert_{P_{\ms H}} = 0\). This implies that we can obtain \(K \cdot \df A^I(u')\) by parallel-transporting \(K\cdot \df A^I(u)\) along \(\Gamma\) to always point in the same Lie algebra direction, and further since \(K\cdot \df A^I\) is covariantly constant on \(P_{\ms H}\), the result is independent of the choice of path \(\gamma\) on the horizon. Thus, \(K \cdot \df A^I(u') \in \mf h\) and \cref{eq:potential-defn} holds at the point \(u' \in \pi^{-1}(x')\) with the same set of potentials \(\ms V^\Lambda\). Since the chosen points \(x,x'\) are arbitrary and \(K\cdot \df A^I\) is smooth, the potentials \(\ms V^\Lambda\) must be constant on the entire horizon.
	\end{proof}
	\end{thm}

	\begin{remark}[Ambiguity in the horizon potentials]\label{rem:amb-potential}
	The first set of ambiguities in the horizon potentials arises due to our choice of a fixed Cartan subalgebra \(\mf h\). From \cref{pr:CSA-iso} we see that different choices of \(\mf h\) will lead to equivalent sets of horizon potentials.

	The other ambiguity in the horizon potentials arises due to our choice of the vector field \(K^m\) on the bundle. From \cref{lem:aut-conn-unique}, given the horizon Killing field \(K^\mu\) on spacetime \(M\), the ambiguity in the corresponding vector field \(K^m\) on \(P\) is given by \(K^m \mapsto \tilde K^m = K^m + Y^m\) where \(Y^m\) is a vertical vector field so that \(Y\cdot \df A^I \in \mf g\) is covariantly constant everywhere (not just on the horizon) on \(P\). Further, if there are other dynamical charged tensor fields (such as the \(\varphi^A\) in \cref{eq:psi-defn}) in the background, then \(Y^m\) is also required to preserve them i.e. \(Y^m\) must also satisfy \cref{eq:aut-field-unique}.  If a non-trivial \(Y^m\) exists for the given dynamical fields \(\df\psi^\alpha\) (\cref{eq:psi-defn}), the new choice \(\tilde K^m\) will define a new set of potentials \(\tilde{\ms V}^\Lambda\) at the horizon. This ambiguity in the potentials does not affect the zeroth law \cref{eq:zeroth-law} since \(\tilde{\ms V}^\Lambda\) are also constant on the horizon. However, there might exist some \(Y^m\) so that we can reduce the number of linearly independent potentials. From \cref{rem:global-symm,rem:aut-field-unique} we see that this ambiguity \(Y^m\) corresponds to a global symmetry of \emph{all} the dynamical fields \(\df\psi^\alpha\). Thus, the number of linearly independent horizon potentials are ambiguous if the dynamical fields \(\df\psi^\alpha\) have a global symmetry on \(P\). In that case, we can use \(Y^m\) to redefine the vector field \(K^m\) on \(P\) so that some, or all, of the horizon potentials vanish.  This redefinition of the horizon potentials also changes the terms at infinity in the first law (see \cref{rem:YM-potential-amb} for the case of Einstein-Yang-Mills theory). Note however, for some given choice of \(\df\psi^\alpha\) that there might not exist any global symmetries and in general one cannot set the horizon potentials to vanish.
	Also from \cref{lem:aut-e-unique}, \(K^m\) is uniquely determined on the Lorentz bundle part of \(P\) (\cref{eq:P-split}) and so this ambiguity does not affect the potentials due to the gravitational Lorentz connection.
	\end{remark}

	\begin{remark}[Independent potentials]\label{rem:independent-potentials}
	From \cref{pr:CSA-iso}, the maximum number of non-zero horizon potentials \(\ms V^\Lambda\) is the dimension of \(\mf h\) i.e. the rank of \(\mf g\). Thus, there are at most \(l\) non-zero potentials for each of the simple Lie algebras of rank \(l\) in Cartan's classification (see Theorem A \S~2.14 \cite{Sam-book}). For \(SU(2)\)-Yang-Mills theory, this reduces to the case considered by Sudarsky and Wald \cite{SW1}, where they find only one Yang-Mills potential. The analysis can be easily extended to include abelian Lie algebras (which are, by definition, neither simple nor semisimple) to find \(l\) horizon potentials for an abelian Lie algebra of dimension \(l\).
	\end{remark}
	\hr

	Using the horizon potentials defined in \cref{prop:potential-defn} we show that the perturbed boundary Hamiltonian \(\delta H_K\) on \(B\) associated to the infinitesimal automorphism \(K^m\) can be put into a ``potential times perturbed charge" form for any theory under consideration, when the dynamical fields \(\df\psi^\alpha\) satisfy the equations of motion and the perturbation \(\delta\df\psi^\alpha\) satisfies the linearised equations of motion.

	\begin{corollary}[Perturbed Hamiltonian and charges on the bifurcation surface]\label{cor:potential-charge}
	The perturbed Hamiltonian on the bifurcation surface \(B\) associated to \(K^m\) can be written as a ``potential times perturbed charge" term of the form
	\be\label{eq:potential-charge-H}
		\lb. \delta H_K\rb\vert_B = \ms V^\Lambda \delta \ms Q_\Lambda
	\ee
	where the horizon potentials \(\ms V^\Lambda\) are as defined in \cref{prop:potential-defn} and the charges \(\ms Q_\Lambda\) are defined by
	\be\label{eq:charge-defn}
		\ms Q_\Lambda \defn \int_B \dfM{ \df Z_I h^I_\Lambda } 
	\ee
	where \(\dfM{ \df Z_I h^I_\Lambda }\) is the gauge-invariant \((d-2)\)-form on \(M\) such that \(\df Z_I h^I_\Lambda = \pi^* \lb( \dfM{ \df Z_I h^I_\Lambda } \rb) \), with \(\df Z_I\) given by \cref{eq:ZI}.
	\begin{proof}
	We first evaluate the perturbed Hamiltonian on the bifurcation surface using \cref{eq:var-H}.  By \cref{lem:hor-theta}, the symplectic potential \(\df\theta\) is horizontal and the second term in \cref{eq:var-H} vanishes at the bifurcation surface since \(\lb.K^m\rb\vert_{P_B}\) is vertical. Similarly, the second term in the form of the Noether charge \cref{eq:noether-charge-form} for \(K^m\) vanishes. Thus, the perturbed horizon Hamiltonian associated to \(K^m\) is
	\[
		\lb.\delta H_K\rb\vert_B = \int_B \delta \dfM{\df Q}_K = \int_B \dfM{ \delta \df Z_I ~( K \cdot \df A^I )}
	\]
	where in the last equality we again use the fact that \(K^m\) is vertical and \(\delta \df A^I\) is horizontal.

	Then, using the definition of the horizon potentials (\cref{eq:potential-defn}) and the zeroth law (\cref{eq:zeroth-law}) we get the form of the perturbed Hamiltonian in \cref{eq:potential-charge-H} with the charges \(\ms Q_\Lambda\) given by \cref{eq:charge-defn}.
	\end{proof}
	\end{corollary}

	\begin{remark}[Independent charges]\label{rem:independent-charges}
	We note, from \cref{pr:WC-basis}, that only the projection of \(\df Z_I\) to the chosen Cartan subalgebra contributes to the charges in \cref{eq:charge-defn}. Further, from \cref{pr:CSA-non-degen}, the maximum number of non-zero charges \(\ms Q_\Lambda\) is the dimension of \(\mf h\) i.e. the rank of \(\mf g\). For \(SU(2)\)-Yang-Mills theory there is only one Yang-Mills charge as found in \cite{SW1}.
	\end{remark}

	We can show that the ambiguities in the symplectic potential \cref{eq:mu-amb,eq:lambda-amb} and the Noether charge \cref{eq:J-Q-amb} do not affect the perturbed Hamiltonian \(\delta H_K\vert_B\) (the following argument also apply to the perturbed Hamiltonian at spatial infinity). These ambiguities give rise to the following change
	\[\begin{split}
		\delta \df Q_K -  K \cdot \df\theta(\delta\psi) \mapsto & ~\quad \delta \df Q_K -  K \cdot \df \theta(\delta\psi) + \delta \df\lambda(\delta_K\psi) + d\delta\df\rho - K \cdot d\df\lambda(\delta\psi)  \\
	& = \delta \df Q_K -  K \cdot \df\theta(\delta\psi) + \delta \df\lambda(\delta_K\psi) - \Lie_K \df\lambda(\delta\psi) + d \lb[ K \cdot \df\lambda(\delta\psi) +  \delta\df\rho \rb]
	\end{split}\]
	Since \(K^m\) is an infinitesimal automorphism which preserves the background dynamical fields \(\df\psi^\alpha\) we have
	\[
		\delta\df\lambda(\delta_K\psi) = \delta\df\lambda[\psi;\Lie_K\psi] = \df\lambda[\psi;\Lie_K\delta\psi] = \Lie_K \df\lambda(\delta\psi)
	\]
	and thus
	\[
	\delta \df Q_K -  K \cdot \df\theta(\delta\psi) \mapsto \delta \df Q_K -  K \cdot \df\theta(\delta\psi) + d \lb[ K \cdot \df\lambda(\delta\psi) +  \delta\df\rho \rb]
	\]
	Since \(\df\lambda\) and \(\df\rho\) are local and covariant horizontal forms, the integral of \(\delta \dfM{\df Q}_K - \dfM{K} \cdot \dfM{\df\theta}(\delta\psi)\) over a closed surface (the bifurcation surface \(B\)), and consequently the perturbed boundary Hamiltonian \(\delta H_K\vert_B\) (from \cref{eq:var-H}), is unambiguous. Since the potentials are defined independently of these ambiguities, from \cref{eq:potential-charge-H} we see that the charges \(\ms Q_\Lambda\) are also unaffected by these ambiguities.

\subsection{Temperature and entropy as the horizon potential and charge for gravity}\label{sec:temp-entropy}

	In the following, we show that the gravitational potential and the perturbed gravitational charge corresponding to the Lorentz connection \(\df\omega^a{}_b\) can be identified (up to conventions of numerical factors) with the temperature and perturbed entropy of the black hole respectively. Thus, the first-order formulation of gravity in terms of the coframes \(\df e^a\) and the Lorentz connection \(\df\omega^a{}_b\) on the Lorentz bundle \(P_O\) gives a new point of view on the temperature and perturbed entropy of the horizon. In particular, the temperature and perturbed entropy can be viewed on the same footing as the potentials and perturbed charges of any matter gauge fields in the theory. Further, we give an explicit formula for the gravitational charge which is a direct parallel of the Wald entropy formula \cite{W-noether-entropy, IW-noether-entropy}.

	\begin{thm}[Temperature and perturbed entropy]\label{thm:temp-entropy}
	The gravitational potential (corresponding to the Lorentz connection \(\df\omega^a{}_b\) on the Lorentz bundle \(P_O\) in \cref{eq:P-split}) at the bifurcation surface \(B\) only has non-vanishing values in a \(1\)-dimensional vector space spanning the abelian Lie algebra \(\mf{so}(1,1)\) corresponding to local Lorentz boosts of the frames \(\tilde E^m_a\) normal to \(B\) in \(M\). Let \(\tilde\epsilon^{ab} = \tilde E^b \cdot \tilde E^a \cdot \df{\tilde\varepsilon}_2\) be the frame components of the binormal to \(B\) along normal frames \(\tilde E^m_a\). Then, \(-\tilde\epsilon^{ab}\) forms a basis of \(\mf{so}(1,1)\) so that the gravitational potential and charge can be written as 
	\be\label{eq:grav-potential-charge}
		\ms V_{grav} = \surgrav \eqsp \ms Q_{grav} = -\int_B \dfM{ \df Z_{ab} \tilde\epsilon^{ab}}
	\ee
	where \(\surgrav\) is the surface gravity of \(B\). In \cref{eq:grav-potential-charge} \(\dfM{ \df Z_{ab} \tilde\epsilon^{ab}}\) is the unique gauge-invariant form on spacetime that pullsback to \(\df Z_{ab} \tilde\epsilon^{ab}\) on the bundle. \(\df Z_{ab}\) can be computed from \cref{eq:ZI} for the Lorentz connection \(\df\omega^a{}_b\), and thus, the gravitational charge formula is in direct parallel to the Wald entropy formula \cite{W-noether-entropy, IW-noether-entropy}. 

	Thus, we can define the temperature and the perturbed entropy of the bifurcate Killing horizon as
	\be\label{eq:temp-entropy-defn}
	T_{\ms H} \defn \frac{1}{2\pi}\ms V_{grav} \eqsp \delta S \defn 2\pi \delta \ms Q_{grav}
	\ee
	\begin{proof}
	Since, for a stationary axisymmetric black hole, \(K^m \in \mf{aut}(P;\df e^a)\) is an infinitesimal automorphism which preserves the coframes, we have from \cref{lem:aut-e-unique}
	\[
		K \cdot \df \omega^{ab} = \tfrac{1}{2}E_a \cdot E_b \cdot d\df\xi + (K \cdot \df e^c) C_{cab}
	\]
	where \(\df\xi = (K\cdot \df e^a)\df e_a\) is the pullback to the bundle of the Killing form \(\dfM{\df \xi} \equiv \xi_\mu = g_{\mu\nu}K^\nu\) and the contorsion \(C_{abc}\) is defined in \cref{eq:contorsion}. The vector field \(K^m\) is vertical on \(P_B\) and using usual definition of the surface gravity \(\surgrav\) we can write the Killing form on the bifurcation surface as \(d\dfM{\df\xi}\vert_B = 2\surgrav \dfM{\df{\tilde\varepsilon}}_2\), where  \(\dfM{\df{\tilde\varepsilon}}_2\) is the binormal to \(B\) (see \S~12.5 \cite{Wald-book}). Thus we have
	\be\label{eq:A-surgrav}
		K \cdot \df \omega^{ab}\vert_{P_B} = -\surgrav E^b \cdot E^a \cdot \df{\tilde\varepsilon}_2
	\ee
	where \(\df{\tilde\varepsilon}_2\) is the binormal \(\dfM{\df{\tilde\varepsilon}}_2\) lifted to the bundle \(P_B\). Note that the torsion terms vanish due to \(K^m\) being a vertical vector field on \(P_B\).

	The surface gravity is constant on a bifurcate Killing horizon (see \cite{KayWald} and also \cref{rem:other-zeroth-law} below) i.e. \(d\surgrav\vert_{P_B} = 0\), and using \cref{eq:Lie-A-H} for the Lorentz connection, we get \(D (E^b \cdot E^a \cdot \df{\tilde\varepsilon}_2)\vert_{P_B} = 0\). Thus the Lorentz bundle \(P_O\vert_B\), identified with the orthonormal frame bundle \(F_OM\vert_B\) of the background solution metric, can be reduced to the \emph{bundle of adapted orthonormal frames} (see \S~VII.1 \cite{KN-book2}) i.e. \(P_O\vert_B \cong F_OB \oplus F_NB \) where \(F_OB\) is the orthonormal frame bundle of \(B\) and \(F_NB\) is the \(O(1,1)\)-bundle of frames normal to \(B\) in \(M\). Denote the adapted normal frames in \(F_NB\) by \(\tilde E_a^m\).

	From Prop.~1.4 in \S VII.1 \cite{KN-book2} we see that the connection \({\df \omega^a}_b\vert_{P_B}\) can be written as a direct sum of a \(O(d-2)\)-connection on \(F_OB\) and an abelian \(O(1,1)\)-connection \(\df \omega_N\) on \(F_NB\). The invariant tensor \(\tilde\epsilon^{ab} = \tilde E^b \cdot \tilde E^a \cdot \df{\tilde\varepsilon}_2 \) acts as a projector to this abelian \(O(1,1)\)-connection as \(\df \omega_N \defn \tfrac{1}{2}\df \omega_{ab} \tilde\epsilon^{ab}\). Thus, \cref{eq:A-surgrav} becomes \(K \cdot \df \omega_N = \surgrav\) i.e. the surface gravity is the vertical part of \(K^m\) (with respect to the background connection) in the normal frame bundle of \(B\).\\

	A choice of Cartan subalgebra \(\mf h\) of the Lorentz Lie algebra is spanned by boosts in \(\mf{so}(1,1)\) normal to \(B\) and some choice of commuting rotations in \(\mf{so}(d-2)\). Since, \(K\cdot \df \omega^{ab}\) only points in the \(\mf{so}(1,1)\)-part we have (choosing \(-\tilde\epsilon^{ab}\) as a basis of \(\mf{so}(1,1)\))
	\[
		\ms V_{grav} = K \cdot \df \omega_N = \surgrav
	\]
	and the corresponding charge, using \cref{eq:charge-defn}
	\[
		\ms Q_{grav} = -\int_B \dfM{ \df Z_{ab} \tilde\epsilon^{ab}}
	\]
	and \(\df Z_{ab}\) is given by \cref{eq:ZI} for the Lorentz connection \(\df\omega^a{}_b\). Thus, the gravitational charge is determined in direct parallel to the Wald entropy formula \cite{W-noether-entropy, IW-noether-entropy} for any theory of gravity formulated in terms of the coframes and a Lorentz connection.
	\end{proof}
	\end{thm}

	Note that since the coframes completely fix the form of the infinitesimal automorphism \(K^m\) (see \cref{lem:aut-e-unique}) we cannot eliminate the temperature and the perturbed entropy by a redefinition of \(K^m\) in any spacetime. For General Relativity, the gravitational charge can be computed to be \( \ms Q_{grav} = \tfrac{1}{8\pi}{\rm Area}(B) \) (see \cref{sec:palatini-holst}) and we get the usual notion of temperature and perturbed entropy for black holes.

	\begin{remark}[Zeroth law]\label{rem:zeroth-law}
	The gravitational potential i.e. the surface gravity \(\surgrav\), was shown to be constant on bifurcate Killing horizons in \cite{KayWald} without using the Einstein equations. Thus, in view of \cref{thm:temp-entropy}, we can see that \cref{thm:zeroth-law} can be seen as a ``generalised zeroth law for bifurcate Killing horizons" (analogous to the result of \cite{KayWald}) showing that the potentials defined in \cref{prop:potential-defn} are always constant on a bifurcate Killing horizon without using any equations of motion.
	\end{remark}

	\begin{remark}[Other versions of the zeroth law]\label{rem:other-zeroth-law}
Rácz and Wald \cite{RW} showed that if the surface gravity has non-vanishing gradient on any null geodesic generator of a (not necessarily bifurcate) Killing horizon then there necessarily exists a parallel-propagated curvature singularity on the horizon, without using the Einstein equations. If one uses the Einstein equations and the dominant energy condition on matter fields, then the surface gravity was shown to be constant on any (not necessarily bifurcate) Killing horizon in \cite{BCH}. The results of \cite{RW, BCH} can also be viewed as different versions of ``the zeroth law".
	\end{remark}

\section{The first law for gauge-invariant Lagrangians}\label{sec:first-law}

	Next, we formulate the first law of black hole mechanics for a stationary-axisymmetric black hole solution described at the beginning of \cref{sec:zeroth-law}. The first law is obtained by evaluating the symplectic form on any Cauchy surface \(\Sigma\) for \(X^m = K^m\) where \(K^m\) is the infinitesimal automorphism that projects to the horizon Killing field \(K^\mu\).  Since the black hole is stationary and axisymmetric \(\Lie_K\df\psi^\alpha = 0\), using \cref{eq:W-boundary} for the symplectic form, the first law is an equality of the perturbed boundary Hamiltonians \(\delta H_K\) evaluated at the bifurcation surface and at spatial infinity.

	The perturbed Hamiltonian on the bifurcation surface was already put into a ``potential times perturbed charge" form in \cref{cor:potential-charge}. Near spatial infinity, we can lift the asymptotic Minkowski radial coordinate \(r\) (viewed as a gauge-invariant function) to the bundle \(P\). We choose the dynamical fields and their perturbations to fall-off suitably in \(1/r\) so that the symplectic form \(W_{\Sigma}\) is finite. The particular choice of fall-off in general depends on the specific Lagrangian theory under consideration. For the infinitesimal automorphisms \(t^m\) and \(\phi^m_{(i)}\) we define the \emph{canonical energy} and \emph{canonical angular momentum} as the corresponding Hamiltonians (whenever they exist) at infinity (see \cite{IW-noether-entropy}).
	\be\label{eq:E-J-can}\begin{split}
		E_{can} &\defn \lb. H_t\rb\vert_\infty = \int_\infty \dfM{\df Q}_t - \dfM t \cdot \dfM{\df \Theta} \\
		J_{(i),can} &\defn - \lb. H_{\phi_{(i)}}\rb\vert_\infty = -\int_\infty \dfM{\df Q}_{\phi_{(i)}} - \dfM \phi_{(i)} \cdot \dfM{\df \Theta}
	\end{split}\ee
	where \(\dfM{\df\Theta}\) is as described by \cref{eq:Theta}. Thus, the perturbed Hamiltonian at infinity associated to \(K^m = t^m + \Omega^{(i)}_{\ms H} \phi^m_{(i)}\) becomes \( \delta H_K\vert_\infty = \delta E_{can} - \Omega_{\ms H}^{(i)} \delta J_{(i),can} \).	In all the examples we consider in \cref{sec:examples}, only the Einstein-Hilbert Lagrangian contributes a non-zero \(\dfM{\df \Theta}\) at infinity and for all other cases \(\dfM K \cdot \dfM{\df \theta}\) falls off fast enough that we can choose \(\dfM{\df \Theta} = 0\). Note that in \(4\)-dimensions, we might need to impose faster fall-off conditions or some suitable generalisation of the Regge-Teitelboim parity conditions \cite{RT} on the dynamical fields and their perturbations for the \(J_{(i),can}\) to be well-defined; we assume that such choices have been made to get a well-defined canonical angular momentum.\\

	This leads to our main result in the following theorem which gives us a general formulation of the first law of black hole mechanics.

	\begin{thm}[The first law of black hole mechanics]\label{thm:first-law-gen}
	Consider any theory with a local, covariant and gauge-invariant Lagrangian of the form \cref{eq:L}. Let \(\df\psi^\alpha\) be a solution corresponding to a stationary axisymmetric black hole with a bifurcate Killing horizon (described at the beginning of \cref{sec:zeroth-law}) and \(\delta\df\psi^\alpha\) be an arbitrary linearised solution. Then the first law of black hole mechanics takes the form
	\be\label{eq:first-law}
	T_{\ms H}\delta S + \ms V'^\Lambda \delta\ms Q'_\Lambda = \delta E_{can} - \Omega_{\ms H}^{(i)}~ \delta J_{(i),can}
	\ee
	where, on left-hand-side the first term consists of the temperature and perturbed entropy of the black hole as described in \cref{thm:temp-entropy} and the second term is the potential and perturbed charge of the connection \(\df A'^{I'}\) (see \cref{eq:conn-split}) described in \cref{prop:potential-defn} and \cref{eq:charge-defn}, and the terms on the right-hand-side being defined at infinity by \cref{eq:E-J-can}.
	\begin{proof}
	The first law is obtained by evaluating the expression \cref{eq:W-boundary} with \(X^m = K^m\) on a hypersurface \(\Sigma\) which goes from the bifurcation surface \(B\) to spatial infinity. For a stationary axisymmetric black hole \(\Lie_K \df\psi^\alpha = 0\) and the left-hand-side of \cref{eq:W-boundary} vanishes and then the first law equates the perturbed boundary Hamiltonian of \(K^m\) at \(B\) to the one at infinity.
	\be
		\delta H_K \vert_B = \delta H_t \vert_\infty - \Omega^{(i)}_{\ms H} \delta H_{\phi_{(i)}}\vert_\infty
	\ee
	Using \cref{cor:potential-charge}, \cref{eq:conn-split}, \cref{thm:temp-entropy} and \cref{eq:E-J-can} we get the first law \cref{eq:first-law}.
	\end{proof}
	\end{thm}
	At this point, we emphasise that we have not assumed any choice of gauge in the above form of the first law, and in fact this form holds even when the principal bundle is non-trivial and no choice of gauge can be made.

	As discussed after \cref{cor:potential-charge}, the perturbed Hamiltonian at \(B\), and also at infinity is not affected by the ambiguities in the Lagrangian and the symplectic potential. However there is an ambiguity in choosing the vector field \(K^m\) on the bundle \(P\) (see \cref{lem:aut-conn-unique} and \cref{rem:global-symm,rem:aut-field-unique}) corresponding to a global symmetry of the background dynamical fields \(\df\psi^\alpha\) if any exists. Note that the vector field \(K^m\) is uniquely determined over the Lorentz bundle \(P_O\) part (from \cref{lem:aut-e-unique}) and thus, the temperature and perturbed entropy, as well as, the gravitational contributions to the canonical energy and canonical angular momentum are unambiguous. Thus, the possible ambiguity in choosing \(K^m\) leads to a simultaneous redefinition of the horizon potentials \(\ms V'^\Lambda\) and charges \(\ms Q'_\Lambda\), and the contributions to the canonical energy and angular momenta at infinity of any non-gravitational fields. One can use such an ambiguity to set some, or possibly all, of the horizon potentials at the horizon to vanish (see \cref{rem:amb-potential}) at the cost of changing the contributions to the canonical energy and canonical angular momenta (see also \cref{rem:YM-potential-amb} for Einstein-Yang-Mills theory).  Even though this ambiguity affects the individual terms, the form of the first law \cref{eq:first-law} holds for any choice of \(K^m\) on the bundle.\\

	Even though \(\delta H_K\) is unambiguously defined (given a choice of \(K^m\)), the Hamiltonian \(H_K\) (if it exists) is not. Consider first the Hamiltonian at the bifurcation surface \(B\) (the analysis proceeds similarly for spatial infinity). Since the \(\df\theta \) contribution vanishes at \(B\), the choice of \(\dfM{\df\Theta}\) has the ambiguity
	\[
		\int_B \dfM K\cdot \dfM{\df\Theta} \mapsto \int_B \dfM K\cdot \dfM{\df\Theta} - \int_B \dfM{K} \cdot \dfM{\df\Lambda}
	\]
	where \(\int_B \dfM{K} \cdot \dfM{\df\Lambda}\) is some topological invariant of \(B\) (possibly depending on the embedding of \(B\) in \(M\)) and does not change under variations. Similarly for the Noether charge at \(B\) we have the ambiguities
	\[
		\int_B \dfM{\df Q}_K \mapsto \int_B \dfM{\df Q}_K + \int_B \dfM{K \cdot \df\mu}
	\]
	Thus the ambiguity in the Hamiltonian \(H_K\) is of the form
	\[
		H_K \mapsto H_K + \int_B \dfM{K} \cdot \dfM{\df\Lambda} +\int_B \dfM{K \cdot \df\mu}
	\]
	But we have already shown that \(\delta H_K\) is unambiguous. Thus, any contribution of the ambiguities \(\df\mu\) and \(\dfM{\df\Lambda}\) to the boundary Hamiltonian can be considered as a \emph{topological charge}. Similarly, we can have topological charge contributions to the Hamiltonian at spatial infinity. In the gravitational case, the topological charges at spatial infinity can be fixed by requiring that flat Minkowski spacetime have vanishing ADM mass and ADM angular momentum. Nevertheless, it is possible to have topologically non-trivial solutions to Yang-Mills theory which result in non-trivial topological charges (e.g. magnetic monopole charges) at the horizon. In fact such topological charges do arise in Yang-Mills theory when we add the \(\df\mu\)-ambiguity to the Lagrangian (\cref{sec:yang-mills}). Similarly, the addition of the Euler density to the Einstein-Hilbert Lagrangian corresponds to the \(\df\mu\)-ambiguity \cref{eq:mu-amb} where \(\df\mu\) is not horizontal\footnote{Iyer and Wald \cite{IW-noether-entropy} only considered \(\df\mu\) that were gauge-invariant in which case the \(\df\mu\)-ambiguity does not affect the Hamiltonian at \(B\).} (see \cref{sec:palatini-holst}) and does contribute a topological term to the Noether charge at \(B\) \cite{JM-lovelock}. Even though these topological charges do not affect the first law of black hole mechanics for stationary black holes, they do affect any attempt to define a total entropy and charge for stationary black holes purely from the first law, as we shall discuss later.\\

	Since the perturbed entropy is given by the perturbed gravitational charge, Iyer and Wald \cite{W-noether-entropy, IW-noether-entropy} prescribe that the total entropy (known as the \emph{Wald entropy}) for a stationary axisymmetric black hole be defined as the gravitational charge
	\be\label{eq:IW-entropy}
		S_{\rm Wald} \defn \ms Q_{grav} = - \int_B \dfM{ \df Z_{ab} \tilde\varepsilon^{ab}}
	\ee
	This prescription has the advantage that the entropy \(S_{\rm Wald}\) satisfies the first law and is the same on any cross-section of the horizon of a stationary axisymmetric black hole since, \(\df J_K = 0\) on the horizon \cite{JKM} . But this prescription is not unambiguous. For instance, an alternative definition of the entropy as \(S = S_{\rm Wald} + C(B)\) where \(C(B)\) is a topological invariant of the bifurcation surface \(B\) also satisfies the first law, since \(C(B)\) does not change under linearised variations \cite{Sar-Wall}. In fact the \(\df\mu\)-ambiguity \cref{eq:mu-amb} in the Lagrangian contributes a topological charge of precisely this nature. In the case of General Relativity, as pointed out by \cite{JM-lovelock, Sar-Wall}, the gravitational charge does acquire such a topological contribution (the Euler number of \(B\)) when the Euler density is added to the Einstein-Hilbert Lagrangian in \(4\)-dimensions even though the equations of motion are unaffected (also see \cref{sec:palatini-holst}). The \emph{area theorem} for General Relativity in \(4\)-dimensions (along with an energy condition) guarantees that the entropy defined as the area of a horizon cross-section always increases and thus satisfies the \emph{second law of black hole mechanics} (see \cite{Haw}, Prop.~9.2.7 \cite{Hawking-Ellis} or Theorem 12.2.6 \cite{Wald-book}). If one includes the topological charges of the horizon in the definition of the entropy then the second law is violated even for General Relativity (see \cite{JM, Sar-Wall, Liko}). Thus, it seems that a version of the second law is needed to fix at least some of the ambiguities in defining the total entropy even for a stationary black hole. To consider the second law one has to consider non-stationary black hole configurations where the entropy prescription \cref{eq:IW-entropy} can have more ambiguities which vanish only in the stationary case \cite{JKM}.  Unfortunately, a general formulation of the second law for an arbitrary theory of gravity including arbitrary matter fields remains out of reach; though it has been investigated in special situations for higher curvature gravity \cite{JKM2, Wall}. In the absence of a second law, the first law \cref{eq:first-law} only determines the perturbed entropy \(\delta S\). In light of this we will refrain from giving a prescription for the total entropy \(S\) for stationary black holes or for a dynamical entropy for non-stationary ones.

\section{Examples}\label{sec:examples}

	In this section we use the formalism described above to derive the first law of black hole mechanics for the first-order coframe formulation of General Relativity, Einstein-Yang-Mills theory and Einstein-Dirac theory. The Lagrangians considered in this section are of the form \(\df L = \df L_{\rm grav} + \df L_{\rm matter}\), where the gravitational Lagrangian \(\df L_{\rm grav}\) only depends on the coframes \(\df e^a\) and a Lorentz connection \({\df \omega^a}_b\) on the Lorentz bundle \(P_O\) and not on the matter fields.\footnote{For the case of dilaton gravity considered in \cite{IW-noether-entropy} the gravitational Lagrangian does depend on an additional scalar field. We will not consider such examples in this section but they are covered in the more general formulation in \cref{sec:first-law}.} It will be convenient to work in the \emph{first-order formalism} where we consider the coframes \(\df e^a\) and the Lorentz connection \(\df\omega^a{}_b\) as independent fields. We will write the equations of motion obtained by varying the Lagrangian with respect to the coframes and the Lorentz connection as
	\be\label{eq:eom-examples}
		\df{\mc E}_a - \df{\mc T}_a = 0 \eqsp \df{\mc E}_{ab} - \df{\mc S}_{ab} = 0
	\ee
	where the gravitational contributions to the equations of motion are
	\be\begin{split}
		\df{\mc E}_a \equiv ({\mc E}_a)_{m_1\ldots m_{d-1}} &= \frac{1}{d} \frac{\delta (L_{\rm grav})_{m_1\ldots m_{d-1}l}}{\delta e_n^a} \delta^{l}_{n} \\
		\df{\mc E}_{ab} \equiv ({\mc E}_{ab})_{m_1\ldots m_{d-1}} &= \frac{1}{d} \frac{\delta (L_{\rm grav})_{m_1\ldots m_{d-1}l}}{\delta \omega_n^{ab}} \delta^{l}_{n}
	\end{split}\ee
	and the matter contributions are
	\be\begin{split}
		\df{\mc T}_a \equiv ({\mc T}_a)_{m_1\ldots m_{d-1}} 
&= - \frac{1}{d} \frac{\delta (L_{\rm matter})_{m_1\ldots m_{d-1}l}}{\delta e_n^a} \delta^{l}_{n} \\
		\df{\mc S}_{ab} \equiv ({\mc S}_{ab})_{m_1\ldots m_{d-1}} &= -\frac{1}{d} \frac{\delta (L_{\rm matter})_{m_1\ldots m_{d-1}l}}{\delta \omega_n^{ab}} \delta^{l}_{n}
	\end{split}\ee
	i.e. \(\df{\mc T}_a\) is the \emph{energy-momentum} and \(\df{\mc S}_{ab}\) is the \emph{spin current} of the matter fields, both written as \((d-1)\)-forms. We note here that when the matter Lagrangian depends on the gravitational Lorentz connection used, \(\df{\mc T}_a\) does not give the usual symmetric energy-momentum tensor \cite{Hehl, BH}.

	\begin{remark}[Second-order formalism]\label{rem:second-order}
 If one insists on having vanishing torsion from the outset (i.e. one works in the \emph{second-order formalism}) then the Lorentz connection is completely determined by the coframes (see \cref{rem:LC-conn}) and one can use \cref{eq:LC-conn} (and \cref{eq:var-frame}) to convert all the variations of the torsionless connection to variation of the coframes. In that case, the second equation of motion in \cref{eq:eom-examples} is deleted but the ``new" energy-momentum tensor \(\df{\mc T}_a\) gets contributions from the spin current \cite{Hehl, BH}.
	\end{remark}

	We consider the first-order formulation of General Relativity with the gravitational Lagrangian to be given by the Palatini-Holst Lagrangian. We show that the gravitational charge \cref{eq:grav-potential-charge} is given by the area of the bifurcation surface and thus we reproduce the usual identification between the perturbed entropy and perturbed area of the bifurcation surface. Similarly, (up to terms involving torsion) \cref{eq:E-J-can} reproduces the ADM mass and ADM angular momentum and we get the usual first law of black hole mechanics for General Relativity. For the matter Lagrangian we consider the two cases of Yang-Mills Lagrangian for gauge fields of any semisimple group, and the free Dirac Lagrangian for spinor fields. These examples can be generalised to include chiral spinor fields with Yang-Mills charge, charged scalar fields such as the Higgs field, and thus, the entire Standard Model of particle physics. We shall work out the details in \(4\)-spacetime dimensions but the computations can be easily generalised to other dimensions. We also illustrate the topological charge ambiguities that arise in the definitions of the Hamiltonian and the entropy.

\subsection{Palatini-Holst}\label{sec:palatini-holst}

	To start let us consider the \emph{first-order formulation} of General Relativity in \(4\)-spacetime dimensions in vacuum. A derivation of the first law in this case was recently given in \cite{JM} by using a generalised notion of Lie derivatives of the coframes called the \emph{Lorentz-Lie derivative}. As discussed in \cref{sec:fields-L-bundle} the Lorentz-Lie derivative defined in \cite{JM} depends non-linearly on the coframes and does not form a Lie algebra for diffeomorphisms of spacetime. Thus, the Lorentz-Lie derivative is not the generator of any group action on the coframes. We will write the first-order Palatini-Holst action on the oriented Lorentz bundle i.e. for this section \(P = P_{SO}\) over spacetime \(M\) and use the notion of the Lie derivative on the bundle to obtain a first law. When a vector field \(X^m\) is an automorphism which preserves the coframes the bundle notion of Lie derivative coincides with the Lorentz-Lie derivative defined by \cite{JM}. Thus, even though the Noether charge we define for arbitrary automorphisms of the frame bundle in not equivalent to that of \cite{JM} we get the same results for the first law for stationary axisymmetric black holes.\\

	The dynamical fields for the first-order formulation of General Relativity are the coframes \(\df e^a \in \Omega^1_{hor}P(\bb R^4)\) and the Lorentz connection \({\df \omega^a}_b \in \Omega^1P(\mf{so}(3,1))\). The \emph{Palatini-Holst Lagrangian} for General Relativity can be written as:
	\be\label{eq:L-PH}
		\df L_{PH} = \tfrac{1}{32\pi}\phi_{abcd}~ \df e^a \wedge \df e^b \wedge \df R^{cd} \in \Omega^4_{hor}P
	\ee
	where \(\phi_{abcd} \defn \epsilon_{abcd} + \tfrac{2}{\gamma}\eta_{a[c}\eta_{d]b} \) is an \(SO(3,1)\)-invariant tensor with \(\gamma\) being the \emph{Barbero-Immirzi parameter}. This tensor satisfies the index symmetries \(\phi_{abcd} = \phi_{cdab}\) and \(\phi_{abcd} = - \phi_{bacd} = -\phi_{abdc}\). The \(\epsilon\)-term is the usual \emph{Palatini-Einstein-Hilbert Lagrangian} and the \(\gamma\)-term corresponds to the \emph{Holst Lagrangian} \cite{Holst}. The corresponding Lagrangian form on spacetime is
	\[
		\dfM{\df L}_{PH} = \frac{1}{16\pi} \dfM{\df\varepsilon}_4 \left( R - \frac{1}{2\gamma}  \varepsilon^{\mu\nu\lambda\rho} R_{\mu\nu\lambda\rho} \right)
	\]
	Note that the last term vanishes (using \cref{eq:torsion-bianchi}) when one restricts, a priori, to a torsionless connection.\\

	A variation of the Lagrangian gives \( \delta \df L_{PH} = \df{\mc E}_a \wedge \delta \df e^a + \df{\mc E}_{ab} \wedge \delta \df \omega^{ab} + d\df\theta \) where:
	\be\begin{split}
		\df{\mc E}_a & = -\tfrac{1}{16\pi} \phi_{abcd}~ \df e^b \wedge \df R^{cd} \\
		\df{\mc E}_{ab} & = \tfrac{1}{16\pi}\phi_{abcd}~ \df e^c \wedge \df T^d \\
		\df\theta & = \tfrac{1}{32\pi}\phi_{abcd}~ \df e^a \wedge \df e^b \wedge \delta \df \omega^{cd}
	\end{split}\ee
	while the symplectic current is:
	\be
		\df\omega = \tfrac{1}{16\pi}\phi_{abcd}~ \df e^a \wedge ( \delta_1 \df e^b \wedge \delta_2 \df \omega^{cd} - \delta_2 \df e^b \wedge \delta_1 \df \omega^{cd} )
	\ee
	In vacuum the equation of motion \(\df{\mc E}_{ab} = 0\) implies that \(\df T^a = 0\) \cite{Holst}. Since we define the Noether current and Noether charge ``off-shell" we will not make use of this fact. Also, in \cref{sec:dirac} the torsion can be sourced by the spin current of matter fields and so it will be useful to keep this term to analyse the effects of torsion.

	A straightforward computation gives the Noether current \cref{eq:noether-current-defn} for \( X^m \in \mf{aut}(P) \) as:
	\be
		\df J_X = \df\theta_X - X \cdot \df L = d\df Q_X + \df{\mc E}_a (X \cdot \df e^a) + \df{\mc E}_{ab} (X \cdot \df \omega^{ab})
	\ee
	with the Noether charge \(\df Q_X = \tfrac{1}{32\pi}\phi_{abcd}~ \df e^a \wedge \df e^b (X \cdot \df \omega^{cd}) \in \Omega^2_{hor}P\).\\

	Now, consider \( X^m \in \mf{aut}(P; \df e^a, {\df \omega^a}_b)\) i.e. an infinitesimal automorphism that preserves the coframes and the connection. From \cref{lem:aut-e-unique} such an automorphism is uniquely determined by a Killing field of the spacetime metric determined by the coframes. The Noether charge for such an \(X^m\) becomes:
	\be\begin{split}
		\df Q_X & = \frac{1}{16\pi}\frac{1}{4}\phi_{ab}{}^{cd} \df e^a \wedge \df e^b  ~\left( E_c \cdot E_d \cdot d\df\xi + 2(X \cdot \df e^e) C_{ecd} \right) \\
	&= -\frac{1}{16\pi} \lb( \star \left[ d\df\xi - 2(X \cdot \df e^a) \df C_a \right] + \frac{1}{\gamma} \left[ d\df\xi - 2(X \cdot \df e^a) \df C_a \right] \rb)
	\end{split}\ee
	where \(\df\xi = (X \cdot \df e^a) \df e_a\), \(\star\) is the horizontal Hodge dual operation on differential forms on the Lorentz bundle \cref{eq:hor-Hodge} and we have written the contorsion \cref{eq:contorsion} as a horizontal \(2\)-form as \(\df C_a \defn \tfrac{1}{2}C_{abc} \df e^b \wedge \df e^c\).

	To get the first law for a stationary axisymmetric black hole solution (see \cref{sec:first-law}) we use the infinitesimal automorphism \(X^m = K^m = t^m + \Omega_{\ms H} \phi^m\) which projects to the horizon Killing field \(K^\mu\) on \(M\). Since \(K^\mu\) vanishes on the bifurcation surface the \(\df\theta\)-term and the contorsion terms do not contribute. Using \(*d\dfM{\df\xi}\vert_B = -2\surgrav\dfM{\df\varepsilon}_{2}\) where \(\dfM{\df\varepsilon}_{2}\) is the volume form on \(B\) we have:
	\be
		\int_B \dfM{\df Q}_X = -\frac{1}{16\pi} \int_B\left(* d\dfM{\df\xi} + \frac{1}{\gamma} d\dfM{\df\xi} \right) = \frac{\surgrav}{8\pi} \int_B \dfM{\df\varepsilon}_2 = \frac{\surgrav}{2\pi} \frac{1}{4}{\rm Area}(B)
	\ee
	Thus, following \cref{thm:temp-entropy}, the perturbed entropy for General Relativity  is
	\be
		\delta S =  \frac{1}{4}\delta {\rm Area}(B)
	\ee

	Next we compute the canonical energy and angular momentum using \cref{eq:E-J-can} and show that they correspond to the ADM mass and ADM angular momentum up to torsion terms. Since the spacetime is asymptotically flat, near spatial infinity the spacetime is asymptotically Minkowskian, and the Lorentz bundle is asymptotically trivial. Then, near infinity there is a section (i.e. choice of gauge) \(s_\infty : M \to P\) such that the pullbacks \( \dfM{\df e}^a \equiv e^a_\mu = (s_\infty^*)^m_\mu e^a_m \) and \({\dfM{\df \omega}^a}_b \equiv {\omega^a{}_b}_\mu = (s_\infty^*)^m_\mu {\omega^a{}_b}_m \) satisfy the asymptotic conditions
	\be\label{eq:asymp-fall-off}
		\dfM{\df e}^a = \dfM{\df e}_{\bb M}^a + O(1/r) \quad;\quad
		{\dfM{\df \omega}^a}_b = O(1/r^2)
	\ee
	where the asymptotic coframes \(\dfM{\df e}_{\bb M}^a\) are adapted to the asymptotic Minkowskian coordinates as
	\be
		\dfM{\df e}_{\bb M}^a = \begin{pmatrix} dt, dx, dy, dz \end{pmatrix}
	\ee

	To compute the canonical energy \cref{eq:E-J-can}, consider the pullback of \(t \cdot \df\theta(\delta\varphi)\) through the section \(s_\infty\) at infinity. Using the fall-off conditions on the pullback of the dynamical fields at infinity \cref{eq:asymp-fall-off} we can write \( s_\infty^*(t \cdot \df\theta) = \delta (\dfM t \cdot \dfM{\df\Theta} ) \) where:
	\be
		\dfM t \cdot \dfM{\df\Theta} =  \dfM t\cdot \left(\frac{1}{32\pi}\phi_{abcd} \dfM{\df e}^a \wedge \dfM{\df e}^b \wedge \dfM{\df \omega}^{cd}\right) = \frac{1}{16\pi}\phi_{abcd} (\dfM t\cdot \dfM{\df e}^a) \dfM{\df e}^b \wedge \dfM{\df \omega}^{cd} + \dfM{\df Q}_X
	\ee
	Note, that \(\dfM{\df\Theta}\) is not gauge-invariant under the action of the Lorentz group \(O(d-1,1)\). Using \cref{eq:E-J-can} The canonical energy at infinity is given by 
	\[
	E_{can} = -\tfrac{1}{16\pi}\int_\infty \phi_{abcd} (\dfM t\cdot \dfM{\df e}^a) \dfM{\df e}^b \wedge \dfM{\df \omega}^{cd}
	\]

	Note here that the Noether charge contribution to the canonical energy actually cancels out. As discussed in \cite{IW-noether-entropy} this accounts for the ``factor of \(2\)" discrepancy in the Komar formula for the ADM mass relative to the one for ADM angular momentum. Also note that this cancellation is much more easily obtained in the computation presented here than the corresponding one in \cite{IW-noether-entropy}. We next show that the above expression for the canonical energy reproduces the well-known ADM mass formula.

	Let \(r_\mu\) be the outward pointing spatial conormal to a \(2\)-sphere at infinity with the volume form \(\dfM {\df\varepsilon}_2\) and \(h_{\mu\nu}\) be the asymptotic spatial metric on the Cauchy surface \(\Sigma\). The Einstein-Hilbert term gives

	\be\begin{split}
		-\tfrac{1}{16\pi}\int_\infty \epsilon_{abcd} (\dfM t\cdot \dfM{\df e}^a) \dfM{\df e}^b \wedge \dfM{\df \omega}^{cd} & = -\tfrac{1}{16\pi} \int_\infty \dfM{\df\varepsilon}_2~ \epsilon_{abcd} t^\sigma e^a_\sigma e^b_\mu \omega^{cd}_\nu t_\lambda r_\rho \varepsilon^{\lambda\rho\mu\nu} \\
	& = -\tfrac{3!}{16\pi}\int_\infty \dfM{\df\varepsilon}_2~ t^\sigma t_\lambda r_\rho  \omega^{cd}_\nu \delta^{[\lambda}_\sigma E^\rho_c E^{\nu]}_d \\
	& = \tfrac{1}{8\pi}\int_\infty \dfM{\df\varepsilon}_2~ r_\rho E^\rho_c (E^{d\eta}\hat\nabla_\nu e^c_\eta ) ( t^\lambda t_\lambda E^\nu_d - t^\nu t_\lambda E^\lambda_d   )\\
	& = \tfrac{1}{8\pi}\int_\infty \dfM{\df\varepsilon}_2~ r_\rho E^\rho_c (\hat\nabla_\nu e^c_\eta ) ( - g^{\nu\eta} - t^\nu t^\eta )\\
	& = -\tfrac{1}{8\pi}\int_\infty \dfM{\df\varepsilon}_2~ r^{[\lambda} h^{\nu]\mu} e_{\lambda a} \left( \partial_\mu e^a_\nu - {\Gamma^\sigma}_{\mu\nu}e^a_\sigma \right)
\end{split}\ee
	where in the third line we written the Lorentz connection in terms of the derivatives of the coframes and in the last line in terms of the Christoffel symbols.

	Using the asymptotic conditions \cref{eq:asymp-fall-off} the first term can be written as
	\be
		-\tfrac{1}{8\pi}\int_\infty \dfM{\df\varepsilon}_2~~ \partial_\mu \left( r^\lambda h_{\bb M}^{\nu\mu} {e_{\bb M}^a}_{[\lambda}e_{\nu]a}\right) = -\tfrac{1}{16\pi}\int_\infty d *_2\left[ r \cdot \left( \dfM{\df e}_{\bb M}^a \wedge \dfM{\df e}_a \right) \right]  = 0
\ee
	where \( h_{\bb M}^{\nu\mu}\) is the flat spatial Cartesian metric on the asymptotic Cauchy surface and \(*_2\) is the \(2\)-dimensional Hodge dual on the asymptotic sphere. The second term, depending on the Christoffel symbols, can be written as
	\be
		\tfrac{1}{8\pi}\int_\infty \dfM{\df\varepsilon}_2~ r^\lambda h^{\nu\mu} \Gamma_{[\lambda\nu]\mu} = \tfrac{1}{16\pi} \int_\infty \dfM{\df\varepsilon}_2~ r^\lambda h^{\nu\mu} \left( \partial_\nu h_{\lambda\mu} - \partial_\lambda h_{\nu\mu} - T_{\mu\nu\lambda} \right)
	\ee
	using the definition of the Christoffel symbols (with torsion)
	\[
		\Gamma_{\lambda\nu\mu} = \tfrac{1}{2} \left( \partial_\mu g_{\nu\lambda} + 2\partial_{[\nu}g_{\lambda]\mu} \right) + \tfrac{1}{2}\left( - T_{\mu\nu\lambda} + 2T_{(\nu\lambda)\mu} \right)
	\]

	Now computing the Holst-term contribution to the canonical energy we have
	\be\begin{split}
		-\tfrac{1}{8\pi\gamma}\int_\infty (\dfM t\cdot \dfM{\df e}^a) \dfM{\df e}^b \wedge \dfM{\df \omega}_{ab} &= -\tfrac{1}{8\pi\gamma}\int_\infty (\dfM t\cdot \dfM{\df e}^a) ( d\dfM{\df e}_a - \dfM{\df T}_a) \\
		& = \tfrac{1}{16\pi\gamma} \int_\infty \dfM{\df\varepsilon}_2~ t^\mu \varepsilon^{\nu\lambda} T_{\mu\nu\lambda}
	\end{split}\ee

	Thus we get the total canonical energy as:
	\be\label{eq:E-can-PH}
		E_{can} = \tfrac{1}{16\pi} \int_\infty \dfM{\df\varepsilon}_2~ r^\lambda h^{\nu\mu} \left( \partial_\nu h_{\lambda\mu} - \partial_\lambda h_{\nu\mu} \right) + \tfrac{1}{16\pi} \int_\infty \dfM{\df\varepsilon}_2~ \left(- r^\lambda h^{\nu\mu} +\tfrac{1}{\gamma}t^\mu \varepsilon^{\nu\lambda} \right)~ T_{\mu\nu\lambda}
	\ee
	The first term can be recognised as the usual formula for the \emph{ADM mass} \(M_{ADM}\), while the second is the canonical energy contributed by the presence of any torsion at infinity.

	Now for the canonical angular momentum, the \(\phi\cdot\df\theta\)-term does not contribute when pulled back to the sphere at infinity and we get (here \(\dfM{\df \phi} \equiv \phi_\mu\))
	\be\label{eq:J-can-PH}\begin{split}
		J_{can} & = - \int_\infty \dfM{\df Q}_\phi = \tfrac{1}{16\pi}\int_\infty * \left[ d\dfM{\df\phi} - 2\phi^\mu \dfM{\df C}_\mu \right] + \tfrac{1}{\gamma} \left[ d\dfM{\df\phi} - 2\phi^\mu \dfM{\df C}_\mu \right] \\
		& = \tfrac{1}{16\pi}\int_\infty  *  d\dfM{\df \phi} -  \tfrac{1}{16\pi}\int_\infty \dfM{\df \varepsilon}_2 \left[-  \tilde\varepsilon^{\nu\lambda} + \tfrac{1}{\gamma} \varepsilon^{\nu\lambda} \right] ~ \phi^\mu C_{\mu\nu\lambda}  
	\end{split}\ee
	where the first term is the Komar formula for the \emph{ADM angular momentum} \(J_{ADM}\) and second is the angular mometum due to any torsion at infinity (\(\tilde{\varepsilon}^{\mu\nu}\) is the binormal to the \(2\)-sphere at infinity).

	As noted before in vacuum, the equation of motion \(\df{\mc E}_{ab} = 0\) ensures that the torsion vanishes everywhere and the canonical energy and angular momentum are exactly the ones given by the ADM quantities. This is also the case if any matter sources for torsion fall-off suitably at infinity, as happens in the case of the Dirac field \cref{sec:dirac}. We note that, when the torsion due to matter sources does not fall-off fast enough the Barbero-Immirizi parameter \(\gamma\) does contribute to the canonical energy and angular momentum at infinity.

	Thus, we get the usual first law for vacuum General Relativity
	\be\label{eq:first-law-PH}
		 \frac{\surgrav}{2\pi} \frac{1}{4}\delta{\rm Area}(B) = \delta M_{ADM} - \Omega_{\ms H} \delta J_{ADM}
	\ee\\

	To illustrate the \(\df\mu\)-ambiguity \cref{eq:mu-amb}, we consider the following three topological terms, that can be added to the Palatini-Holst Lagrangian in \(4\) dimensions.
	\[\begin{split}
		\df L_{\rm E} & = \tfrac{1}{2}\epsilon_{abcd} \df R^{ab} \wedge \df R^{cd} \\
		\df L_{\rm P} & = {\df R^a}_b \wedge {\df R^b}_a = -\df R^{ab} \wedge \df R_{ab} \\
		\df L_{\rm NY} & = \df T^a \wedge \df T_a - \df e^a \wedge \df e^b \wedge \df R_{ab}
	\end{split}\]

	The first is the \emph{Euler character} of the Lorentz bundle over \(M\) (also known as the \emph{Gauss-Bonnet invariant}), the second is the corresponding \emph{Pontryagin character}, and the third is the \emph{Nieh-Yan character} \cite{Nieh-Yan, CZ, Nieh}, which exists only for a connection with torsion. Each of these terms are exact forms \(\df L = d\df\mu\) where the corresponding \(\df\mu\)'s can be computed to be\footnote{The explicit expression for \(\df\mu_{\rm E}\) seems to be largely absent from the literature except in \cite{CRGV}, and in \cite{Nieh-GB} where it is given in terms of the Dirac matrices.}
	\[\begin{split}
		\df\mu_{\rm E} & = \tfrac{1}{2} \epsilon_{abcd} \df \omega^{ab} \wedge \left( \df R^{cd} - \tfrac{1}{3} {\df \omega^c}_e \wedge \df \omega^{ed} \right) \\
		\df\mu_{\rm P} & = {\df \omega^a}_b \wedge \left( {\df R^b}_a - \tfrac{1}{3} {\df \omega^b}_c \wedge {\df \omega^c}_a \right) \\
		\df\mu_{\rm NY} & = \df e^a \wedge \df T_a
	\end{split}\]
	Note that \(\df\mu_{\rm P}\) is just the Chern-Simons term for the Lorentz bundle, \(\df\mu_{\rm E}\) is similar but with the ``trace" taken with a \(\epsilon_{abcd}\), and \(\df\mu_{\rm NY}\) can be viewed as the Chern-Simons term for torsion. These terms contribute the following additional terms to the Noether charge at the bifurcation surface
	\[\begin{split}
		\lb. \df Q_K \rb\vert_{\rm E} & = \epsilon_{cdab} \df R^{cd} (K \cdot \df \omega^{ab}) = \tfrac{1}{2} \epsilon_{cdab} \df R^{cd} (E^a \cdot E^b \cdot d\df\xi) \\
		\lb. \df Q_K \rb\vert_{\rm P} & = -2 \df R_{ab} (K \cdot \df \omega^{ab})  = -\df R_{ab} (E^a \cdot E^b \cdot d\df\xi)\\
		\lb. \df Q_K \rb\vert_{\rm NY} & = - \df e_a \wedge \df e_b (K \cdot \df \omega^{ab})  = -\tfrac{1}{2} \df e_a \wedge \df e_b(E^a \cdot E^b \cdot d\df\xi)\\
	\end{split}\]
	and hence integrating the corresponding gauge-invariant forms over \(B\) gives
	\be\label{eq:PH-top-Q}\begin{split}
		\int_B \lb. \dfM{\df Q}_K \rb\vert_{\rm E} & = - 2 \surgrav \int_B \star \tilde{\epsilon}_{ab} \dfM{\df R}^{ab} = 2 \surgrav \int_B \epsilon_{ab} \dfM{\df R}^{ab} \\
		\int_B \lb. \dfM{\df Q}_K \rb\vert_{\rm P} &  = 2 \surgrav \int_B \tilde\epsilon_{ab} \dfM{\df R}^{ab} \\
		\int_B \lb. \dfM{\df Q}_K \rb\vert_{\rm NY} &  =  \int_B d\dfM{\df\xi} = 0
	\end{split}\ee
	The Euler contribution is the Euler class of the tangent bundle \(TB\) i.e. the Euler characteristic of \(B\). The Pontryagin contribution is, similarly, the Euler class of the normal bundle of \(B\) in \(M\). Since, we have smooth, no-where vanishing normals to \(B\), the Euler class of the normal bundle must vanish (see Prop.~11.17 in \cite{BT-book}).\footnote{This can also be shown by an explicit computation of this term, and the fact that the extrinsic curvature of \(B\) in \(M\) vanishes (see 
\cite{JM}).} Further, we see that the Holst Lagrangian is \(\df L_{Holst} \sim -\df L_{NY} + \df T^a \wedge \df T_a\), and is thus an exact form up to terms not involving the curvature. This explains why the Holst term does not contribute to the Noether charge at the horizon.\footnote{See \cite{Liko-BI} for an Euclidean path integral approach to the Holst and Nieh-Yan terms in the Lagrangian.} The Noether charge contributions \cref{eq:PH-top-Q} are purely topological and do not contribute to the perturbed entropy, and due to the asymptotic fall-off conditions they do not contribute to the canonical energy and angular momentum at infinity. Hence none of the above terms affect the first law. But they do affect a prescription for a total entropy as discussed towards the end of \cref{sec:first-law}. As shown in \cite{JM-lovelock,Sar-Wall}, the Euler term in the Noether charge leads to violations of the second law in General Relativity if one prescribes that the total entropy be given by the gravitational charge.

\subsection{Yang-Mills}\label{sec:yang-mills}

	Next let us consider Einstein-Yang-Mills theory where the Yang-Mills connection \(\df A'^{I'}\) in \cref{eq:conn-split} is a dynamical field governed by the Yang-Mills Lagrangian \cref{eq:L-YM}. Since we have worked out the gravitational contribution in \cref{sec:palatini-holst} in this section we only deal with the Yang-Mills connection, and so, for simplicity omit the ``primes" and write the Yang-Mills connection as \(\df A^I\).

	The contribution of Yang-Mills fields to the first law was worked out by Sudarsky and Wald \cite{SW1, SW2} under the assumption that one can pick a global choice of gauge i.e. an everywhere smooth section \(s : M \to P\) (see \cref{sec:bundles}).\footnote{Sudarsky and Wald also assume that the Yang-Mills group \(G\) is compact, and is \(SU(2)\), but this restriction can be easily removed.} They then consider the pullback \(A^I_\mu = (s^*)^m_\mu A^I_m\) of the Yang-Mills connection as a tensor field on spacetime. They further assume that the section \(s\) can be chosen so that \(A^I_\mu\) is stationary with respect to the horizon Killing field \(K^\mu\).

	As discussed in \cref{sec:fields-L-bundle}, these assumptions are too restrictive to cover all Yang-Mills fields that are of interest. As noted before, sections on a principal bundle exist if and only if the bundle is trivial. For a non-trivial bundle there exist no global sections and the Yang-Mills connection cannot be considered as a globally well-defined tensor field on spacetime. It is far from clear that even for a trivial bundle a section \(s\) can be chosen so that for a given connection \(\df A^I\), the pullback \(s^*\df A^I \equiv A_\mu^I\) is both smooth everywhere and is annihilated by the Lie derivative with the horizon Killing field \(K^\mu\). Using this assumption Sudarsky and Wald conclude that the Yang-Mills fields do not contribute to the first law at the bifurcation surface as the Yang-Mills potential on \(B\) vanishes since \(K^\mu A^I_\mu\vert_B = 0\). In fact, as argued by Gao \cite{Gao-YM}, the Maxwell gauge field on the Reissner-Norst\"orm spacetime, in the standard choice of gauge, is singular on the bifurcation surface. Working instead on some other cross-section of the horizon where the Maxwell vector potential is smooth, Gao finds a ``potential times perturbed charge" term at the horizon for Maxwell fields. \cite{Gao-YM} also finds that Yang-Mills fields do contribute at the horizon but their contribution cannot be put into a ``potential times perturbed charge" form without additional gauge choices which might be incompatible with the assumed stationarity of \(A_\mu^I\) (see \S~4 \cite{Gao-YM}).

	Our formalism allows us to work on arbitrary non-trivial bundles for Yang-Mills fields without making any choice of gauge, and we only assume that the Yang-Mills connection on the principal bundle is Lie-derived up to a gauge transformation i.e. \(\Lie_K \df A^I = 0\) where \(K^m\) is an infinitesimal automorphism of the bundle which projects to the horizon Killing field \(K^\mu\). Using this, we find potentials for the Yang-Mills fields that are constant on the horizon (\cref{thm:zeroth-law}) and that the Yang-Mills fields do contribute a ``potential times perturbed charge" term at the bifurcation surface (\cref{cor:potential-charge}) without assuming any choice of gauge. Thus, the following will be a generalisation of the results of \cite{SW1, SW2, Gao-YM}.\\

	The Yang-Mills Lagrangian can be written for any structure group \(G\) using a non-degenerate, symmetric bilinear form on its Lie algebra \(\mf g\), which is invariant under the adjoint action of \(G\) on its Lie algebra. Since we have assumed that the Lie algebra \(\mf g\) of the structure group is semisimple we will use the Killing form \(k_{IJ}\) (\cref{eq:killing-form}) as such a non-degenerate, symmetric, invariant bilinear form.\footnote{Our convention for the Killing form differs from that of \cite{SW1} by a sign and a factor of \(2\).} Further, any semisimple Lie algebra can be decomposed uniquely into a direct sum of simple Lie algebras (see \S~1.10 \cite{Sam-book} or \S~11.2 \cite{Corn-book}), and thus the Yang-Mills Lagrangian for \(\mf g\) can be written as a sum of Yang-Mills Lagrangians of the same form for each simple factor (with possibly different coupling constants). We can also include abelian groups (which, by definition, are neither simple nor semisimple) into the theory by using the natural product on their Lie algebra \(\bb R^n\). For instance, to get Maxwell electromagnetism we can use \( k_{IJ} \to -2 \) and \(g^2 \to \mu_0 \) in \cref{eq:L-YM}.\\

	With the above discussion, we write the Yang-Mills Lagrangian on the bundle as:
	\be\label{eq:L-YM}
		\df L_{YM} = \frac{1}{4g^2} \lb(\star \df F_I \rb) \wedge \df F^I = \frac{1}{8g^2} \df\varepsilon_4 \lb( F^2 \rb) \in \Omega^4_{hor}P
	\ee
	where \(F^2 \defn (E^a \cdot E^b \cdot \df F^I)(E_a \cdot E_b \cdot \df F_I)\) and \(g^2\) is the Yang-Mills coupling constant. On spacetime \(M\) we get the usual Lagrangian \(\dfM{\df L}_{YM} = \frac{1}{8g^2}\dfM{\df\varepsilon}_4 F^I_{\mu\nu} F_I^{\mu\nu} \).

	The first variation gives \(\delta \df L_{YM} = -\df{\mc T}_a \wedge \delta \df e^a + \df{\mc E}_I \wedge \delta \df A^I + d\df\theta^{\rm (YM)}\) with:
	\be\begin{split}
		\df{\mc E}_I & = -\tfrac{1}{2g^2} D\star \df F_I \\
		\df\theta^{\rm (YM)} & = \tfrac{1}{2g^2} \star \df F_I \wedge \delta \df A^I
	\end{split}\ee
	where we have used the first form of the Lagrangian. The symplectic current contribution takes the form \( \df\omega^{\rm (YM)} = \tfrac{1}{2g^2} \left[ \delta_1 (\star \df F_I) \wedge \delta_2 \df A^I - \delta_2 (\star \df F_I) \wedge \delta_1 \df A^I \right] \).

	To compute the energy-momentum \(3\)-form \(\df{\mc T}_a\), it is convenient to use the second form of the Lagrangian \cref{eq:L-YM}. Varying with the tetrad we have
	\[\begin{split}
		\tfrac{1}{8g^2}\delta_e (\df\varepsilon_4 F^2) & = \tfrac{1}{8g^2}\tfrac{1}{3!} \epsilon_{abcd} \delta \df e^a \wedge \df e^b \wedge \df e^c \wedge \df e^d ~F^2 + \tfrac{1}{8g^2} \df\varepsilon_4 \delta_e \left[ (E^a \cdot E^b \cdot \df F^I) (E_a \cdot E_b \cdot \df F_I) \right] \\
		& = \tfrac{1}{8g^2} \left[ -\tfrac{1}{3!} \epsilon_{abcd} \df e^b \wedge \df e^c \wedge \df e^d ~F^2 \right] \wedge \delta \df e^a + \df\varepsilon_4 \tfrac{1}{2g^2} (E^a \cdot E^b \cdot \df F^I) (\delta E_a \cdot E_b \cdot \df F_I)
	\end{split}\]

	The first term can be written in the form
	\[
		\tfrac{1}{8g^2} \left[ -\tfrac{1}{3!} \epsilon_{abcd} \df e^b \wedge \df e^c \wedge \df e^d ~F^2 \right] \wedge \delta \df e^a = \tfrac{1}{2g^2} \left[ \star \df e_a \left( -\tfrac{1}{4} F^2 \right) \right] \wedge \delta \df e^a
	\]
	and the second term as (using \cref{eq:var-frame})
	\[\begin{split}
		\df\varepsilon_4 \tfrac{1}{2g^2} (E^a \cdot E^b \cdot \df F^I) (\delta E_a \cdot E_b \cdot \df F_I) &= - \df\varepsilon_4 \tfrac{1}{2g^2} (E^c \cdot E^b \cdot \df F^I) ( E_a \cdot E_b \cdot \df F_I) E_c \cdot \delta \df e^a \\
			& = \tfrac{1}{2g^2} \left[ (E_c \cdot \df\varepsilon_4) (E^c \cdot E^b \cdot \df F^I) (E_a \cdot E_b \cdot \df F_I) \right] \wedge \delta \df e^a \\
			& = \tfrac{1}{2g^2} \left[ \star \df e_c (E^c \cdot E^b \cdot \df F^I) (E_a \cdot E_b \cdot \df F_I) \right] \wedge \delta \df e^a 
	\end{split}\]
	putting these together we have
	\[
		\df{\mc T}_a  = - \tfrac{1}{2g^2} \star \df e^c \left[ (E_a \cdot E_b \cdot \df F_I) (E_c \cdot E^b \cdot \df F^I) -  \tfrac{1}{4} \eta_{ac} F^2 \right] 
	\]\\

	To find the Noether current consider the following computation for some \(X^m \in \mf{aut}(P)\):
	\be\begin{split}
		X \cdot \df L_{YM} & = \tfrac{1}{8g^2} X \cdot (\df\varepsilon_4 F^2) = \tfrac{1}{8g^2} \tfrac{1}{3!} \epsilon_{abcd} (X \cdot \df e^a)~ \df e^b \wedge \df e^c \wedge \df e^d F^2 \\
		& = -\tfrac{1}{2g^2} \left[ \star \df e_a \left( -\tfrac{1}{4} F^2 \right) \right] (X\cdot \df e^a)
	\end{split}\ee
	\be\begin{split}
		\df\theta_X^{\rm (YM)} & = \tfrac{1}{2g^2}\star \df F_I \wedge \Lie_X \df A^I =  \tfrac{1}{2g^2}\star \df F_I \wedge \left( X \cdot \df F^I + D(X \cdot \df A^I ) \right) \\
			& = \tfrac{1}{2g^2}\star \df F_I \wedge X \cdot \df F^I + \tfrac{1}{2g^2}D\left( \star \df F_I (X \cdot \df A^I) \right) - \tfrac{1}{2g^2} D \star \df F_I (X \cdot \df A^I)
	\end{split}\ee

	The first term in \(\df\theta_X^{\rm (YM)}\) can be written as
	\[\begin{split}
		\tfrac{1}{2g^2}\star \df F_I \wedge X \cdot \df F^I & = \tfrac{1}{2g^2}\star \df F^I \wedge \tfrac{1}{2!} X \cdot (\df e^a \wedge \df e^b) (E_b \cdot E_a \cdot \df F_I) \\
	& = \tfrac{1}{2g^2} \tfrac{1}{2! 2!} \epsilon_{efcd} \df e^c \wedge \df e^d (E^f \cdot E^e \cdot \df F^I) \wedge \df e^b (E_b \cdot E_a \cdot \df F_I) ( X \cdot \df e^a) \\
	& = \tfrac{1}{2g^2} \star \df e^c (E^b \cdot E_c \cdot \df F^I) (E_b \cdot E_a \cdot \df F_I) ( X \cdot \df e^a)
	\end{split}\]

	This gives the Noether current:
	\be\begin{split}
		\df J_X^{\rm (YM)} & = \tfrac{1}{2g^2} \star \df e^c \left[ (E^b \cdot E_c \cdot \df F^I) (E_b \cdot E_a \cdot \df F_I) -\tfrac{1}{4} \eta_{ac} F^2 \right] (X\cdot \df e^a) \\
		& ~\quad + \df{\mc E}_I (X \cdot \df A^I) + \tfrac{1}{2g^2}D\left( \star \df F_I (X \cdot \df A^I) \right) \\
		& = d\df Q_X^{\rm (YM)} + \df{\mc E}_I (X \cdot \df A^I) - \df{\mc T}_a (X \cdot \df e^a)
	\end{split}\ee
	The terms with \(\df{\mc E}_I\) and \(\df{\mc T}_a\) contribute to the constraints of Einstein-Yang-Mills theory (see \cref{eq:J-C-Q}) and the Noether charge contribution is:
	\be
		\df Q_X^{\rm (YM)} = \tfrac{1}{2g^2}\star \df F_I (X \cdot \df A^I)
	\ee
	which is of the general form given in \cref{lem:noether-charge-form}.\\

	Now consider a stationary-axisymmetric connection \(\df A^I\) which satisfies the Einstein-Yang-Mills equations on a stationary axisymmetric black hole spacetime. This means that the horizon Killing field \(K^m\) is an infinitesimal automorphism that preserves the Yang-Mills conection \(\Lie_K \df A^I = 0\). The extent to which \(K^m\)  is determined by its projection \(K^\mu\) on \(M\) is given by \cref{lem:aut-conn-unique,lem:aut-e-unique}. Following the computations in \cref{thm:zeroth-law}--\cref{cor:potential-charge} we can write the Noether charge of Yang-Mills fields at the horizon as
	\[
		\int_B \dfM{\df Q}_K^{\rm (YM)} = \ms V^\Lambda \ms Q_\Lambda
	\]
	where \(\ms V^\Lambda\) is the Yang-Mills potential and \(\ms Q_\Lambda \) is the Yang-Mills electric charge (also see \cite{CK}) given by
	\be\label{eq:YM-charge}
		\ms Q_\Lambda = \tfrac{1}{2g^2}\int_B * \dfM{\df F_I h^I_\Lambda }
	\ee
	where the \(h^I_\Lambda\) are some fixed basis of a fixed choice of Cartan subalgebra on the horizon as defined in \cref{sec:zeroth-law}, and the contribution to the first law at \(B\) becomes
	\be
		\delta\int_B \dfM{\df Q}_K^{\rm (YM)} = \ms V^\Lambda \delta \ms Q_\Lambda
	\ee

	To compute the contribution at infinity, first consider the vector field \(t^m\) which gives the canonical energy \(E_{can}\). We choose the Yang-Mills fields to satisfy the asymptotic conditions
	\[
		\df F^I = O(1/r^2) 
	\]
	which also gives \(\delta \df A^I = O(1/r)\). We immediately see that \(\df\theta^{\rm (YM)} = \tfrac{1}{2g^2}\star \df F_I \wedge \delta \df A^I = O(1/r^3)\) and does not contribute to the first law.

	Since \(t^m\) is an infinitesimal automorphism which preserves the connection we see that
		\[\begin{split}
		0 &= \left(\Lie_t \df F^I\right)\vert_{P_\infty} = D(t \cdot \df F^I) - {c^I}_{JK} (t \cdot \df A^J) \df F^K \\
		& = - {c^I}_{JK} (t \cdot \df A^J) \df F^K + O(1/r^3) \\
		0 & = (\Lie_t \df A^I)\vert_{P_\infty} = t\cdot \df F^I + D(t\cdot \df A^I) \\
		& = D (t\cdot \df A^I) + O(1/r^2)
	\end{split}\]
	Up to higher order terms in \(1/r\), we can repeat the procedure in \cref{thm:zeroth-law}--\cref{cor:potential-charge} to get contribution of Yang-Mills fields to the canonical energy as
	\[
		E_{can}^{\rm (YM)} = \int_\infty \dfM{\df Q}_t^{\rm (YM)} = \ms V^\Lambda \ms Q_\Lambda
	\]
	with the Yang-Mills electric charge \cref{eq:YM-charge} but evaluated at infinity. Note that aymptotically the Yang-Mills equation of motion at infinity becomes 
	\[\begin{split}
		0 = \df{\mc E}_I\vert_{P_\infty} = -\tfrac{1}{2g^2} D\star \df F_I & = O(1/r^3)  \\
	\implies D(\star \df F_I h^I_\Lambda) = d (\star \df F_I h^I_\Lambda) & = O(1/r^3)
	\end{split}\]
	where we used the basis \(h^I_\Lambda\) of the fixed Cartan subalgebra defined in \cref{sec:zeroth-law}. Thus, the Yang-Mills charge can be computed over any ``sufficiently large" surface homologous to a sphere. Further 
	\be
		\delta (t \cdot \df A^I)\vert_\infty =  (t \cdot \delta \df A^I)\vert_\infty = O(1/r)
	\ee
	and the variation of the potential term falls off faster at infinity and we have the first law contribution at infinity as
	\be
		\delta E_{can}^{\rm (YM)} = \delta\int_\infty \dfM{\df Q}_t^{\rm (YM)} = \ms V^\Lambda \delta \ms Q_\Lambda
	\ee

	In a similar manner (assuming faster fall-off for the asymptotic fields if necessary), the Yang-Mills contribution to the canonical angular momentum is
	\be
		J_{can}^{\rm (YM)} = -\int_\infty \dfM{\df Q}_\phi^{\rm (YM)} = - \tfrac{1}{2g^2} \int_\infty (\dfM{\phi \cdot \df A^I}) * \dfM{\df F}_I
	\ee

	The first law of black hole mechanics in Einstein-Yang-Mills then can be written as
	\be\label{eq:first-law-EYM}
		T_{\ms H}\delta S + \lb.(\ms V^\Lambda \delta \ms Q_\Lambda)\rb\vert_B =  \delta M_{ADM} + \lb.(\ms V^\Lambda \delta \ms Q_\Lambda)\rb\vert_\infty - \Omega_{\ms H} \lb(  \delta J_{ADM} +  \delta J_{can}^{\rm (YM)}  \rb)
	\ee

	\begin{remark}[Yang-Mills potentials at horizon and infinity]\label{rem:YM-potential-amb}
	When the Yang-Mills structure group is abelian the charge at \(B\) and infinity are equal (using the abelian Yang-Mills equation of motion). Thus, the abelian Yang-Mills term in the first law can be written as a ``difference in potentials times perturbed charge" \(\lb( \lb.\ms V^\Lambda\rb\vert_\infty - \lb.\ms V^\Lambda\rb\vert_B \rb) \delta \ms Q_\Lambda\). Further, the ambiguity in the choice of the vector field \(K^m\) (for a given horizon Killing field \(K^\mu\)) is given by a \(\bb R^n\)-valued constant function \(\lambda\); here \(n\) is the dimension of the abelian structure group (see \cref{rem:global-symm}). By a suitable choice of \(\lambda\) we can always set the potentials at the horizon to vanish, while shifting the potentials at infinity by a constant.

	Even for a non-abelian structure group, if there exists a vertical vector field \(Y^m \in VP\) corresponding to a global symmetry of \(\df A^I\) (which is a solution to the Einstein-Yang-Mills equations) as described in \cref{lem:aut-conn-unique}, we can use \(Y^m\) to redefine our choice of the horizon Killing field \(K^m\) on the bundle. Using, this freedom  we can set some, or possibly all, of the horizon potentials to zero but at the cost of changing the potentials at infinity. The cases where we can set all potentials at the horizon to vanish correspond to the analysis of Sudarsky and Wald \cite{SW1, SW2}. For arbitrary connections \(\df A^I\), which solve the Einstein-Yang-Mills equations, there may not exist any such global symmetry (which corresponds to the existence of global solutions to \(D\lambda^I = 0 \), see \cref{rem:global-symm}). Thus, it seems that the first law as derived in \cite{SW1, SW2} applies only to special cases and in general, the first law for Einstein-Yang-Mills takes the form above (also see the related discussion in \cite{Gao-YM}).
	\end{remark}

	As an illustration of the \(\df\mu\)-ambiguity (\cref{eq:mu-amb}) we add a topological term to the Yang-Mills Lagrangian using the Chern character of the bundle as \footnote{In particle physics literature this is also known as the \(\theta\)-term; where the \(\theta\) refers to the conventional coupling constant in front of \(\df L_C\) and is unrelated to the symplectic potential.}
	\be
		\df L_{\rm C} \defn \tfrac{1}{2} \df F_I \wedge \df F^I = d\df\mu_{\rm C}
	\ee
	where \(\df\mu_{\rm C}\) is the Chern-Simons form
	\be
		\df\mu_{\rm C} = \tfrac{1}{2}\lb(\df A_I \wedge \df F^I - \tfrac{1}{6}c_{IJK} \df A^I \wedge \df A^J \wedge \df A^K\rb)
	\ee
	The corresponding Noether charge for \(K^m\) then gets an additional contribution
	\be
		\int_B \dfM{\df Q}_K = \ms V^\Lambda \tilde{\ms Q}_\Lambda
	\ee
	where the Yang-Mills magnetic charge is
	\be
		\tilde{\ms Q}_\Lambda = \int_B \dfM{\df F_I h^I_\Lambda}
	\ee
	The magnetic charge is purely a topological charge that does not vary under perturbations and hence does not contribute to the first law.

\subsection{Dirac spinor}\label{sec:dirac}

	As the third case of interest we consider spinor matter fields in Einstein-Dirac theory. The principal bundle of interest in this case is \(P_{Spin}\) with the structure group \(Spin^0(3,1)\) (see \cref{sec:bundles} for details). The Dirac Lagrangian can be written as:
	\be\label{eq:L-Dirac}
		\df L_{Dirac} \defn \df\varepsilon_4 \left( \frac{1}{2}\adj\Psi \dirac \Psi - \frac{1}{2}\dirac \adj\Psi \Psi - m \adj\Psi\Psi \right)
	\ee
	where \(\df\varepsilon_4\) is the horizontal volume \(4\)-form on \(P_{Spin}\) \cref{eq:hor-volume} and \(\dirac\) is the Dirac operator \cref{eq:dirac-op}. Note that we have admitted a connection with torsion and so this Lagrangian is \emph{not} equivalent to usual Dirac Lagrangian which uses the torsionless Levi-Civita connection \cref{eq:LC-conn} (see \S~V.B.4 \cite{HVKN}); the dynamics of the Dirac field \emph{does} depend on the choice of spin connection used. If one assumes that the connection is torsionless from the outset, then one has to work in the ``second-order formulation". Since the following  computations are easier in the first-order formalism, we will continue to use an independent connection with torsion to obtain a first law for Einstein-Dirac theory. The computations in the second-order formalism can be performed in exactly the same manner (see \cref{rem:second-order}).

	To get the Dirac equation and the symplectic potential compute first the variation with the Dirac spinor fields:
	\be\begin{split}
		\df\varepsilon_4 \left(\adj\Psi \dirac \delta\Psi \right) & = \df\varepsilon_4 \lb( \adj\Psi \gamma^a E_a \cdot D\delta\Psi \rb) = - ( E_a \cdot \df\varepsilon_4 ) \adj\Psi \wedge \gamma^a D\delta\Psi \\
		& = D \left[ (E_a \cdot \df\varepsilon_4) \adj\Psi \gamma^a \delta\Psi \right] - D\left[ (E_a\cdot \df\varepsilon_4) \adj\Psi \gamma^a \right] \delta\Psi
	\end{split}\ee
	where the second equality in the first line uses the vanishing of a horizontal \(5\)-form. The first term in the last line contributes to the symplectic potential \(\df\theta\) while, using \cref{eq:hor-volume} the second term can be written as:
	\be\begin{split}
		- D\left[ (E_a\cdot \df\varepsilon_4) \adj\Psi \gamma^a \right] \delta\Psi &= - \frac{1}{3!} D \left( \adj\Psi \gamma^a \epsilon_{abcd} \df e^b \wedge \df e^c \wedge \df e^d  \right) \delta \Psi \\
		& = - \df\varepsilon_4 \left(\dirac\adj\Psi \delta\Psi \right) - \frac{1}{3!} \adj\Psi \gamma^a \epsilon_{abcd} \df T^b \wedge \df e^c \wedge \df e^d \delta \Psi
	\end{split}\ee\\

	Similarly computing the variation with respect to the Dirac cospinor field \(\adj\Psi\) we have \( \delta_\Psi \df L_{Dirac} = \adj{\df{\mc E}} \delta\Psi + \delta\adj\Psi~ \df{\mc E} + d\df\theta^{\rm (Dirac)} \) with
	\be\begin{split}
		\df{\mc E} & = \left[ (\dirac - m )\Psi  - \tfrac{1}{3!} T^b{}_{ba} (\gamma^a \Psi) \right] \df\varepsilon_4  \\
		\adj{\df{\mc E}} & = \left[ (- \dirac - m )\adj\Psi + \tfrac{1}{3!} T^b{}_{ba} (\adj\Psi \gamma^a) \right] \df\varepsilon_4  \\
		\df\theta^{\rm (Dirac)} & = \tfrac{1}{2} ( E_a\cdot \df\varepsilon_4) \left( \adj\Psi \gamma^a \delta\Psi - \delta\adj\Psi \gamma^a \Psi \right) \\
			& = \tfrac{1}{2} ( \star \df e_a) \left( \adj\Psi \gamma^a \delta\Psi - \delta\adj\Psi \gamma^a \Psi \right)
	\end{split}\ee
	where we have used the frame components of the torsion \(T^c{}_{ab} \defn E^b \cdot E^a \cdot \df T^c\) and the horizontal Hodge dual \cref{eq:hor-Hodge}.
	On spacetime \(M\), the Dirac equation \(\dfM{\df{\mc E}}\) takes the form
	\[
		\dfM{\df{\mc E}} =  \left[ (\dirac - m )\Psi  - \tfrac{1}{3!} {T^\nu}_{\nu\mu}~ \gamma^\mu \Psi \right]\dfM{\df\varepsilon}_4
	\]
	We again, note that this is not equivalent to the usual Dirac equation since we chose a spin connection with torsion \cite{HVKN} --- setting the torsion to vanish however does give us the standard Dirac equation.

	For the energy-momentum form we have to compute the variation with the tetrad. For this we rewrite
	\[\begin{split}
		\df\varepsilon_4  \adj\Psi \dirac\Psi & = \tfrac{1}{4!} \epsilon_{abcd} \df e^a \wedge \df e^b \wedge \df e^c \wedge \df e^d \adj \Psi \gamma^e E_e \cdot D\Psi \\
		& = - \tfrac{1}{3!} \epsilon_{abcd} \df e^b \wedge \df e^c \wedge \df e^d \wedge (\adj\Psi \gamma^a D \Psi)
	\end{split}\]
	Then
	\[
		\delta_e (\df\varepsilon_4  \adj\Psi \dirac\Psi) = -\tfrac{1}{2} \epsilon_{abcd} \df e^b \wedge \df e^c \wedge (\adj\Psi \gamma^d D\Psi) \wedge \delta \df e^a
	\]
	and 
	\[
	- \delta_e (\df\varepsilon_4  m \adj\Psi \Psi) = \tfrac{1}{3!} \epsilon_{abcd} \df e^b \wedge \df e^c \wedge \df e^d (m \adj\Psi \Psi) \wedge \delta \df e^a
	\]

	Thus, we have \(\delta_e \df L_{Dirac} = - \df{\mc T}_a \wedge \delta \df e^a\) with the \emph{energy-momentum}
	\[\begin{split}
		\df{\mc T}_a &= \epsilon_{abcd} \df e^b \wedge \df e^c \wedge \left[ \tfrac{1}{4} \left( \adj\Psi \gamma^d D\Psi - D\adj\Psi \gamma^d \Psi \right) - \tfrac{1}{3!} \df e^d m \adj\Psi \Psi \right] \\
		& = (\star \df e_a) \left( \tfrac{1}{2}\adj\Psi \dirac \Psi - \tfrac{1}{2}\dirac \adj\Psi \Psi - m \adj\Psi\Psi \right) - \tfrac{1}{2} (\star \df e_b) \left( \adj\Psi \gamma^b E_a \cdot D\Psi - E_a \cdot D\adj\Psi \gamma^b \Psi \right) 
	\end{split}\]\\

	For the spin current compute the variation due to the connection:
	\be
	\delta_\omega \frac{1}{2}\left( \df\varepsilon_4 \adj\Psi \dirac \Psi \right) = -\frac{1}{16}~ \df\varepsilon_4 \lb( \adj\Psi \gamma^c [\gamma_a, \gamma_b] \Psi E_c \cdot \delta \df \omega^{ab} \rb) = \frac{1}{16} ( E_c \cdot \df\varepsilon_4 ) \left( \adj\Psi \gamma^c [\gamma_a, \gamma_b] \Psi \right) \wedge \delta \df \omega^{ab} 
\ee
where the last equality uses the vanishing of a horizontal \(5\)-form. Thus we have \( \delta_\omega \df L = -\df{\mc S}_{ab} \wedge \delta \df \omega^{ab} \) where the \emph{spin current} is:
\be\begin{split}
	\df{\mc S}_{ab} & \defn -\frac{1}{16} (E_c \cdot \df\varepsilon_4 ) \left( \adj\Psi \gamma^c [\gamma_a, \gamma_b] \Psi + \adj\Psi [\gamma_a, \gamma_b]\gamma^c \Psi \right) \\
			& = -\frac{1}{16} (\star \df e_c) ~ \left( \adj\Psi \gamma^c [\gamma_a, \gamma_b] \Psi + \adj\Psi [\gamma_a, \gamma_b]\gamma^c \Psi \right) 
\end{split}\ee\\

	To compute the Noether current and Noether charge we need a notion of a ``Lie derivative" for spinor fields. As discussed in \cref{sec:intro} the prescriptions for defining a Lie derivative on spinors in the existing literature \cite{FF, GM, LRW} are not satisfactory. Since we view spinors as fields defined on the spin bundle \(P_{Spin}\) we can use the natural notion of Lie derivatives with respect to a vector field \(X^m \in \mf{aut}(P)\) on the bundle which we will show (see \cref{eq:get-LK-Lie}) reduces to the definition given by Lichnerowicz \cite{Lich} in the case \(X^m\) projects to a Killing vector field. Using \cref{eq:Lie-cov-D} the Lie derivative on spinor fields on the bundle can be written as  \( \Lie_X\Psi \defn X \cdot d\Psi = X \cdot D\Psi + \tfrac{1}{8} X \cdot \df \omega_{ab}[\gamma^a, \gamma^b]\Psi  \). The Noether current then is
	\be
		\df J_X^{\rm (Dirac)} = - \df{\mc T}_a (X \cdot \df e^a) - \df{\mc S}_{ab} (X \cdot \df \omega^{ab})
	\ee
	The energy-momentum and spin current terms on the right-hand-side contribute to the constraints. Thus the Noether charge contribution of the Dirac fields can be chosen to vanish i.e. \(\df Q_X^{\rm (Dirac)} = 0\), as we expect from the general formula in \cref{lem:noether-charge-form}.\\

	For the first law, stationary axisymmetric Dirac fields i.e. \(\Lie_K \Psi = 0 \) do not explicitly contribute to the black hole entropy since the Dirac field contribution to the Noether charge vanishes identically. Near infinity, the Dirac field \(\Psi\) falls off faster than \(1/r^{3/2}\), in which case \(t \cdot \df\theta^{\rm (Dirac)}\) falls-off faster than \(1/r^3\). Since the Noether charge contribution of the Dirac field vanishes, the Dirac field does not explicitly contribute to boundary integral defining the canonical energy (\cref{eq:E-J-can}). These fall-offs also ensure that the torsion terms in the gravitational canonical energy in \cref{eq:E-can-PH} vanish. Similarly, there is no explicit Dirac contribution to the boundary integral for the canonical angular momentum and the torsion terms in the gravitational canonical angular momentum \cref{eq:J-can-PH} also vanish. Thus, the first law of black hole mechanics with spinor fields governed by the Einstein-Dirac Lagrangian \cref{eq:L-Dirac} retains the form \cref{eq:first-law-PH}.

\section*{Acknowledgements}

I would like to thank Robert M. Wald for numerous insightful discussions throughout the course of this work, and Caner Nazaroglu for very helpful comments on Lie groups and Lie algebras. I would also like to thank Arif Mohd, Ted Jacobson, Igor Khavkine and Alexander Grant for comments on the initial draft of this work. This research was supported in part by the NSF grants PHY~12-02718 and PHY~15-05124 to the University of Chicago.

\appendix

\section{Mathematical aside}\label{sec:math}

	In this section we collect some useful mathematical formulae and new results needed in the main arguments of the paper.\\

	First we recall the definition of a Cartan subalgebra and the properties of the corresponding Weyl-Chevalley basis for a semisimple Lie algebra \(\mf g\), which are needed to define the horizon potentials in \cref{sec:zeroth-law}.

	\begin{definition}[Cartan subalgebra]
	A \emph{Cartan subalgebra} \(\mf h\) of a complex semisimple Lie algebra \(\mf g\) is a maximal abelian Lie subalgebra such that the adjoint action of \(\mf h\) on \(\mf g\) (given by the Lie bracket) is diagonalisable (see \S 2.1 \cite{Sam-book}).
	\end{definition}

	\begin{property}[Properties of a Cartan subalgebra]\label{pr:CSA} We list below the key properties of Cartan subalgebras we will need in the following.
	\begin{propertylist}
	\item Any two Cartan subalgebras of a semisimple complex Lie algebra \(\mf g\) are isomorphic under the adjoint action of some element of the corresponding group \(G\) (Theorem F. \S 2.10 \cite{Sam-book}); the dimension of the Cartan subalgebras is called the \emph{rank} of \(\mf g\). \label{pr:CSA-iso}
	\item The Killing form \(k_{IJ}\) is non-degenerate on any Cartan subalgebra of \(\mf g\) (\S~2.3 \cite{Sam-book}). \label{pr:CSA-non-degen}
	\item For any given choice of Cartan subalgebra \(\mf h\) (of dimension \(l\)), there exists a choice of basis for \(\mf g\) (of dimension \(n\)) --- the \emph{Weyl-Chevalley basis} --- given by \(\{h^I_\Lambda, a^I_i\}\). Here, \(h^I_\Lambda\) with \(\Lambda = 1,2,\ldots,l\) are a choice of \emph{simple coroots} (see \S~2 \cite{Sam-book}) and form a basis of \(\mf h\). The remaining basis elements \(a^I_i\) for \(i = 1,2,\ldots, n-l\) are orthogonal to \(\mf h\) with respect to the Killing form \(k_{IJ}\) (see \S~2.8 and 2.9 \cite{Sam-book} for details). \label{pr:WC-basis}
	\item Given a choice of the simple coroots \(h^I_\Lambda\) any other choice can be obtained by the action of a \emph{finite} subgroup of \(G\) --- called the \emph{Weyl group of \(\mf g\)} (\S~2.11 \cite{Sam-book}). The action of the Weyl group elements on the \(h^I_\Lambda\) is generated by certain permutations and sign changes (see \S~2.14 \cite{Sam-book} for a description of the simple coroots and the Weyl group for simple Lie algebras). \label{pr:weyl-group}
	\item Any given element \(X^I \in \mf g\) can be mapped into a chosen Cartan subalgebra \(\mf h\) by the adjoint action of some element of the group \(G\) \cite{Bump-book}. For a given \(X^I \in \mf g\), all possible choices of the corresponding element in \(\mf h\) under the above map, are related by the action of the Weyl group on \(\mf h\); since the Weyl group is a finite group there are only \emph{finitely many} possible choices.\footnote{This last statement can be proved by writing an element of \(\mf h\) in a basis given by the simple coroots, and then applying the results of the first theorem in \S~10.3 \cite{Hump-book}.} \label{pr:CSA-map-into}
	\end{propertylist}
	\end{property}

	The above properties of a Cartan subalgebra strictly hold for a complex semisimple Lie algebra. When, the Lie algebra \(\mf g\) of the theory under consideration is a real semisimple Lie algebra, we first take its complexification (which is also semisimple; see \S~11.3 \cite{Corn-book}) to apply the Cartan subalgebra construction above, and then in the end take the real form corresponding to the original real Lie algebra \(\mf g\) (see \S~11.10 \cite{Corn-book}).

\subsection{Principal fibre bundles}\label{sec:bundles}

	Consider a \(G\)-principal bundle \(\pi : P \to M\). Let \(\df A^I \in \Omega^1P(\mf g, {\rm Ad})\) be a \emph{connection} on \(P\) and let the corresponding \emph{covariant exterior derivative} \(D : \Omega^kP(\bb V;R) \to \Omega^{k+1}_{hor}P(\bb V;R)\) on equivariant differential forms valued in a vector space \(\bb V\) on \(P\) (see \S~II.5. \cite{KN-book1} and \S~Vbis.A.4 \cite{CDD-book}). If \(\df\sigma^A \in \Omega^k_{hor}P(\bb V;R)\) is a horizontal equivariant form (corresponding to gauge covariant fields on spacetime) then
	\be\label{eq:cov-D-hor}
		D\df\sigma^A = d\df\sigma^A + ( \df A^I {r_I}^A{}_B )\wedge \df\sigma^B
	\ee
	where \(r\) is the representation of the Lie algebra \(\mf g\) on \(\bb V\). The covariant derivative \(D\) acting on the connection itself defines the \emph{curvature} \(2\)-form \(\df F^I \in \Omega^2_{hor}P(\mf{g}, {\rm Ad})\) as
	\be\label{eq:F-defn}
		\df F^I \defn D\df A^I = d\df A^I + \tfrac{1}{2}{c^I}_{JK} \df A^J \wedge \df A^K
	\ee
	where \({c^I}_{JK}\) are the \emph{structure constants} of the Lie algebra. The curvature further satisfies the \emph{Bianchi identity}
	\be\label{eq:F-bianchi}
		D\df F^I = d\df F^I + {c^I}_{JK} \df A^J \wedge \df F^K = 0
	\ee
	which can be directly checked using \cref{eq:F-defn}.\\

	If \(f : P \to P\) is an \emph{automorphism of the principal bundle} \(P\) we denote the corresponding diffeomorphism of \(M\) by \(\dfM f\) so that \( \pi \circ f = \dfM f \circ \pi\). Let the group of automorphisms of \(P\) be \({\rm Aut}(P)\) and the corresponding Lie algebra of vector fields \(\mf{aut}(P) \subset TP\). A \emph{vertical automorphism} is an \(f \in {\rm Aut}(P)\) which projects to the identity on the base space \(M\) i.e. \(\dfM f = \id_M\). The vector fields in the Lie algebra \(\mf{aut}(P)\) act on equivariant differential forms by the usual Lie derivative. Using \cref{eq:cov-D-hor,eq:F-defn} we can write the Lie derivative of equivariant forms in terms of the covariant exterior derivative as
	\be\label{eq:Lie-cov-D}\begin{split}
		\Lie_X \df\sigma^A & = X \cdot D\df\sigma^A + D( X \cdot \df\sigma^A) - (X \cdot \df A^I)~ {r_I}^A{}_B  \df\sigma^B \\
		\Lie_X \df A^I & = X \cdot \df F^I + D(X \cdot \df A^I)
	\end{split}\ee
	for \(\df\sigma^A \in \Omega^k_{hor}P(\bb V;R)\) and the connection \(\df A^I\). Note both \(\Lie_X \df\sigma^A\) and \(\Lie_X \df A^I\) are horizontal forms (and hence gauge covariant). When \(X^m \in \mf{aut}_V(P)\) this corresponds to infinitesimal internal gauge transformations.

	Given some equivariant differential form \(\df\sigma^A\) there might exist some bundle automorphisms which keep \(\df\sigma^A\) invariant.

	\begin{definition}[Automorphism which preserves \(\df\sigma^A\)]\label{def:auto-psi}
		An automorphism of the bundle \(f \in {\rm Aut}(P)\) is an automorphism which preserves some given equivariant differential form \(\df\sigma^A \in \Omega^kP(\bb V;R)\) if \(f^*\df\sigma^A = \df\sigma^A\). Similarly \(X^m \in \mf{aut}(P)\) is an \emph{infinitesimal automorphism} which preserves \(\df\sigma^A\) if \(\Lie_X \df\sigma^A = 0\). 
	For a given \(\df\sigma^A\), denote the subgroup of the automorphisms preserving \(\df\sigma^A\) by \({\rm Aut}(P;\df\sigma^A) \subseteq {\rm Aut}(P)\), and the corresponding Lie subalgebra of infinitesimal automorphisms \(\mf{aut}(P;\df\sigma^A) \subseteq \mf{aut}(P)\).
	\end{definition}

	Since for the first law we are interested in stationary and axisymmetric field configurations, we define a notion of stationarity (axisymmetry) for a charged field \(\df\sigma^A\) defined on a bundle \(P\) over a base space \(M\) with a stationary (axisymmetric) metric as bundle automorphisms which preserve \(\df\sigma^A\) and project to the stationary (axisymmetric) isometries of some given metric \(g_{\mu\nu}\) on \(M\).

	\begin{definition}[Stationary and/or axisymmetric field]\label{def:stationary-axisymm}
		For a stationary and/or axisymmetric spacetime \(M\) with a metric \(g_{\mu\nu}\), a charged field \(\df\sigma^A\) defined on a bundle \(P \to M\) is called \emph{stationary} if there exists a one-parameter family \(f_t \in {\rm Aut}(P;\df\sigma^A) \) which projects to a one-parameter family \(\dfM f_t\) of stationary isometries of \(g_{\mu\nu}\). Similarly, \(\df\sigma^A\) is \emph{axisymmetric} if there exists a one-parameter family \(f_\phi \in {\rm Aut}(P;\df\sigma^A) \) which projects to a one-parameter family \(\dfM f_\phi\) of axisymmetric isometries of \(g_{\mu\nu}\).

	The corresponding stationary (and axial) infinitesimal automorphism vector field \(t^m\) (and \(\phi^m\)) projects to the stationary (and axial) Killing vector \(t^\mu\) (and \(\phi^\mu\)) of \(g_{\mu\nu}\), respectively.
	\end{definition}

	A diffeomorphism of the base space does not uniquely determine an automorphism of the bundle. But if we require that the automorphism preserve the connection (for example, if the Yang-Mills connection is stationary) then we can classify this ambiguity as follows.

	\begin{lemma}[Uniqueness of an infinitesimal automorphism that preserves a connection]\label{lem:aut-conn-unique}
	For a given connection \(\df A^I\) on \(P\), any \(X^m \in \mf{aut}(P;\df A^I)\) is uniquely determined by its projection \((\pi_*)^\mu_m X^m \in TM\) up to a vertical vector field \(Y^m \in \mf{aut}_V(P)\) such that
	\be\label{eq:cov-const-comm}
		D(Y \cdot \df A^I) = 0
	\ee
	if any such non-trivial \(Y^m\) exists on \(P\).
	\begin{proof}
	Any \(X^m \in \mf{aut}(P;\df A^I)\) satisfies (using \cref{eq:Lie-cov-D})
	\be\label{eq:iso-A}
		0 = \Lie_X \df A^I = X \cdot \df F^I + D(X\cdot \df A^I)
	\ee

	If \(X^m \in VP\) i.e. \((\pi_*)^\mu_m X^m = 0\) this means \(X\cdot \df A^I\) is a covariantly constant \(\mf g\)-valued function (since \(\df F^I\) is a horizontal form). So if \(X^m\) and \(X'^m\) are such that \(X^\mu = (\pi_*)^\mu_m X^m = (\pi_*)^\mu_m X'^m\), then \(Y^m = X^m - X'^m \in VP\) and \(\lambda^I = Y \cdot \df A^I \) satisfies \cref{eq:cov-const-comm}. Since the connection is an isomorphism between \(V P\) and \(\Omega^0P(\mf g)\), any such choice of \(\lambda^I\) uniquely fixes the ambiguity \(Y^m\).
	\end{proof}
	\end{lemma}

	\begin{remark}\label{rem:global-symm}
	The ambiguity \(Y^m\) in the above lemma corresponds to a \emph{global symmetry} of the chosen connection in the following sense. Taking Lie derivative of the connection with respect to \(Y^m\) we have 
	\[
		\Lie_Y \df A^I = D(Y\cdot \df A^I) = 0
	\]
	i.e. the gauge transformation \(f_Y \in {\rm Aut}_V(P)\) generated by \(Y^m\) keeps the connection invariant and hence \(\df A^I\) and \(f_Y^* \df A^I\) correspond to the same physical field configuration at every point. Connections with such a non-trivial automorphism \(f_Y\) are called \emph{reducible} while connections for which no such non-trivial automorphism exists are called \emph{irreducible} (see \S~4.2.2 \cite{DK-book}). For a compact structure group, the space of irreducible connections is known to be an open and dense subspace of the space of all connections \cite{Sing, GM1, GM2}.
	\end{remark}

	\begin{remark}[Infinitesimal automorphism that preserves some charged fields]\label{rem:aut-field-unique}
	Consider the case where we have both a connection \(\df A^I\) and some charged field \(\sigma^A\) (we consider a scalar field for simplicity) which transforms under a representation \(r\) of the Lie algebra \(\mf g\) on the bundle \(P\). If \(X^m\) is an infinitesimal automorphism that preserves both \(\df A^I\) and \(\sigma^A\) then, in addition to \cref{eq:iso-A}, we have
	\be
		0 = \Lie_X \sigma^A = X \cdot D\sigma^A - (X \cdot \df A^I)~ {r_I}^A{}_B  \sigma^B
	\ee
	This imposes a further restriction on the ambiguity \(Y^m \in \mf{aut}_V(P) \) in the choice of \(X^m\) given by \cref{lem:aut-conn-unique} i.e. \(Y^m\) has to satisfy the additional condition
	\be\label{eq:aut-field-unique}
		0 = (Y \cdot \df A^I) {r_I}^A{}_B  \sigma^B
	\ee
	i.e. the gauge transformation generated by \(Y^m\) keeps both the connection \(\df A^I\) and the field \(\sigma^A\) invariant at every point. The question of whether any non-trivial \(Y^m\) exists depends on the chosen connection \(\df A^I\) and field \(\sigma^A\). As we will see in \cref{lem:aut-e-unique}, for the Lorentz connection \(\df\omega^a{}_b\) and the coframes \(\df e^a\), there is no non-trivial ambiguity.
	\end{remark}

\hr

	To derive a first law for a coframe formulation of gravity we write the orthonormal coframes and the Lorentz gauge field on a principal bundle over spacetime \(M\). Since our treatment is a bit non-standard, we review the relevant constructions in this section.

	Consider the \emph{linear frame bundle} \(FM\) which is a \(GL(d, \bb R)\)-principal bundle over \(M\); the details can be found in \S~I.5 \cite{KN-book1}, \S~III.B.2 and Vbis.A.5. \cite{CDD-book}. The fibre of \(FM\) over any point \(x \in M\) is the set of all possible choices of linearly-independent frames \(E^\mu_a\) at \(x\). When writing Lagrangians that depend explicitly on the frames \(E^\mu_a\) it would be inconvenient to have an explicit dependence on the point in \(FM\). To avoid this, we consider the frames as vector fields (instead of points) on the frame bundle as follows. Locally, at any point \(u = (x, E^\mu_a) \in FM\) and for any \(X^a \in \bb R^d\) define a \({\bb R^d}^*\)-valued vector field \(E^m_a \in T_uFM({\bb R^d}^*)\) by
	\be
		(\pi_*)^\mu_m (X^a E^m_a) = X^a E^\mu_a
	\ee
	This construction can be extended globally to define frames \(E^m_a\) as smooth vector fields on \(FM\). Note we consider two frames \(E^m_a\) and \(E'^m_a\) as equivalent (defined as vector fields on \(FM\)) iff \(E'^m_a - E^m_a \in VP({\mathbb R^d}^*)\) since they project to the same frame \(E^\mu_a\) on \(M\). The frames \(E^m_a\) are non-degenerate on horizontal forms in the sense
	\be\label{eq:non-degen-frame}
	E_a \cdot \df\sigma = 0 \iff \df\sigma = 0 \quad\forall~ \df\sigma \in \Omega^k_{hor}P
	\ee
	We define the coframes \(\df e^a\) on \(FM\) in the standard way as the \emph{canonical form} or \emph{soldering form} (see Prop.~2.1 \S~3.2 \cite{KN-book1}). The coframes are non-degenerate in the sense
	\be\label{eq:non-degen-coframe}
		X \cdot \df e^a = 0 \iff X^m \in VP
	\ee

	Choosing a preferred Lorentzian metric \(\eta_{ab} = {\rm diag}(-1,1,\ldots,1)\) on \(\bb R^d\) the bundle \(FM\) can be reduced to an \emph{orthonormal frame bundle} \(F_OM\) with structure group \(O(d-1,1)\) i.e the \emph{Lorentz group}. Any choice of a reduction of the frame bundle \(FM\) to an orthonormal frame bundle \(F_OM\) gives rise to a metric on \(M\) and conversely a choice of metric \(g_{\mu\nu}\) on \(M\) gives a reduction of the frame bundle to some subbundle of orthonormal frames (see Example~5.7 Ch. I \cite{KN-book1}). The orthonormal frame bundle \(F^{(g)}_OM\) determined by \(g_{\mu\nu}\) then consists precisely of those frames that are orthonormal in the sense
	\be
		g_{\mu\nu} E^\mu_a  E^\nu_b = \eta_{ab}
	\ee

	Thus, to formulate a theory of gravity in terms of orthonormal coframes, it seems one should work on some choice of orthonormal frame bundle \(F^{(g)}_OM\), but such a choice will necessarily give us a fixed metric \(g_{\mu\nu}\) on spacetime. Since the orthonormal coframes \(e^a_\mu\) are dynamical fields of the theory, we do not have an a priori fixed metric on spacetime. Moreover, consider an automorphism \(f\) of the frame bundle \(FM\). In general, the corresponding projection \(\dfM f \in {\rm Diff}(M)\) need not preserve a given metric \(g_{\mu\nu}\) on \(M\). Thus, an arbitrary automorphism \(f \in {\rm Aut}(FM)\) will map the subbundle \(F^{(g)}_OM\) determined by \(g_{\mu\nu}\) to \(F^{(g')}_OM\) determined by \(g'_{\mu\nu} = ({\dfM f^*})^{-1} g_{\mu\nu}\). Thus, it will be problematic to pick a particular orthonormal frame bundle if one wants to consider the coframes as dynamical fields and the action of diffeomorphisms on the dynamical fields of the theory.

	We circumvent this issue as follows. We will consider an abstract principal \emph{Lorentz bundle} \(P_O\) with structure group \(O(d-1,1)\) which has globally well-defined coframes \(\df e^a = e^a_m\) and frames \(E^m_a\), which are non-degenerate in the sense of \cref{eq:non-degen-frame,eq:non-degen-coframe}, similar to the ones defined on the frame bundle above. Once, we have some specific choice of coframes \(\df e^a\) obtained by solving the equations of motion of the theory, we can identify the abstract bundle \(P_O\) with the orthonormal frame bundle \(F^{(g)}_OM\) determined by the solution metric \(g_{\mu\nu} = \eta_{ab} e^a_\mu e^b_\nu\).

	We can similarly construct the \emph{oriented Lorentz bundle} \(P_{SO}\) (by choosing a preferred orientation \(\epsilon_{{a_1}\ldots {a_d}}\)) and the \emph{proper Lorentz bundle} \(P_{SO}^0\) (by choosing a time-orientation) as abstract principal bundles with structure groups \(SO(d-1,1)\) and \(SO^0(d-1,1)\). Henceforth, we will always work on the above defined Lorentz bundle instead of the frame bundle. We will use the Lorentz bundle \(P_O\) to formulate general theories of gravity in terms of the coframes. For the Lagrangian of General Relativity (\cref{sec:palatini-holst}) we will need to introduce an orientation and hence we work on the oriented Lorentz bundle \(P_{SO}\). Similarly to define spinor fields we will need the proper Lorentz bundle \(P_{SO}^0\) (\cref{sec:dirac}).\\

	On the oriented Lorentz bundle \(P_{SO}\) we have a preferred orientation \(\epsilon_{a_1\ldots a_d}\) given by the completely anti-symmetric symbol with \( \epsilon_{01\ldots d-1} \defn 1 \). Using this orientation we can define the \emph{horizontal volume form} on \(P_{SO}\) as
	\be\label{eq:hor-volume}
		\df\varepsilon_d \defn \tfrac{1}{d!}\epsilon_{a_1\ldots a_d} \df e^{a_1} \wedge \ldots \wedge \df e^{a_d} \in \Omega^d_{hor}P_{SO}
	\ee
	which is the lift through \(\pi\) of the volume form \(\dfM{\df\varepsilon}_d \equiv \varepsilon_{\mu_1 \ldots \mu_d}\) on \(M\). For a horizontal form \(\df\sigma^A \in \Omega^k_{hor}P_{SO}(\bb V, R) \) define the \emph{horizontal Hodge dual} \(\star: \Omega^k_{hor}P_{SO}(\bb V, R) \to \Omega^{d-k}_{hor}P_{SO}(\bb V, R)\) as:
	\be\label{eq:hor-Hodge}
		\star\df\sigma^A \defn \frac{1}{(d-k)!k!}\epsilon^{a_1\ldots a_k}{}_{b_1\ldots b_{d-k}}~ \lb(E_{a_k} \cdot \ldots \cdot E_{a_1} \cdot \df\sigma^A \rb)~ \df e^{b_1}\wedge\ldots\wedge \df e^{b_{d-k}}
	\ee
	or in terms of the frame components as
	\be\label{eq:hor-Hodge-comp}
		(\star \sigma)^A_{b_1\ldots b_{d-k}} = \frac{1}{k!}\epsilon^{a_1\ldots a_k}{}_{b_1\ldots b_{d-k}}~ \sigma^A_{a_1\ldots a_k}
	\ee
	Note, that the horizontal Hodge dual maps equivariant horizontal forms to equivariant horizontal forms. It is straightforward to verify that for an invariant form \(\df \sigma\), if \(\df\sigma = \pi^*\dfM{\df\sigma}\) then \(\star\df\sigma = \pi^*(*\dfM{\df\sigma})\) where \(*\) is the Hodge dual acting on differential forms on \(M\).\\

	The \(O(d-1,1)\)-connection on the Lorentz bundle \(P_O\) is \( {\df \omega^a}_b \in \Omega^1P_O(\mf{g}) \) with \(\df \omega^{ab} = - \df \omega^{ba}\). The curvature and torsion of \({\df \omega^a}_b\) are defined by
	\begin{subequations}\begin{align}
		{\df R^a}_b & \defn D{\df \omega^a}_b = d{\df \omega^a}_b + {\df \omega^a}_c \wedge {\df \omega^c}_b \label{eq:F-frame-defn} \\
		\df T^a & \defn D\df e^a = d\df e^a + {\df \omega^a}_b \wedge \df e^b \label{eq:torsion-defn}
	\end{align}\end{subequations}
	and satisfy
	\be\label{eq:torsion-bianchi}
	D\df T^a = {\df R^a}_b \wedge \df e^b
	\ee

	\begin{remark}[Levi-Civita connection]\label{rem:LC-conn}
	In the torsionless case \(\df T^a = 0\), there is a unique Lorentz connection, the \emph{Levi-Civita connection} \(\tilde{\df \omega}^a{}_b\) which can be expressed as
	\be\label{eq:LC-conn}
	\tilde{\df \omega}^a{}_b = - E^{[a}\cdot d \df e^{b]} + \frac{1}{2} \lb( E^a \cdot E^b \cdot d \df e^c \rb) \df e_c
	\ee
	Any Lorentz connection with torsion can be written in terms of the Levi-Civita connection as
	\be
		\df \omega^a{}_b = \tilde{\df \omega}^a{}_b + C_c{}^a{}_b \df e^c
	\ee
	where \emph{contorsion} \(C_{cab} = C_{c[ab]}\) is defined by
	\be\label{eq:contorsion}
		C_{cab} \defn \tfrac{1}{2} \lb(T_{cab} - T_{abc} - T_{bca}  \rb)
	\ee
	with \(T_{cab} = E_b \cdot E_a \cdot \df T_c\) being the frame components of the torsion form \cref{eq:torsion-defn}. One can work with coframes and the contorsion (or the torsion) as independent dynamical fields instead of the coframes and the connection \(\df\omega^a{}_b\). However, the computations are much simpler in the latter case. Thus, in the main body of the paper we will always consider the Lorentz connection \(\df\omega^a{}_b\) as independent of the coframes \(\df e^a\) i.e. we work in the first-order formalism, but we provide \crefrange{eq:LC-conn}{eq:contorsion} for readers interested in the second-order formalism for gravity.
	\end{remark}

	Next we consider the possible automorphisms of the Lorentz bundle \(P_O\) that preserve the orthonormal coframes. From \cref{lem:aut-conn-unique} we know that an infinitesimal automorphism preserving the connection is determined only up to a covariantly constant function but we can show that if \(X^m\) is an infinitesimal automorphism that preserves the coframes, it is completely determined by a Killing vector field on \(M\) as follows (see \cite{RS} for a spacetime version of this result).

	\begin{lemma}[Uniqueness of the infinitesimal automorphism which preserves orthonormal coframes]\label{lem:aut-e-unique}
	Given an automorphism which preserves some chosen orthonormal coframes \(X^m \in \mf{aut}(P_O; \df e^a)\) so that \(\Lie_X \df e^a = 0\), the projection \(X^\mu = (\pi_*)^\mu_m X^m\) is a Killing vector field for the metric \(g_{\mu\nu} = \eta_{ab}e^a_\mu e^b_\nu\) on \(M\) determined by the chosen coframes.

	Further, given any connection \({\df \omega^a}_b\) on \(P_O\), \(X^m\) is uniquely determined by \(X^\mu\) as follows. Denote the Killing form corresponding to the Killing vector field \(X^\mu \) as \(\dfM{\df \xi} \equiv g_{\mu\nu}X^\nu\) and its lift to \(P_O\) as \(\df\xi \defn \pi^*\dfM{\df\xi} = (X \cdot \df e^a)\df e_a \) then the vertical part of \(X^m\) with respect to the chosen connection \({\df \omega^a}_b\) is given by
	\be\label{eq:aut-e-vert}
	X \cdot \df \omega_{ab} = \tfrac{1}{2}E_a \cdot E_b \cdot d\df\xi + (X \cdot \df e^c) C_{cab}
	\ee
	where the contorsion \(C_{cab}\) is defined by \cref{eq:contorsion}.
	\begin{proof}
	For some chosen coframes \(\df e^a\) let, \(X^m \in \mf{aut}(P_O;\df e^a)\) be an infinitesimal automorphism preserving \(\df e^a\) i.e. \(\Lie_X \df e^a = 0\). It immediately follows that the projection \(\dfM X \equiv X^\mu = (\pi_*)^\mu_m X^m\) satisfies
	\be
	\Lie_{\dfM X} (\eta_{ab}e^a_\mu e^b_\nu) = \Lie_{\dfM X} (g_{\mu\nu}) =  0
	\ee
	Thus, \(X^m\) projects to a Killing vector for \(g_{\mu\nu}\).

	Using \cref{eq:Lie-cov-D,eq:torsion-defn} we have
	\be\label{eq:iso-tetrad}
	0 = \Lie_X \df e^a = X \cdot \df T^a + D(X \cdot \df e^a ) - (X \cdot {\df \omega^a}_b ) \df e^b 
	\ee
	and taking the interior product of \cref{eq:iso-tetrad} with \(E^m_c\) we immediately get
	\be\label{eq:iso-tetrad-vert}
		X \cdot \df \omega^{ab} = E^{[b} \cdot D(X \cdot \df e^{a]}) + E^{[b} \cdot X \cdot  \df T^{a]}
	\ee
	 For the Killing form \(\df\xi = (X\cdot \df e^a)\df e_a\) we have
	\be\begin{split}
	& d\df\xi = D\df\xi = D(X\cdot \df e^a)\wedge \df e_a + (X \cdot \df e^a)~ \df T_a \\
\implies & E^a \cdot E^b \cdot d\df\xi  = 2 E^{[b} \cdot D(X \cdot \df e^{a]}) + (X \cdot \df e^c) E^a \cdot E^b \cdot \df T_c 
	\end{split}\ee
	Substituting this into \cref{eq:iso-tetrad-vert} we can write
	\be
		X \cdot \df \omega^{ab} = \tfrac{1}{2}E^a \cdot E^b \cdot d\df\xi - \tfrac{1}{2} (X \cdot \df e^c) E^a \cdot E^b \cdot \df T_c  - E^{[a} \cdot X \cdot \df T^{b]}
	\ee
	Using the frame components of the torsion \(2\)-form and \cref{eq:contorsion}, we get \cref{eq:aut-e-vert}. The right-hand-side depends only on \(X^\mu\) (and its first derivative) and the torsion of the chosen connection. So we see that any \(X^m \in \mf{aut}(P_O;\df e^a)\) is uniquely determined by its projection.
	\end{proof}
	\end{lemma}

	Using \cref{lem:aut-e-unique} we show that for an automorphism preserving some orthonormal coframes, the Lie derivative on the bundle coincides with the Lorentz-Lie derivative of \cite{BG,JM}. We consider a scalar field \(\sigma^a\) that transforms under the local Lorentz transformations for simplicity. The Lie derivative on the bundle with respect to \(X^m \in \mf{aut}(P_O; \df e^a)\) is then (using \cref{eq:Lie-cov-D,eq:aut-e-vert})
	\be\begin{split}
		\Lie_X \sigma^a & = X \cdot D\sigma^a - \lb(\tfrac{1}{2}E_a \cdot E_b \cdot d\df\xi + (X \cdot \df e^c) C_{cab}\rb) \sigma^b
	\end{split}\ee
	Picking a local section and denoting the projection \(\dfM{X} \equiv X^\mu = (\pi_*)^\mu_m X^m\) we get (in the torsionless case)
	\be\label{eq:get-Lor-Lie}\begin{split}
		\hat{\Lie}_{\dfM X} \sigma^a & = X^\mu D_\mu \sigma^a + \lb( E^{a \mu} E^{b \nu} \nabla_{[\mu}X_{\nu]} \rb) \sigma_b \\
		& = \Lie_{\dfM X} \sigma^a + \lb( X^\mu \omega_{\mu}{}^{ab} + E^{\mu [a} e^{b]}_\nu \nabla_{\mu}X^\nu \rb) \sigma_b
	\end{split}\ee

	The first line is (up to differences in sign and factor conventions) the Lie derivative on Lorentz tensors defined in \cite{BG}. In the second line, \(\Lie_{\dfM X}\) is the Lie derivative computed by ignoring the internal index \(a\) and the second term coincides with \(\lambda^a{}_b\) (\cref{eq:Lor-Lie}) used by \cite{JM}.

	\hr

	Next we collect our notational conventions for spinor fields (see \cite{Ko, G1, G2}, Problem 4 of \S Vbis. \cite{CDD-book}, and Ch.1 \cite{CD-book} for details of the construction of spinor fields). We note that these references use the ``mostly minus" signature for the Lorentzian metric but to conform to the earlier sections we stick to the ``mostly plus" signature making appropriate changes in signs according to Remark 3.8 \cite{Ko}. We will stick to the case of a \(3+1\)-dimensional spacetime (for general dimensions and signature see \cite{CD-book}).

	Consider the \emph{Clifford algebra} generated by an identity element \(\id\) and the \emph{Dirac matrices} \( \gamma^a \) which satisfy (see \cite{Ko}):
	\be
		\{ \gamma^a, \gamma^b \} \defn \gamma^a \gamma^b + \gamma^b \gamma^a = -2\eta^{ab}\id
	\ee
	The group \(Spin^0(3,1)\) is embedded in the Clifford algebra according to Definition 2.4 of \cite{Ko}. The Dirac matrices also implement the double cover homomorphism \(Spin^0(3,1) \to SO^0(3,1)\) as detailed in Prop.~2.6 of \cite{Ko}. The complex representation of the Clifford algebra as a matrix group on \(\bb C^4\) (see Theorem 2.2 of \cite{Ko}), induces a representation of \(Spin^0(3,1)\) on \(\bb D \cong \bb C^4\) which is the vector space of \emph{Dirac spinors}. Similarly, we denote the dual vector space of \emph{Dirac cospinors} by \( \bb D^* \cong \bb C^4 \). To avoid a proliferation of indices, we will use the standard ``matrix-type notation" for spinors.

	We denote the \emph{Dirac adjoint} map or \emph{Dirac conjugation} by \(\adj{\phantom{x}}\), so that in our conventions for \(u \in \mathbb D^*\) and \(v \in \mathbb D\)
	\be
		\adj{(uv)} = \adj{v}~\adj{u} \eqsp \adj{(\gamma^a)} = - \gamma^a \eqsp \adj{(\adj{v}v)} = \adj{v}v \in \mathbb R
	\ee

	To consider spinor fields on spacetime we need the notion of a \emph{spin structure} on \(M\) that is a \(Spin^0(3,1)\)-principal fibre bundle \(F_{Spin}M\) of \emph{spin frames} together with a \(2\)-to-\(1\) bundle homomorphism to the proper orthonormal frame bundle \(F^0_{SO}M\) which is equivariant with respect to the double cover map on the respective structure groups. To formulate Dirac fields on a spacetime \(M\) with a fixed metric \(g_{\mu\nu}\) and orientation, we can choose some spin structure \(F_{Spin}M\) corresponding to the proper frame bundle \(F_{SO}^0M\) determined by the given metric and orientation on \(M\). However, as discussed before for the Lorentz bundle, this is problematic when considering theories where the metric (or the coframes) themselves are dynamical fields. As before we circumvent this, by considering a choice of spin bundle \(P_{Spin}\) corresponding to a proper Lorentz bundle \(P_{SO}^0\) in a manner similar to the spin structure \(F_{Spin}M\) corresponding to \(F_{SO}^0M\). Once we have solved the equations of motion to get a metric, we can identify \(P_{Spin}\) with a spin structure \(F_{Spin}M\) given by that metric.

	On the spin bundle \(P_{Spin}\), we can lift the Lorentz connection \( {\df \omega^a}_b \) on \(P^0_{SO}\) to a connection on \(P_{Spin}\) as \( \df \omega_{spin} \defn -\frac{1}{8}\df \omega_{ab}[\gamma^a, \gamma^b] \). We can similarly lift other structures such as the coframes and frames, the horizontal volume form \cref{eq:hor-volume}, and the horizontal Hodge dual \cref{eq:hor-Hodge}.\\

A \emph{Dirac spinor field} \(\Psi \in \Omega^0 P_{Spin}(\bb D)\) is a function on the spin bundle \(P_{Spin}\) valued in the Dirac spinor representation \(\bb D\) and similarly, a \emph{Dirac cospinor field} is \(\Phi \in \Omega^0F_{Spin}M(\bb D^*)\). The spin covariant exterior derivative for \( \Psi \in \Omega^0 P_{Spin}(\mathbb D) \) and \( \Phi \in \Omega^0 P_{Spin}(\mathbb D^*) \) is given by:
	\be
		D\Psi = d\Psi - \frac{1}{8}(\df \omega_{ab}[\gamma^a, \gamma^b])~ \Psi \eqsp	D\Phi = d\Phi + \frac{1}{8}\Phi~ (\df \omega_{ab}[\gamma^a, \gamma^b])
	\ee
	and the \emph{Dirac operator} \(\dirac \) on Dirac spinor fields and cospinor fields is given by
	\be\label{eq:dirac-op}
	\dirac\Psi \defn \gamma^a E_a \cdot D\Psi \eqsp \dirac\Phi \defn (E_a \cdot D\Phi)\gamma^a
	\ee
	Note that in our conventions \(\adj{\dirac\Psi} = - \dirac\adj\Psi\).

	Any infinitesimal automorphism \(X^m \in \mf{aut}(P_{Spin})\) then acts on the Dirac spinor field through the Lie derivative 
	\be\label{eq:Lie-spinor}
	\Lie_X\Psi \defn X \cdot d\Psi = X \cdot D\Psi + \tfrac{1}{8} (X \cdot \df \omega_{ab})[\gamma^a, \gamma^b]\Psi
	\ee

	Consider now an \(X^m\) preserving some orthonormal frames. By \cref{lem:aut-e-unique}, such an \(X^{m}\) always projects to a Killing field \(X^\mu\) of the metric on \(M\) determined by the chosen coframes, and is uniquely determined using \cref{eq:aut-e-vert}. For such vector fields the Lie derivative \cref{eq:Lie-spinor} on the bundle of the Dirac spinor field reads (in the torsionless case)
	\be
		\Lie_X \Psi = X \cdot D\Psi + \tfrac{1}{8} \left( \tfrac{1}{2}E^a \cdot E^b \cdot d\df\xi \right)[\gamma_a, \gamma_b]\Psi
	\ee
	Viewed on spacetime with \(\dfM{X} \equiv X^\mu \) being a Killing field of the given metric, this becomes
	\be\label{eq:get-LK-Lie}
		\Lie_{\dfM X} \Psi = X^\mu D_\mu\Psi - \tfrac{1}{8} \left( \nabla_{[\mu}X_{\nu]} \right)[\gamma^\mu, \gamma^\nu]\Psi
	\ee
	which is \cref{eq:LK-Lie}, the Lie derivative of spinors with respect to the Killing field \(X^\mu\) as defined by Lichnerowicz \cite{Lich} (see also \cite{Kos, BG}).

\subsection{Local and covariant functionals on a principal bundle}\label{sec:loc-cov}

	In constructing physical theories on the bundle we will require that the Lagrangian be a locally and covariantly constructed functional of the dynamical fields on the bundle which we define as follows.
	\begin{definition}[Local and covariant functional]\label{def:loc-cov-func}
	A functional  \(\mc F[\Phi] \) on a \(G\)-principal bundle \(P\) depending on a set of fields \(\Phi\) and finitely many of its derivatives (with respect to an arbitrary derivative operator which is taken to be part of \(\Phi\)) is a local and covariant functional if for any \(f \in {\rm Aut}(P)\) we have
	\be
	(f^* \mc F)[\Phi] = \mc F[f^*\Phi]
	\ee
	where it is implicit that on the right-hand-side \(f\) also acts on the derivatives of \(\Phi\). If \(X^m\) is the vector field generating the automorphism \(f\) then the above equation implies that
	\be
	\Lie_X \mc F[\Phi] = \mc F[\Lie_X \Phi]
	\ee

	Each of \(\mc F\) and \(\Phi\) can have arbitrary tensorial structure on \(P\) and be valued in some representation of the structure group.
	\end{definition}

	For many of the crucial results in the main body we will need to ensure that a closed differential form on \(P\) is in fact, (globally) exact. For instance, such a result is used in the classification of the ambiguities in the symplectic potential (\cref{eq:lambda-amb}), and is needed to ensure that a horizontal (i.e. gauge-invariant) Noether charge exists for any Noether current (\cref{lem:noether-charge}). Under certain assumptions on a differential form \(\df\sigma\) and its dependence on certain fields \(\Phi\), which we detail next, we show that if \(\df \sigma\) is closed then it is exact. We shall show this in direct analogy to Lemma~1 \cite{W-closed} (this result can also be derived using jet bundle methods and the \emph{variational bicomplex}; see Theorem 3.1 \cite{CFT-Anderson}) and the assumptions on \(\df\sigma\) given below are geared towards generalising the algorithm of Lemma~1 \cite{W-closed} to work with differential forms on the bundle \(P\).

	\begin{ass}[Assumptions for \cref{lem:W-lemma}]\label{ass:W-lemma}
		Let \(\Phi = \{ \phi, \psi \}\) be a collection of two types of fields --- \(\phi\) are the ``dynamical fields" and \(\psi\) are the ``background fields" (distinguished by the assumptions listed below), and let \(\df\sigma[\Phi] \in \Omega^p P\) with \(p<d\) (where \(d\) is the dimension of the base space \(M\)) be a \(p\)-form on a principal bundle \(P\) so that
	\begin{asslist}
	\item \(\df\sigma\) is a horizontal form on \(P\) which is invariant under the action of the structure group \(G\) on \(P\) i.e. \(\df\sigma[\Phi] \in \Omega^p_{hor} P\). \label{ass:horizontal}
	\item \(\df\sigma[\Phi]\) is a local and covariant functional of the fields \(\Phi = \{\phi,\psi\}\) as in \cref{def:loc-cov-func}. \label{ass:loc-cov}
	\item The ``dynamical fields" \(\phi\) are sections of a vector bundle over \(P\) which is equivariant under the group action \(G\) on \(P\), and the action of any automorphism \(f \in {\rm Aut}(P)\) on \(\phi\) is linear (i.e. preserves the vector bundle structure). \label{ass:linear-phi}
	\item \(\df\sigma[\Phi]\) depends \emph{linearly} on up to \(k\)-derivatives of the ``dynamical fields" \(\phi\). \label{ass:linear-phi-dphi}
	\end{asslist}
	\end{ass}

	With the above assumptions on \(\df\sigma\) we can prove the following (generalising Lemma~1 \cite{W-closed})

	\begin{lemma}[Generalised Lemma~1 \cite{W-closed}]\label{lem:W-lemma}
	Let \(\df\sigma[\Phi] \in \Omega^p_{hor}P\) and \(\Phi = \{\phi,\psi\}\) be as assumed in \cref{ass:W-lemma}. If \(d\df\sigma[\Phi] = 0\) for all ``dynamical fields" \(\phi\) and any given ``background fields" \(\psi\), then there exists (globally) a differential form \(\df\eta[\Phi] \in \Omega^{p-1}_{hor}P\) which, similarly satisfies \cref{ass:W-lemma} but depends linearly on at most \((k-1)\)-derivatives of \(\phi\) such that \(\df\sigma = d\df\eta\).

	Further, if \(k=0\) i.e. \(\df\sigma\) does not depend on derivatives of the ``dynamical fields" \(\phi\), then \(\df\sigma=0\).
	\begin{proof}
	To begin, we note that connections on a principal bundle \(P\) are not sections of a vector bundle, (see Remark 1, Ch. IV \cite{K-conn}). Thus, any choice of connection on \(P\) would be part of the ``background fields" \(\psi\).\footnote{Note, that even though \cite{W-closed} assumes in the beginning that the ``background fields" \(\psi\) also are sections of a vector bundle, it is not required for the proof of Lemma~1 \cite{W-closed} as discussed towards the end of \S~II \cite{W-closed}.} We assume that a choice of such a connection has been made and denote the corresponding (horizontal) covariant derivative operator on \(P\) by \(D_m\) and the covariant exterior derivative by \(D\).

	By \cref{ass:linear-phi}, the derivative operator \(D_m\) can be used to take horizontal derivatives of \(\phi\); the derivatives of \(\phi\) along the vertical directions are fixed by the equivariance requirement. Furthermore, any antisymmetric derivatives of \(\phi\) can be written in terms of lower order derivatives and the curvature (and possibly torsion on a Lorentz bundle) of the chosen connection; thus, we only need to consider totally symmetrised derivatives of \(\phi\). Thus, using \cref{ass:loc-cov} and \cref{ass:linear-phi-dphi} we can write the \(p\)-form \(\df\sigma\) as
	\be\label{eq:sigma-exp}
		\df\sigma \equiv \sigma_{m_1\ldots m_p} = \sum\limits_{i=0}^{k} S^{(i)}{}_{m_1\ldots m_p}{}^{n_1\ldots n_i}{}_A D_{(n_1} \cdots D_{n_i)} \phi^A
	\ee
	where, each of the tensors \(S^{(i)}\) are local and covariant functionals of the ``background fields" \(\psi\) and we have used an abstract index notation on the ``dynamical fields" \(\phi \equiv \phi^A\). \cref{eq:sigma-exp} is the direct analogue of Eq.~2 \cite{W-closed} on the principal bundle \(P\). Note here that, since all of the \(m\)-indices are horizontal and the \(n,A\)-indices are contracted away, \(\df\sigma\) tranforms as a horizontal form under the action of any automorphism \(f \in {\rm Aut}(P)\) as required in \cref{ass:horizontal}.

	Again using \cref{ass:horizontal} we have
	\be\label{eq:dsigma-exp}
		d\df\sigma = D\df\sigma \equiv (p+1)\sum\limits_{i=0}^{k} D_{[l} \lb\{ S^{(i)}{}_{m_1\ldots m_p]}{}^{n_1\ldots n_i}{}_A D_{(n_1} \cdots D_{n_i)} \phi^A \rb\}
	\ee
	Since, \(d\df\sigma = 0\) for all ``dynamical fields" \(\phi\) and the horizontal symmetrised derivatives of \(\phi\) can be specified independently at any point of \(P\), we can directly apply the arguments of Lemma~1 \cite{W-closed} to \cref{eq:dsigma-exp}. Note, that in each step of the algorithm of Lemma~1 \cite{W-closed}, the \(m\)-indices are always horizontal and the \(n,A\)-indices are contracted away. Thus the algorithm of \cite{W-closed} gives us a horizontal \((p-1)\)-form \(\df\eta\) on \(P\) so that \(\df\sigma = d\df\eta\) where \(\df\eta\) has an expansion similar to \cref{eq:sigma-exp} except that it depends linearly on at most, \((k-1)\)-derivatives of \(\phi\). Thus, the form \(\df\eta\) manifestly satisfies \cref{ass:W-lemma} and the claim of this lemma.

	The algorithm of \cite{W-closed} further shows that when \(k=0\) we have \(\df\sigma = 0\).
	\end{proof}
	\end{lemma}

	We point out that all of \cref{ass:W-lemma} are crucial to prove \cref{lem:W-lemma}. \cref{ass:horizontal} was used in the first equality in \cref{eq:dsigma-exp} to convert the exterior derivative \(d\) to the covariant exterior derivative \(D\) (which does not hold if \(\df\sigma\) is, either not invariant under the \(G\)-action, or not horizontal) to get an expansion in terms of derivatives of \(\phi\) which is the first step in using the algorithm of \cite{W-closed}. As already noted, \cref{ass:loc-cov,ass:linear-phi-dphi} are needed to write down the expansion \cref{eq:sigma-exp}. Finally, \cref{ass:linear-phi} is already used to formulate \cref{ass:linear-phi-dphi}, since a vector bundle structure is necessary for the notion of \(\df\sigma\) depending linearly on \(\phi\) and its derivatives as in \cref{eq:sigma-exp}. Similarly, the assumptions of equivariance of  \(\phi\) and linear action of \(f \in {\rm Aut}(P)\) on \(\phi\) in \cref{ass:linear-phi}, are necessary to ensure that the expansion \cref{eq:sigma-exp} --- and the corresponding expansion for \(\df\eta\) obtained by applying the algorithm of \cite{W-closed} --- give us horizontal forms on \(P\). \cref{ass:linear-phi} also ensures that we can use any connection, as part of the background fields \(\psi\), to define the horizontal covariant derivatives of \(\phi\). This is crucial since only the horizontal derivatives of \(\phi\) can be freely specified at any point of \(P\) and one needs a connection to define horizontal derivatives.

	In our applications of \cref{lem:W-lemma} in the main body of the paper, the ``dynamical fields" \(\phi\) will either be 
	\begin{enumerate*}
	\item the perturbations \(\delta\df\psi^\alpha\) of the dynamical fields \(\df\psi^\alpha\) (\cref{eq:psi-defn}) of the theory which are equivariant horizontal forms valued in some vector space carrying a representation of \(G\), or
	\item infinitesimal automorphisms \(X^m \in \mf{aut}(P)\) of the principal bundle \(P\).
	\end{enumerate*}
	In both cases the ``dynamical fields" \(\phi\) can be considered as sections of vector bundles over \(P\) in accordance with \cref{ass:linear-phi}.



\bibliographystyle{JHEP}
\bibliography{noether-wald-entropy}      
\end{document}